\documentclass[aps,prc,twocolumn,10pt, superscriptaddress, showpacs, floatfix]{revtex4-1}

\usepackage{amssymb,epsfig}

\newcommand{\be}{\begin{equation}}
\newcommand{\ee}{\end{equation}}
\newcommand{\bea}{\begin{eqnarray}}
\newcommand{\eea}{\end{eqnarray}}

\begin{document}

\title{Toward an accurate strongly coupled many-body theory within the equation-of-motion framework}
%
\author{Elena Litvinova}
\affiliation{Department of Physics, Western Michigan University, Kalamazoo, MI 49008, USA}
\affiliation{National Superconducting Cyclotron Laboratory, Michigan State University, East Lansing, MI 48824, USA}
\author{Peter Schuck}
\affiliation{Institut de Physique Nucl\'eaire, IN2P3-CNRS, Universit\'e Paris-Sud, F-91406 Orsay Cedex, France}
\affiliation{Universit\'e Grenoble Alpes, CNRS, LPMMC, 38000 Grenoble, France}


\date{\today}

\begin{abstract}
Non-perturbative aspects of the quantum many-body problem are revisited, discussed and advanced in the equation of motion framework. 
We compare the approach to the two-fermion response function truncated on the two-body level by the cluster expansion of the dynamical interaction kernel to the approach known as time blocking approximation. Such a comparison leads to an extended many-body theory with non-perturbative treatment of high-order configurations.
The present implementation of the advanced theory introduces a new class of solutions for the response functions, which include explicitly beyond-mean-field correlations between up to six fermions. The novel approach, which includes configurations with two quasiparticles coupled to two phonons  (2q$\otimes$2phonon), is discussed in detail for the particle-hole nuclear response and applied to medium-mass nuclei. The proposed developments are implemented numerically on the basis of the relativistic effective meson-nucleon Lagrangian and compared to the models confined by two-fermion and four-fermion configurations, which are considered as state-of-the-art for the response theory in nuclear structure calculations. The results obtained for the dipole response of $^{42,48}$Ca and $^{68}$Ni nuclei in comparison to available experimental data show that the higher configurations are necessary for a successful description of both gross and fine details of the spectra
in both high-energy and low-energy sectors.

%
\end{abstract}

\maketitle

\section{Introduction} 

Response of many-body quantum systems to external perturbations encompasses a large class of problems actively studied in various areas of quantum physics. The observed characteristics of response are associated with quantum  
correlation functions, which encode the complete information about strongly coupled complex media.  The general notion of correlation functions forms the common underlying background connected across the fields of quantum chromodynamics, dynamics of hadrons, astrophysics, condensed matter, solid state, atomic, molecular, nuclear structure physics, and formal aspects of quantum field theory (QFT).  Thus, advancements toward an exact theory of correlations in strongly coupled quantum systems have potentially a broad impact on many fields of research.

Response of atomic nuclei to various external perturbations represents a very rich playground to study correlation functions due to the availability of numerous experimental probes, which was further extended with the advent of rare beam facilities \cite{Tanihata1998,Glasmacher2017}.  It has a long history of theoretical studies based on QFT. It was recognized rather early that the random phase approximation (RPA) \cite{Bohm1951} appears as a good approach to the gross features of nuclear spectra, such as the positions and total strengths of collective excitations, however, the response theory which, in principle, should provide the full spectral composition has to be extended by configurations beyond one-particle one-hole ($1p-1h$) ones considered in RPA. The idea of coupling between single-particle and emergent collective degrees of freedom  in nuclei \cite{BohrMottelson1969,BohrMottelson1975,Broglia1976,BortignonBrogliaBesEtAl1977,BertschBortignonBroglia1983,Soloviev1992}, which explained successfully many of the observed phenomena, was, in fact, linked to the non-perturbative versions of QFT-based and, in principle, exact, equations of motion for correlation functions in nuclear medium \cite{RingSchuck1980}.
Indeed, the equation of motion method developed in Ref. \cite{Rowe1968} and further elaborated, e.g., in Refs. \cite{Schuck1976,AdachiSchuck1989,Danielewicz1994,DukelskyRoepkeSchuck1998,SchuckTohyama2016}, was shown to produce a hierarchy of approximations to the dynamical kernels of the equations for one-fermion and two-time two-fermion propagators. In particular, the non-perturbative versions of those kernels, which include full resummations in the particle-hole and particle-particle channels, can be mapped to the kernels of the phenomenological nuclear field theories (NFT), where such kernels are commonly referred to as particle-vibration coupling (PVC) or quasiparticle-phonon models (QPM) \cite{BohrMottelson1969,BohrMottelson1975,Broglia1976,BortignonBrogliaBesEtAl1977,BertschBortignonBroglia1983,Soloviev1992}. This mapping, although it has to be corrected for the accurate lowest-order limit, provides an understanding of the emergent collective phenomena, explaining the mechanism of their formation from the underlying strongly interacting degrees of freedom. Moreover, the EOM method gives clear insights into the relationship between the bare nucleon-nucleon interaction and its modification in the strongly coupled medium. It reveals, in particular, that the latter is not reducible to static 'potentials', but splits into a static part calculable from the bare interactions beyond the Hartree-Fock approximation \cite{SchuckTohyama2016} and a well-defined dynamical component \cite{AdachiSchuck1989,DukelskyRoepkeSchuck1998}.

\begin{table*}[!htbp]
\caption
{Table of symbols}
\vspace{3mm}
\tabcolsep=0.7em
\renewcommand{\arraystretch}{1.5}%
{\begin{tabular}{@{}|c|c|c|@{}}
\hline\hline 
Symbol & Name & Equations and Figures
\\
\hline\hline
$G_{11'}(t-t')$ & One-fermion in-medium propagator & Eqs. (\ref{spgf},\ref{spgfspec})\\
\hline
$G_{12,1'2'}(t-t'), G^{pp}_{12,1'2'}(\omega)$ & Two-time particle-particle in-medium propagator &Eqs. (\ref{mbgfs},\ref{resppp}) \\
\hline
$H$ & Many-body Hamiltonian & Eq. (\ref{Hamiltonian})\\
\hline
${v}_{1234}, {\bar v}_{1234}$& Non-symmetrized and antisymmetrized two-fermion bare interaction & Eq.(\ref{Hamiltonian2})\\
\hline 
$R_{12,1'2'}(t-t'), R_{12,1'2'}(\omega)$ & Two-time particle-hole in-medium propagator (response function) & Eqs. (\ref{phresp},\ref{respspec}) \\
\hline 
$T_{11'}(t-t')$ & One-fermion T-matrix & Eq. (\ref{Toperator}) \\
\hline 
$\Sigma_{11'}(t-t'),  \Sigma_{11'}(\omega)$ & Irreducible one-fermion self-energy & Eq. (\ref{Somega}) \\
\hline 
$\Sigma_{11'}^{(0)}$ & Static part of the one-fermion self-energy &  Eq. (\ref{MF}) \\
 \hline
$\Sigma_{11'}^{(r)}(t-t'), \Sigma_{11'}^{(r)}(\omega)$ & Dynamical part of the one-fermion self-energy & Eqs. (\ref{SEirr},\ref{SEomega}), Figs. \ref{SEirrd},\ref{SEdyn} \\
\hline
${\cal N}_{121'2'}$  & Two-fermion norm kernel & Eq. (\ref{norm})\\
\hline
$R^{(0)}_{12,1'2'}(\omega)$ & Uncorrelated particle-hole propagator & Eq. (\ref{resp0})\\
\hline
$T_{12,1'2'}(t-t')$ & Two-time particle-hole T-matrix & Eq. (\ref{F1})\\
\hline
$K_{12,1'2'}(t-t')$ & Irreducible two-time particle-hole interaction kernel & Eqs. (\ref{Dyson2},\ref{Kkernel})\\
\hline
$K^{(0)}_{12,1'2'}$ & Static part of the particle-hole interaction kernel & Eqs. (\ref{Womega},\ref{Fstatic1}), Fig. \ref{SE2static}\\
\hline
$K^{(r)}_{12,1'2'}(t-t')$ & Dynamical part of the particle-hole interaction kernel & Eqs. (\ref{Womega},\ref{Fr}-\ref{Fcomponents22}), Figs. \ref{SE2irrtot}-\ref{SE2irrcc}\\
\hline
$\tilde V_{1234}$ &Static effective two-fermion interaction & Eq. (\ref{effint})\\
\hline
 $V^{(e)}$(12,34) &Energy-dependent effective two-fermion interaction & Eq. (\ref{effint})\\
 \hline
 $\tilde\Sigma_{11'}$ & Static part of the one-fermion self-energy in PVC approaches & Eq. (\ref{PVCsigma}) \\
 \hline
$\Sigma^{(e)}_{11'}(\omega)$ & Dynamical part of the one-fermion self-energy in PVC approaches & Eq. (\ref{PVCsigma}) \\
 \hline
$\Phi_{12,1'2'}(\omega)$ & Dynamical particle-hole interaction kernel in PVC-TBA & Eq. (\ref{Phikernel}), Fig. \ref{Phi_PVC}\\
\hline
$\Phi^{(n)}_{12,1'2'}(\omega)$& Dynamical particle-hole interaction kernel in EOM/R(Q)TBA$^{(n)}$& Eqs. (\ref{phires2ex},\ref{respn}), Fig. \ref{Phiiter} \\
\hline
\hline
\end{tabular}}
\label{tab}
\end{table*}

Although these aspects were widely ignored in the majority of semi-phenomenological PVC models based on effective in-medium interactions over the years \cite{Bortignon1978,Bortignon1981a,BertschBortignonBroglia1983,MahauxBortignonBrogliaEtAl1985,Bortignon1986,Bortignon1997,ColoBortignon2001,Tselyaev1989,KamerdzhievTertychnyiTselyaev1997,Ponomarev2001,Ponomarev1999b,LoIudice2012}, these and other approaches provided invaluable knowledge about the importance of coupling between single-nucleon and collective degrees of freedom in nuclear structure. Lately, this type of approaches was linked to the contemporary density functional theories \cite{LitvinovaRing2006,LitvinovaRingTselyaev2008,LitvinovaRingTselyaev2010,Tselyaev2013,AfanasjevLitvinova2015,Tselyaev2016,NiuNiuColoEtAl2015,Niu2018}, advancing the PVC models to self-consistent frameworks, and applied to experimental data analyses \cite{EndresLitvinovaSavranEtAl2010,PoltoratskaFearickKrumbholzEtAl2014,MassarczykSchwengnerDoenauEtAl2012,LanzaVitturiLitvinovaEtAl2014,Oezel-TashenovEndersLenskeEtAl2014}.

However, very little progress has been made on the conceptual advancements of the many-body aspects of the non-perturbative NFT's.  Some specific rare topics, such as PVC with charge-exchange phonons \cite{Litvinova2016,RobinLitvinova2018} or PVC-induced ground state correlations \cite{Kamerdzhiev1991,Tselyaev1989,KamerdzhievTertychnyiTselyaev1997,Robin2019}, have been addressed in the time blocking approximation (TBA), however, relatively little effort has been made on developments and numerical implementations of NFT's beyond the two-particle two-hole $2p-2h$ level \cite{Tselyaev2018,Litvinova2015}, although the phenomenological multiphonon approach \cite{Soloviev1992,Ponomarev2001,SavranBabilonBergEtAl2006,Andreozzi2008}  indicates the possibility to meet the shell-model standards in large model spaces. Also, the problem of consistent linking those approaches to the underlying bare interactions remains unsolved. 

The goal of the present work is to bring together the past and recent developments of the EOM method and to compare them to the PVC approach in the time blocking approximation. As the EOM method starts from the bare interaction between fermions, while the existing versions of PVC models for atomic nuclei are based on the effective static interactions, it only makes sense to compare the dynamical kernels of the equations for the two-fermion propagators. We will show that such a comparison allows one to recognize and justify the use of cluster expansions truncated on the two-body level in the PVC models and to substantiate their extensions to higher-order configurations.  We will also discuss another important feature of the EOM method, namely its capability of deriving the emergence of the long-range correlations and the corresponding collective degrees of freedom from the bare short-range correlations in a parameter-free way. Eventually, by confronting the EOM method with the class of approaches of the PVC type, we develop a non-perturbative, consistent and systematically improvable theory for one- and two-fermionic correlation functions, which is directly based on the bare two-fermion interactions. 

We want to keep the presentation sufficiently general to be useful also in other than just the nuclear physics field. The dynamical aspects of the in-medium fermionic correlations, in particular, the PVC are surely also of prime importance in electronic condensed matter systems where, e.g., we know that particle-phonon coupling can even reverse the sign of the repulsive Coulomb interaction to give rise to superconductivity \cite{50BCSorig,50BCS}. Also, even though in atoms and molecules collective features are not well born out, particle-hole correlations are important in screening the Coulomb force \cite{Tiago2008,Martinez2010,Sangalli2011}. Similar techniques as the one we outline here have indeed been presented recently in the field of chemical physics  \cite{Olevano2018}.

In Sections \ref{Propagators} - \ref{EOM2} the EOM's are reviewed following the formalism of Refs. \cite{Schuck1976,AdachiSchuck1989,Danielewicz1994,DukelskyRoepkeSchuck1998,SchuckTohyama2016}, Section \ref{PVC-TBA} discusses the PVC model in the time blocking approximation (PVC-TBA) in terms of Refs.  \cite{KamerdzhievTertychnyiTselyaev1997,LitvinovaRingTselyaev2007} and its mapping to the EOM, Section \ref{PVCext} is devoted to a non-perturbative advancement of the theory for the particle-hole response beyond $2p-2h$ configurations, and Section \ref{Results} presents the first numerical implementations of the extended theory. Finally, Section \ref{summary} provides conclusions and outlook.
In Table \ref{tab} we collect the most important symbols used in the article with the references to the corresponding equations and figures, where they are introduced and employed, to help navigate throughout the text.

\section{One-fermion and two-fermion propagators in a correlated medium}
\label{Propagators}
A convenient way of describing a strongly correlated many-body system is calculating various correlation functions, or
propagators, which are directly linked to observables.
For instance, the one-fermion in-medium propagator, or Green function, is defined as follows:
\be
G(1,1') \equiv G_{11'}(t-t') = -i\langle T \psi(1){\psi^{\dagger}}(1') \rangle,
\label{spgf}
\ee
where $T$ is the chronological ordering operator, $\psi(1),{\psi^{\dagger}}(1)$ are one-fermion (for instance, one-nucleon) fields:
\be
\psi(1) = e^{iHt_1}\psi_1e^{-iHt_1}, \ \ \ \ \ \ {\psi^{\dagger}}(1) = e^{iHt_1}{\psi^{\dagger}}_1e^{-iHt_1},
\ee
and the subscript '1' denotes the full set of the one-fermion quantum numbers in an arbitrary representation. 
The fermionic fields obey the usual anticommutation relations:
\bea
[\psi_1,{\psi^{\dagger}}_{1'}]_+ \equiv \psi_1{\psi^{\dagger}}_{1'}  +  {\psi^{\dagger}}_{1'}\psi_1 = \delta_{11'}, \nonumber \\
\left[ \psi_1,{\psi}_{1'} \right]_{+} \equiv \psi_1{\psi}_{1'}  +  {\psi}_{1'}\psi_1 = 0,\nonumber\\
\left[ {\psi^{\dagger}}_1,{\psi^{\dagger}}_{1'}\right]_+ \equiv {\psi^{\dagger}}_1{\psi^{\dagger}}_{1'}  +  {\psi^{\dagger}}_{1'}{\psi^{\dagger}}_1 = 0.
\label{anticomm}
\eea
The averaging in Eq. (\ref{spgf}) $\langle ... \rangle$ is performed over the formally exact ground state of the many-body system of $N$ particles described by the Hamiltonian $H$:
\be
H = H^{(1)} + V^{(2)} + W^{(3)} + ...
\label{Hamiltonian}
\ee
Here, the operator $H^{(1)}$ is the one-body contribution to the Hamiltonian:
\be
H^{(1)} = \sum_{12} t_{12} \psi^{\dag}_1\psi_2 + \sum_{12}v^{(MF)}_{12}\psi^{\dag}_1\psi_2 \equiv \sum_{12}h_{12}\psi^{\dag}_1\psi_2,
\label{Hamiltonian1}
\ee
whose matrix elements $h_{12}$, in general, combine the kinetic energy $t$ and the mean-field $v^{(MF)}$ part  of the interaction. The operator $V^{(2)}$ describes the two-body sector associated with the two-fermion interaction
\be
V^{(2)} = \frac{1}{4}\sum\limits_{1234}{\bar v}_{1234}{\psi^{\dagger}}_1{\psi^{\dagger}}_2\psi_4\psi_3,
\label{Hamiltonian2}
\ee
the operator $V^{(3)}$ generates the three-body forces
\be
W^{(3)} = \frac{1}{36}\sum\limits_{123456}{\bar w}_{123456}{\psi^{\dagger}}_1{\psi^{\dagger}}_2{\psi^{\dagger}}_3\psi_6\psi_5\psi_4
\ee
with the antisymmetrized matrix elements ${\bar v}_{1234}$ and ${\bar w}_{123456}$, respectively, and so on. We will make an explicit derivation of the equations of motion assuming that the Hamiltonian is confined by the two-body interaction, however, the theory can be naturally extended to multiparticle forces. The inclusion of a three body force, as may be necessary in nuclear physics for an ab initio approach, can formally be handled straightforwardly, but at the cost of inflating the amount of formulas. Therefore, we will refrain from doing this here, but may consider an extension to the case of the presence of the three-body forces in a future work.


It is often convenient to work in the 
basis, which diagonalizes the one-body (also named single-particle) part of the Hamiltonian (\ref{Hamiltonian1}), so that $h_{12} =  \delta_{12}\varepsilon_1$.
We will use this basis from the beginning, while on the way to the final equations of motion it will be redefined as soon as the mean-field part of the Hamiltonian will absorb additional contributions from the two-body sector. 
Furthermore, for the case of self-bound finite Fermi systems the common practice is to work with intrinsic Hamiltonians which are obtained by isolating the kinematics of the center of mass from the relative motion. In this case the two-body part of the Hamiltonian absorbs the kinetic energy of the relative motion, see for instance, Ref. \cite{Hergert2016} and references therein for details.

The single-particle propagator (\ref{spgf}) depends explicitly on a single time difference $\tau = t-t'$, and the Fourier transform with respect to $\tau$ to the energy domain leads to the spectral (Lehmann) expansion:
\bea
G_{11'}(\varepsilon) = \sum\limits_{n}\frac{\eta^{n}_{1}\eta^{n\ast}_{1'}}{\varepsilon - (E^{(N+1)}_{n} - E^{(N)}_0)+i\delta} +  \nonumber \\
+ \sum\limits_{m}\frac{\chi^{m}_{1}\chi^{ m\ast}_{1'}}{\varepsilon + (E^{(N-1)}_{m} - E^{(N)}_0)-i\delta}. 
\label{spgfspec}
\eea
This expansion is a sum of simple poles with the residues composed of matrix elements of the field operators between the ground state $|0^{(N)}\rangle$ of the $N$-particle system and states $|n^{(N+1)} \rangle$ and $|m^{(N-1)} \rangle$ of the neighboring systems with $N+1$ and $N-1$ particles, respectively:
\be
\eta^{n}_{1} = \langle 0^{(N)}|\psi_1|n^{(N+1)} \rangle , \ \ \ \ \ \ \ \  \chi^{m}_{1} = \langle m^{(N-1)}|\psi_1|0^{(N)} \rangle .
\ee
By definition, these matrix elements give the weights of the given single-particle (single-hole) configuration on top of the ground state $|0^{(N)}\rangle$ in the $n$-th ($m$-th) state of the $(N+1)$-particle ($(N-1)$-particle) system, respectively. The residues are associated with the observable occupancies of the corresponding states, often called spectroscopic factors.
The poles are located at the energies $E^{(N+1)}_{n} - E^{(N)}_0$ and $-(E^{(N-1)}_{m} - E^{(N)}_0)$ of those systems, respectively, i.e., related to the ground state of the $N$-particle system.

The two-fermion, three-fermion and, in general, $n$-fermion propagators,  or Green functions, are defined in analogy to Eq. (\ref{spgf}):
\bea
&G&(12,1'2') = (-i)^2\langle T \psi(1)\psi(2){\psi^{\dagger}}(2'){\psi^{\dagger}}(1')\rangle, 
\label{ppGF} \\
&G&(123,1'2'3') = (-i)^3\langle T \psi(1)\psi(2)\psi(3){\psi^{\dagger}}(3'){\psi^{\dagger}}(2'){\psi^{\dagger}}(1')\rangle, \nonumber \\
&G&(12...n,1'2'...n') = \nonumber \\ &=&(-i)^n\langle T \psi(1)\psi(2)...\psi(n){\psi^{\dagger}}(n')...{\psi^{\dagger}}(2'){\psi^{\dagger}}(1')\rangle.
\label{mbgfs}
\eea

The response of a many-body system to external perturbations, which can be associated with one-body operators,  is expressed via 
the two-time two-fermion particle-hole propagator (response function):
\bea
R(12,1'2') &\equiv& R_{12,1'2'}(t-t') = -i\langle T(\psi^{\dagger}_1\psi_2)(t)(\psi^{\dagger}_{2'}\psi_{1'})(t')\rangle \nonumber \\ &=& -i\langle T\psi^{\dagger}(1)\psi(2)\psi^{\dagger}(2')\psi(1')\rangle,
\label{phresp}
\eea
where we imply that $t_1 = t_2 = t, t_{1'} = t_{2'} = t'$. 
The Fourier transformation of Eq. (\ref{phresp}) to the energy (frequency) domain leads to the spectral expansion
\be
R_{12,1'2'}(\omega) = \sum\limits_{\nu>0}\Bigl[ \frac{\rho^{\nu}_{21}\rho^{\nu\ast}_{2'1'}}{\omega - \omega_{\nu} + i\delta} -  \frac{\rho^{\nu\ast}_{12}\rho^{\nu}_{1'2'}}{\omega + \omega_{\nu} - i\delta}\Bigr]
\label{respspec}
\ee
which, similarly to the one for the one-fermion propagator (\ref{spgfspec}), satisfies the general quantum field theory requirements of locality and unitarity. The residues of this expansion 
are products of the matrix elements
\be
\rho^{\nu}_{12} = \langle 0|\psi^{\dagger}_2\psi_1|\nu \rangle 
\ee
called transition densities, which are properly normalized and represent the weights of the pure particle-hole configurations on top of the ground state $|0\rangle$ in the model (ideally, exact) excited states $|\nu\rangle$ of the same N-particle system. The poles are the excitation energies $\omega_{\nu} = E_{\nu} - E_0$ relative to the ground state.

The response to the probes with pair transfer is associated with the two-time two-fermion Green function (\ref{ppGF}) whose spectral expansion reads
\be
iG^{pp}_{12,1'2'}(\omega) =  \sum\limits_{\mu} \frac{\alpha^{\mu}_{21}\alpha^{\mu\ast}_{2'1'}}{\omega - \omega_{\mu}^{(++)}+i\delta} - \sum\limits_{\varkappa} \frac{\beta^{\varkappa\ast}_{12}\beta^{\varkappa}_{1'2'}}{\omega + \omega_{\varkappa}^{(--)}-i\delta},
\label{resppp}
\ee
where the residues are composed of the matrix elements
\be
\alpha_{12}^{\mu} = \langle 0^{(N)} | \psi_2\psi_1|\mu^{(N+2)} \rangle , \ \ \ \ \ \ \beta_{12}^{\varkappa} = \langle 0^{(N)} | \psi^{\dagger}_2\psi^{\dagger}_1|\mu^{(N-2)} \rangle
\ee
and the poles $\omega_{\mu}^{(++)}$ and  $\omega_{\varkappa}^{(--)}$ are formally exact states of the $(N+2)$- and $(N-2)$-particle systems, respectively.
The spectral expansions of Eqs. (\ref{spgfspec},\ref{respspec},\ref{resppp}) are model independent and valid for any physical approximations to the many-body states $|n\rangle, |m\rangle$, $|\nu\rangle$, $|\mu\rangle$, and $|\varkappa\rangle$. The sums in Eqs. (\ref{spgfspec},\ref{respspec},\ref{resppp}) are formally complete, i.e., run over the discrete spectra and imply continuous integrals over the continuum states.

Thus, one can see that, indeed, due to their direct links to the observables, both one-fermion and two-fermion  Green functions as well as the particle-hole response function are of high importance for the characterization of strongly correlated quantum many-body systems, in particular, of atomic nuclei.
In the next two sections we consider equations of motion for the one-fermion  Green function $G_{11'}(t-t')$ and for  the particle-hole response $R_{12,1'2'}(t-t')$  and outline possible strategies on the way to their accurate descriptions. As the correlation functions provide, in principle, complete characteristics of quantum many-body systems, essentially the same theory is applicable to atomic physics, quantum chemistry, condensed matter and other areas of physics dealing with fermionic systems.

\section{Equation of motion for one-fermion propagator}
\label{EOM1}
\subsection{General formalism}
Let us consider the time evolution of the  one-fermion propagator (\ref{spgf}). Taking the derivative with respect to $t$, one obtains
\bea
\partial_t G_{11'}(t-t') = -i\delta(t-t')\langle [\psi_1(t),{\psi^{\dagger}}_{1'}(t')]_+\rangle + \nonumber \\
+ \langle T[H,\psi_1](t){\psi^{\dagger}}_{1'}(t')\rangle, \nonumber\\ 
\label{dtG}                           
\eea
where we defined
\be
[H,\psi_1](t) = e^{iHt}[H,\psi_1]e^{-iHt}.
\ee
With the help of the commutator
\be
[H,\psi_1] = -\varepsilon_1\psi_1 + [V,\psi_1],
\ee
assuming that $V \equiv V^{(2)}$ and discarding the three-body interaction,
Eq. (\ref{dtG}) leads to
\be
(i\partial_t - \varepsilon_1)G_{11'}(t-t') = \delta_{11'}\delta(t-t') + i\langle T[V,\psi_1(t)]{\psi^{\dagger}}_{1'}(t')\rangle.
\label{spEOM}
\ee
Let us introduce a function $R_{11'}(t-t')$ corresponding to the last term on the right hand side of Eq. (\ref{spEOM}):
\be
R_{11'}(t-t') =  i\langle T[V,\psi_1](t){\psi^{\dagger}}_{1'}(t')\rangle .
\ee
As we will see in the following, it is useful to determine the equation of motion  for this function with respect to $t'$:
\bea
R_{11'}(t-t')\overleftarrow{\partial_{t'}} 
&=& -i\delta(t-t') \langle \bigl[[V,\psi_1](t),{\psi^{\dagger}}_{1'}(t')\bigr]_+\rangle - \nonumber \\ 
&-& \langle T[V,\psi_1](t)[H,{\psi^{\dagger}}_{1'}](t')\rangle.
\eea
Using the commutator
\be
[H,{\psi^{\dagger}}_{1'}] = \varepsilon_{1'}{\psi^{\dagger}}_{1'} + [V,{\psi^{\dagger}}_{1'}],
\ee
one arrives at the EOM for $R_{11'}(t-t')$:
\bea
R_{11'}(t-t')(-i\overleftarrow{\partial_{t'}}  - \varepsilon_{1'}) &=& -\delta(t-t')\langle \bigl[ [V,\psi_1](t),{\psi^{\dagger}}_{1'}(t')\bigr]_+\rangle\nonumber \\
&+& i\langle T [V,\psi_1](t)[V,{\psi^{\dagger}}_{1'}](t')\rangle.
\label{EOMR}
\eea
Combining it with the first EOM (\ref{spEOM}) and performing the Fourier transformation to the energy (frequency) domain with respect to the time difference $t-t'$, we obtain
\be
G_{11'}(\omega) 
= G^{(0)}_{11'}(\omega) + 
\sum\limits_{22'}G^{(0)}_{12}(\omega)T_{22'}(\omega)G^{(0)}_{2'1'}(\omega),
\label{spEOM3}
\ee
where we introduced the free (uncorrelated) one-fermion propagator $G^{(0)}_{11'}(\omega) = \delta_{11'}/(\omega - \varepsilon_1)$
and the interaction kernel (one-body T-matrix, not to be confused with the time ordering operator):
\bea
T_{11'}(t-t') &=& T^{(0)}_{11'}(t-t') + T^{(r)}_{11'}(t-t'), \nonumber\\
T^{(0)}_{11'}(t-t') &=& -\delta(t-t')\langle \bigl[ [V,\psi_1](t),{\psi^{\dagger}}_{1'}(t')\bigr]_+\rangle, \nonumber \\ 
T^{(r)}_{11'}(t-t') &=&  i\langle T [V,\psi_1](t)[V,{\psi^{\dagger}}_{1'}](t')\rangle.
\label{Toperator}
\eea
Here and in the following we use the superscript "(0)" to denote the static parts of the interaction kernels and "(r)" for their dynamical time-dependent parts, which  are associated 
with {\it retardation} effects in our approach. 
The EOM (\ref{spEOM3}) which, in the operator form, is
\be
G(\omega) = G^{(0)}(\omega) + G^{(0)}(\omega)T(\omega)G^{(0)}(\omega),
\label{spEOM4}
\ee
can be transformed to the Dyson equation: 
\be
G(\omega) = G^{(0)}(\omega) + G^{(0)}(\omega)\Sigma(\omega) G(\omega)
\label{Dyson}
\ee
with the interaction kernel $\Sigma(\omega)$, such as
\be
T(\omega) = \Sigma(\omega) + \Sigma(\omega) G^{(0)}(\omega)T(\omega),
\label{DysonT}
\ee
from which it follows that the operator $\Sigma$ represents the one-fermion self-energy (also called mass operator) as the irreducible (with respect to one-fermion line) part of the kernel $T$: $\Sigma = T^{irr}$. Analogously to Eq. (\ref{Toperator}), the self-energy is decomposed into the instantaneous mean-field part $\Sigma^{(0)}$ and the energy-dependent dynamical part  $\Sigma^{(r)}(\omega)$:
\be
\Sigma_{11'}(\omega) = \Sigma_{11'}^{(0)} + \Sigma_{11'}^{(r)}(\omega).
\label{Somega}
\ee
Notice here that the decomposition of the 
kernels (\ref{Toperator},\ref{Somega}) into the static and time- (energy-) dependent, or dynamical, parts is a generic feature 
 and the direct consequence of the time-independence of the bare interaction $V$ of Eq. (\ref{Hamiltonian2}). 

The first static (instantaneous) terms of both kernels coincide and read:
\bea
T^{(0)}_{11'}(t-t') &=& -\delta(t-t')\langle \bigl[ [V,\psi_1](t),{\psi^{\dagger}}_{1'}(t')\bigr]_+\rangle = \nonumber\\
&=& -\delta(t-t')\langle \bigl[ [V,\psi_1],{\psi^{\dagger}}_{1'}\bigr]_+\rangle.
\eea
Here we need first to evaluate the commutator $[V,\psi_1]$ which, with help of the anticommutation relations (\ref{anticomm}), can be obtained as
\be
[V,\psi_1] =  \frac{1}{2}\sum\limits_{ikl}{\bar v}_{i1kl}{\psi^{\dagger}}_i\psi_l\psi_k ,
\label{psiVcomm}
\ee
where the Latin indices have the same meaning as the number indices and the definition of the antisymmetrized interaction matrix elements ${\bar v}_{1234} = v_{1234} - v_{1243}$ was taken into account.
Evaluating the anticommutator
\be
[{\psi^{\dagger}}_j\psi_l\psi_k,{\psi^{\dagger}}_{1'}]_+ = {\psi^{\dagger}}_j\psi_l\delta_{1'k} - {\psi^{\dagger}}_j\psi_k\delta_{1'l}, 
\ee
one gets:
\be
[[V,\psi_1],{\psi^{\dagger}}_{1'}]_+ = - \sum\limits_{il} {\bar v}_{1i1'l}\psi_{i}^{\dagger}\psi_l. 
\ee
Thus, the first (instantaneous) part $\Sigma^{(0)}$ of the mass operator (\ref{Somega}) is associated with the mean field contribution:
\be
 \Sigma^{(0)}_{11'} = -\langle[[V,\psi_1],{\psi^{\dagger}}_{1'}]_+\rangle = \sum\limits_{il}{\bar v}_{1i1'l}\rho_{li}, 
 \label{MF}
\ee
where  $\rho_{li} = \langle{\psi^{\dagger}}_i\psi_l\rangle$  is the ground-state one-body density
and we have applied the (anti)symmetry properties of the antisymmetrized interaction matrix elements: ${\bar v}_{1234} = -{\bar v}_{1243} = -{\bar v}_{2134} = {\bar v}_{2143}$. 
%
The second (dynamical) part $\Sigma^{(r)}(\omega)$ of the mass operator comprises all retardation effects induced by the nuclear medium. 

In order to understand the dynamical part $\Sigma^{(r)}(\omega)$ of the self-energy $\Sigma(\omega)$, let us first evaluate its reducible counterpart $T^{(r)}_{11'}(t-t')$.
Here we can use the result of Eq. (\ref{psiVcomm}) for the commutator $[V,\psi_1]$, and the following result for the second commutator:
\be
[V,\psi^{\dagger}_{1'}]  =  \frac{1}{2}\sum\limits_{mnq}{\bar v}_{mn1'q}\psi^{\dagger}_m\psi^{\dagger}_n\psi_q, 
\ee
so that
\bea
&T&^{(r)}_{11'}(t-t') =\nonumber \\
&=&  -\frac{i}{4} \sum\limits_{ikl}\sum\limits_{mnq}{\bar v}_{i1kl}\langle T (\psi^{\dagger}_i\psi_l\psi_k)(t) (\psi^{\dagger}_n\psi^{\dagger}_m\psi_q)(t')\rangle
{\bar v}_{mn1'q}\nonumber\\
\eea
or, returning to the number indices,
\bea
T^{(r)}_{11'}(t&-&t') = -\frac{i}{4} \sum\limits_{2'3'4'}\sum\limits_{234}{\bar v}_{1234}\times \nonumber\\
&\times&\langle T \psi^{\dagger}(2)\psi(4)\psi(3)\psi^{\dagger}(3')\psi^{\dagger}(4')\psi(2')\rangle
{\bar v}_{4'3'2'1'},  \nonumber\\
\label{Tr}
\eea
where we assume that $t_{2}=t_{3}=t_{4}=t$ and $t_{2'}=t_{3'}=t_{4'}=t'$, as dictated by the instantaneous interaction. 
With the help of of the three-fermion Green function 
Eq. (\ref{Tr}) can be rewritten as
\be
T^{(r)}_{11'}(t-t') = -\frac{1}{4} \sum\limits_{2'3'4'}\sum\limits_{234}{\bar v}_{1234}G(432',23'4'){\bar v}_{4'3'2'1'}. 
\label{Tr1}
\ee
Here we realize that, although the EOM for one-fermion propagator $G(\omega)$ (\ref{Dyson}) is formally  a closed equation, however,  its interaction kernel $\Sigma(\omega)= T^{irr}(\omega)$ is defined by Eqs. (\ref{Toperator}), (\ref{MF}) and (\ref{Tr1}), where its time-dependent part  contains the three-fermion Green function $G(432',23'4')$.  Up to here the theory is exact, but, in order to calculate the three-body propagator, one would need to generate 
a series of equations of motion for higher-rank propagators. However, with a very good accuracy the problem can be truncated at the two-body level. 
%
The three-fermion Green function, according to Refs. \cite{Martin1959,VinhMau1969,Mau1976,RingSchuck1980}, can be (approximately) decomposed as follows \footnote{Accuracy of this approximation will be examined elsewhere.}:
 \begin{eqnarray}
 G(432^{\prime},23^{\prime}4^{\prime}) &=& G(4,4')G(32',23') + G(3,3')G(42',24') +  \nonumber \\ &+& G(2',2)G(43,3'4') + G(4,2)G(32',3'4') +    \nonumber \\ 
 &+& G(2',4')G(43,23') - G(3,2)G(42',3'4')  - \nonumber \\ 
 &-& G(2',3')G(43,24') - G(4,3')G(32',24')  - \nonumber \\ &-& G(3,4') G(42',23') - 2G^{(0)}(432^{\prime},23^{\prime}4^{\prime}),\nonumber \\
 \label{G3}
 \end{eqnarray}
 where
 \begin{eqnarray}
 &G&^{(0)}(432^{\prime},23^{\prime}4^{\prime}) = \nonumber \\ &=& - G(4,4')G(3,3')G(2',2) + G(4,3')G(3,4')G(2',2) + \nonumber \\ 
 &+& G(4,2)G(3,3')G(2',4')  + G(4,4')G(3,2)G(2',3') - \nonumber \\ 
 &-&G(4,2)G(3,4')G(2',3') - G(4,3')G(3,2)G(2',4') 
 \label{G03}
 \end{eqnarray}
contains all uncorrelated three-body contributions. 
 This kind of truncation of the many-body EOM hierarchies is referred to as cluster-expansion approach in condensed matter physics
\cite{Kira2006,Kira2011}.
Both Eqs. (\ref{G3}) and Eq. (\ref{G03}) include terms with one-fermion propagators connecting the same time moments, either $t$ or $t'$. As these propagators form closed loops, obviously such contributions are connected by one-fermion lines, i.e., are reducible. Thus, the irreducible part of the three-fermion propagator 
of Eq. (\ref{G3}) reads:
 \begin{eqnarray}
 G^{irr}(432^{\prime},23^{\prime}4^{\prime}) = G(4,4')G(32',23') +  \nonumber \\ + G(3,3')G(42',24') +  G(2',2)G(43,3'4')  -  \nonumber \\ -G(4,3')G(32',24')  - G(3,4') G(42',23') + 
  \nonumber \\ 
+  2\Bigl(G(4,4')G(3,3')G(2',2) - G(4,3')G(3,4')G(2',2) \Bigr). \nonumber \\
 \label{G3irr}
 \end{eqnarray}

\begin{figure},
\begin{center}
\includegraphics[scale=0.52]{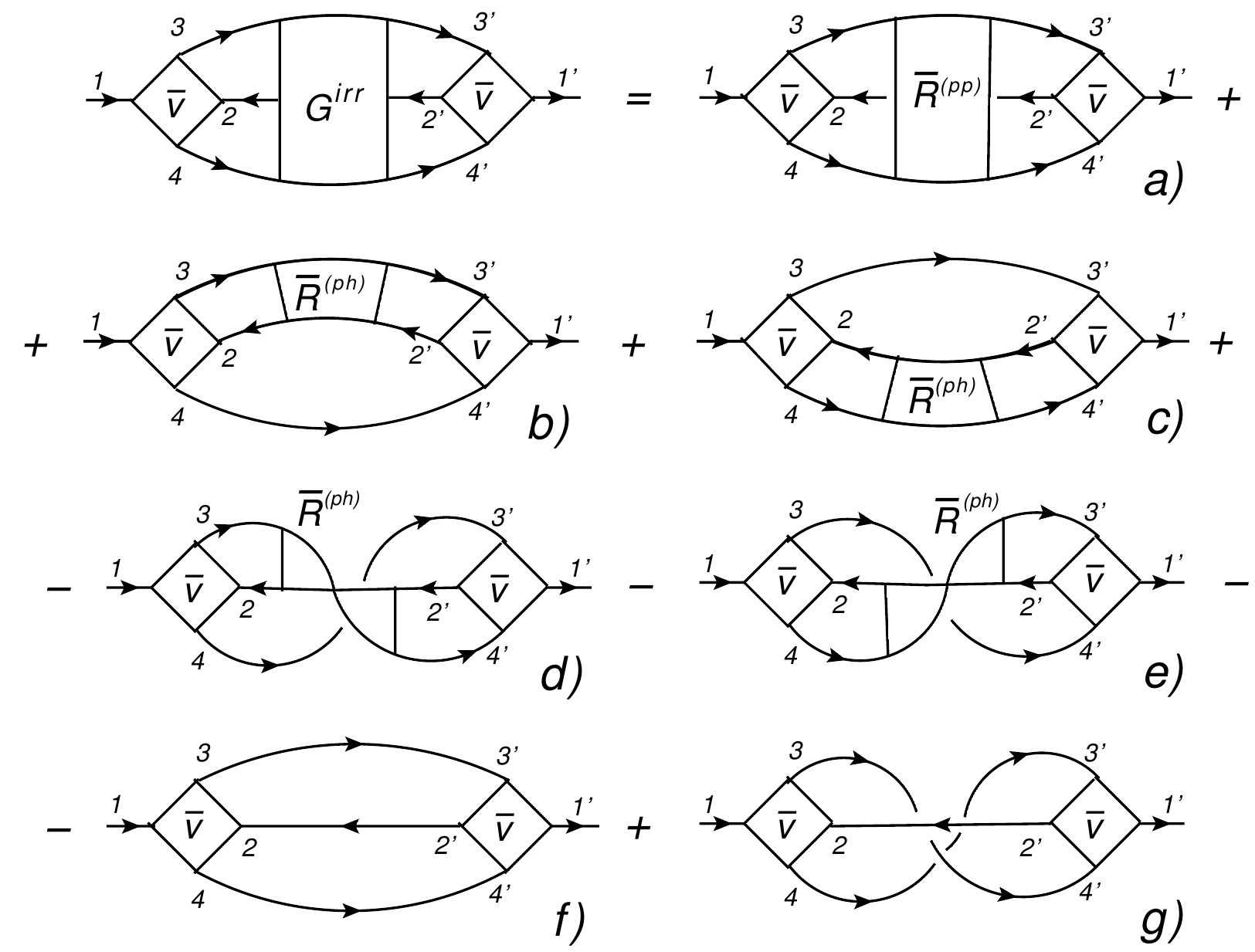}
\end{center}
\caption{Diagrammatic representation of the dynamical self-energy $\Sigma^{(r)}$ of Eq. (\ref{SEirr}) in terms of the uncorrelated one-fermion (lines with arrows) and correlated two-fermion (boxes $\bar R$ together with long lines with arrows) intermediate propagators. Blocks $\bar v$ stand for the bare two-fermion interaction.}
\label{SEirrd}%
\end{figure}

The two-body propagators in Eq. (\ref{G3irr}), such as $G(32',23')$, contain both uncorrelated and correlated terms. It is convenient to introduce its connection to the response functions $R(12',21')$ and to the fully correlated parts of the two-body propagators ${\bar R}(12',21')$:
 \bea
 R^{pp}(12',21') &=& G(12',21') - G(1,2)G(2',1'), \nonumber \\ 
 {\bar R}^{pp}(12',21') &=& G(12',21') - \nonumber \\ &-&\bigl(G(1,2)G(2',1') - G(1,1')G(2',2)\bigr),  \nonumber \\
-iR^{ph}(12,1'2') &=& G(21',12') - G(2,1)G(1',2'), \nonumber\\ 
- i{\bar R}^{ph}(12,1'2') &=& G(21',12') - \nonumber \\ &-&\bigl(G(2,1)G(1',2') - G(1',1)G(2,2')\bigr),  \nonumber \\
\label{RG}
 \eea 
thus related as follows:
\bea
 R^{pp}(12',21') = {\bar R}^{pp}(12',21') - G(1,1')G(2',2),\nonumber \\
  -iR^{ph}(12,1'2') = -i{\bar R}^{ph}(12,1'2') - G(1',1)G(2,2').\nonumber \\
 \label{resp}
\eea
Collecting the terms associated with $\tilde R$ components, one can separate fully correlated and fully uncorrelated $G^{(0)}$ parts of $G^{irr}(432^{\prime},23^{\prime}4^{\prime})$ according to:
 \begin{eqnarray}
 G^{irr}(432^{\prime},23^{\prime}4^{\prime}) = G(2',2){\bar R}^{pp}(43,3'4') - \nonumber \\ - iG(4,4'){\bar R}^{ph}(23,2'3') - iG(3,3'){\bar R}^{ph}(24,2'4')  + \nonumber \\  
 + iG(4,3'){\bar R}^{ph}(23,2'4')  + iG(3,4') {\bar R}^{ph}(24,2'3') - \nonumber \\ 
 - G(4,4')G(3,3')G(2',2) + G(4,3')G(3,4')G(2',2).\nonumber\\
 \label{G3corirr}
 \end{eqnarray}
The corresponding dynamical part of the self-energy shown diagrammatically in Fig. \ref{SEirrd} is
\begin{eqnarray}
 \Sigma^{(r)}_{11'}(t-t')= -\frac{1}{4}\sum\limits_{2342'3'4'}{\bar v}_{1234}G^{irr}(432^{\prime},23^{\prime}4^{\prime}){\bar v}_{4'3'2'1'} = \nonumber\\
 = -\frac{1}{4}\sum\limits_{2342'3'4'}{\bar v}_{1234}\Bigl(G(2',2){\bar R}^{pp}(43,3'4') - \nonumber \\ - iG(3,3'){\bar R}^{ph}(24,2'4') - iG(4,4'){\bar R}^{ph}(23,2'3')   
 + \nonumber \\ + iG(4,3'){\bar R}^{ph}(23,2'4')  + iG(3,4') {\bar R}^{ph}(24,2'3') - \nonumber \\
 - G(4,4')G(3,3')G(2',2) + G(4,3')G(3,4')G(2',2)\Bigr){\bar v}_{4'3'2'1'}. \nonumber \\
 \label{SEirr}
 \end{eqnarray}
Using the symmetry properties of the interaction $\bar v$ and renaming the indices $3\leftrightarrow4,3'\leftrightarrow4'$ in the third term, $4'\leftrightarrow3'$ in the fourth term, 
$4\leftrightarrow3$ in the fifth term and $4\leftrightarrow3$ in the last term, one obtains
 \begin{eqnarray}
  \Sigma^{(r)}_{11'}(t-t')
 = -\frac{1}{4}\sum\limits_{2342'3'4'}{\bar v}_{1234}\Bigl(G(2',2){\bar R}^{pp}(43,3'4') - \nonumber \\ - 4iG(3,3'){\bar R}^{ph}(24,2'4')   
 - 2G(3,3')G(2',2)G(4,4')\Bigr) \times \nonumber\\ \times {\bar v}_{4'3'2'1'}
\nonumber\\
 \label{SEirr1}
 \end{eqnarray}

This form of the self-energy allows for a direct reduction to the second-order (with respect to the bare interaction $\bar v$) approach for the dynamical self-energy. Indeed, by dropping the fully correlated terms with $\tilde R$ the lowest-order approach takes the form
\bea
&\Sigma&^{(r)0}_{11^{\prime}}(t-t^{\prime}) =  \nonumber \\
&=& \frac{1}{2}\sum\limits_{234}\sum\limits_{2^{\prime}3^{\prime}4^{\prime}}{\bar v}_{1234}
G(4,4')G(3,3')G(2',2) 
{\bar v}_{4^{\prime}3^{\prime}2^{\prime}1^{\prime}},\nonumber\\
\label{SEpert}
\eea
which corresponds to the last two terms in Fig. \ref{SEirrd}.
The response theory built on this self-energy leads to an approach called second random phase approximation (SRPA),
which has been implemented within the frameworks employing both microscopic \cite{PapakonstantinouRoth2009,Papakonstantinou2010} and effective \cite{GambacurtaGrassoCatara2010,GambacurtaGrassoCatara2011,Gambacurta2016} interactions and demonstrates a good ability of generating fragmentation effects on nuclear excitation spectra. Closely related are the methods based on a stochastic one-body transport theory, that distinguish and include in-medium nucleon-nucleon collisions of both coherent and incoherent nature \cite{Lacroix2001}.

%

\subsection{Emergent phonons and particle-vibration coupling}
\label{PVCmapping}
The prerequisites for calculating the self-energy beyond the 
perturbation theory are the correlation functions ${\bar R}$ in the particle-hole and particle-particle channels. In practice, however, it is more convenient to deal with the 
particle-particle Green function and particle-hole response function
which contain both correlated and uncorrelated contributions (\ref{resp}). By isolating the particle-hole response function, the self-energy can be reorganized as follows:
\begin{eqnarray}
  \Sigma^{(r)}_{11'}(t-t') =-\sum\limits_{2342'3'4'}{\bar v}_{1234}\Bigl(\frac{1}{4}G(2',2){G}^{pp}(43,3'4') \nonumber \\ - iG(3,3'){R}^{ph}(24,2'4') + G(3,3')G(2',2)G(4,4')\Bigr){\bar v}_{4'3'2'1'}  
\nonumber \\
=  \Sigma^{(r)pp}_{11'}(t-t') +  \Sigma^{(r)ph}_{11'}(t-t') +  \Sigma^{(r)0}_{11'}(t-t'),\nonumber \\
 \label{SEirr2}
 \end{eqnarray}
where we separate the terms with the particle-particle (pp) and particle-hole (ph) 
Green functions coupled to one-hole and one-particle ones, respectively, from the term with the uncorrelated two-particle-one-hole propagator. This equation can serve as foundation for microscopic approaches to the single-particle self-energy, which refer to the phenomenon of particle-vibration coupling, or PVC. The correlation functions $G^{pp}$ and $R^{ph}$ represent the emergent degrees of freedom, phonons, which are the quasibound states of two fermions embedded in the strongly-correlated medium.  In nuclear physics the phonons  are associated with nuclear vibrations and, in relatively large systems, such as medium-mass and heavy nuclei,  can acquire a collective character. They are often called "vibrations" and attributed to the vibrational motion of the nuclear surface, although, as we will see in this subsection, in the formally exact theory they are more general objects.

\begin{figure}
\begin{center}
\includegraphics[scale=0.52]{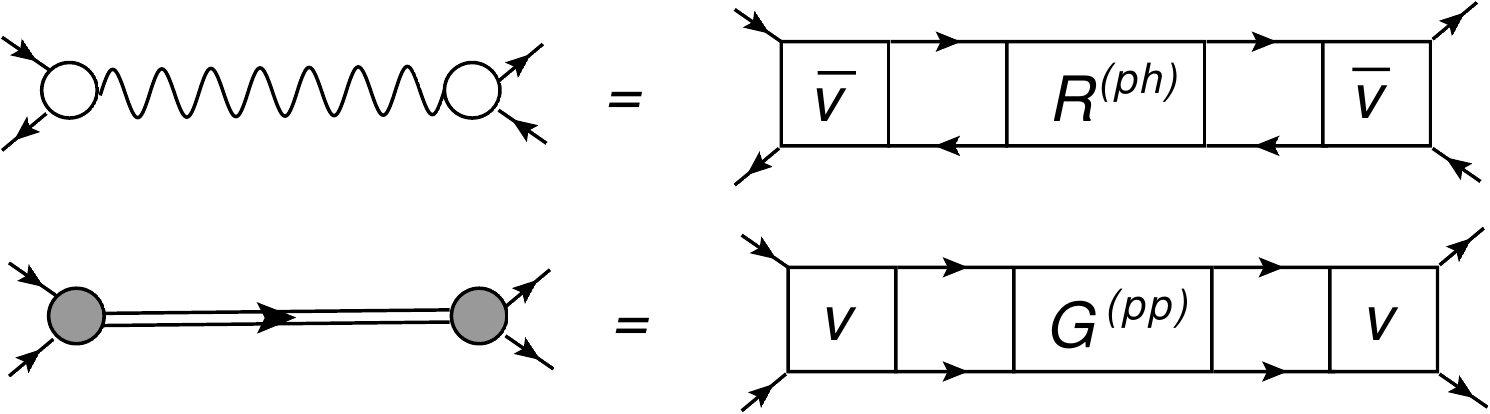}
\end{center}
\caption{Diagrammatic mapping (definition) of the phonon vertices (circles) and propagators (wavy lines and double lines) onto the  bare interaction and two-fermion correlation functions. Top: normal (particle-hole) phonon, bottom: pairing (particle-particle) phonon,  as introduced in Eqs. (\ref{mappingph},\ref{mappingpp}), respectively}.
\label{PVCmap}%
\end{figure}

The connection to the PVC can be seen more explicitly if one identifies the correlation functions contracted with the interaction matrix elements with the phonon propagators and coupling vertices, as displayed diagrammatically in Fig. \ref{PVCmap}.
For this purpose it is convenient to work with the Fourier image of $\Sigma^{(r)}_{11'}(t-t')$ in the energy domain 
\be
\Sigma^{(r)}_{11'}(\omega) = \int\limits_{-\infty}^{\infty} d\tau e^{i\omega\tau} \Sigma^{(r)}_{11'}(\tau).
\label{SEomega}
\ee
The first term of Eq. (\ref{SEirr2}) transforms as follows:
\bea
\Sigma^{(r)pp}_{11'}(\omega) = \sum\limits_{22'} \Bigl[ \sum\limits_{\mu m} \frac{\chi_2^{m\ast} \gamma_{12}^{\mu(+)}\gamma_{1'2'}^{\mu(+)\ast}\chi_{2'}^{m}}{\omega - \omega_{\mu}^{(++)} - \varepsilon_m^{(-)} + i\delta} + \nonumber\\
+ \sum\limits_{\varkappa n}\frac{\eta_2^{n\ast}{\gamma}_{21}^{\varkappa(-)\ast}{\gamma}_{2'1'}^{\varkappa(-)}\eta_{2'}^n}{\omega + \omega_{\varkappa}^{(--)} + \varepsilon_n^{(+)} - i\delta} \Bigr],
\label{FISrpp}
\eea
where we denoted the single-particle energies as $\varepsilon_n^{(+)} = E^{(N+1)}_{n} - E^{(N)}_0$ and $\varepsilon_m^{(-)} = E^{(N-1)}_{m} - E^{(N)}_0$, and defined the pairing phonon vertex functions according to:
\be
\gamma^{\mu(+)}_{12} = \sum\limits_{34} v_{1234}\alpha_{34}^{\mu}, \ \ \ \ \ \ \gamma_{12}^{\varkappa(-)} = \sum\limits_{34}\beta_{34}^{\varkappa}v_{3412}. 
\ee
Then, introducing the amplitude $\Gamma^{pp}_{12,1'2'}(\omega) $,
\bea
i\Gamma^{pp}_{12,1'2'}(\omega) = i\sum\limits_{343'4'}{v}_{1234}G^{pp}_{43,3'4'}(\omega){v}_{4'3'2'1'} = \nonumber \\
= \sum\limits_{\mu,\sigma=\pm1} \gamma^{\mu(\sigma)}_{12}\Delta^{(\sigma)}_{\mu}(\omega)\gamma^{\mu(\sigma)\ast}_{1'2'}
\label{mappingpp}
\eea
with the pairing phonon propagator
\be
\Delta^{(\sigma)}_{\mu}(\omega) = \frac{\sigma}{\omega - \sigma(\omega_{\mu}^{(\sigma\sigma)} - i\delta)},
\ee
Eq. (\ref{FISrpp}) can be alternatively  obtained by the convolution
\be
\Sigma^{(r)pp}_{11'}(\omega) = i\sum\limits_{22'}\int\limits_{-\infty}^{\infty}\frac{d\varepsilon}{2\pi i} \Gamma^{pp}_{12,1'2'}(\omega + \varepsilon)G_{2'2}(\varepsilon).
\ee
\begin{figure*}
\begin{center}
\includegraphics*[scale=0.75]{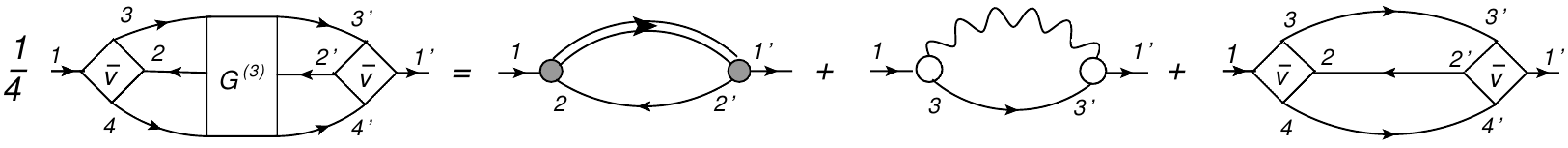}
\end{center}
\caption{Diagrammatic representation of the dynamical part of the kernel $\Sigma^{(r)}$ of Eq. (\ref{SEirr2}) in terms of the particle-vibration coupling.}
\label{SEdyn}%
\end{figure*}
Similarly, the Fourier image of the second term of Eq. (\ref{SEirr2})
\bea
\Sigma^{(r)ph}_{11'}(\omega) = \sum\limits_{33'} \Bigl[ 
\sum\limits_{\nu n}\frac{\eta_3^{n}{g}_{13}^{\nu}{g}_{1'3'}^{\nu\ast}\eta_{3'}^{n\ast}}{\omega - \omega_{\nu} - \varepsilon_n^{(+)} + i\delta} +
\nonumber\\ +
\sum\limits_{\nu m} \frac{\chi_3^{m} g_{31}^{\nu\ast}g_{3'1'}^{\nu}\chi_{3'}^{m\ast}}{\omega + \omega_{\nu} + \varepsilon_m^{(-)} - i\delta} 
\Bigr]
\eea
can be obtained either directly or by introducing the following mapping:
\bea
\Gamma^{ph}_{13',1'3} =  \sum\limits_{242'4'}{\bar v}_{1234}R^{ph}_{24,2'4'}(\omega){\bar v}_{4'3'2'1'} = \nonumber \\ = 
\sum\limits_{\nu,\sigma=\pm1} g^{{\nu}(\sigma)}_{13}D^{(\sigma)}_{\nu}(\omega)g^{\nu(\sigma)\ast}_{1'3'}
\label{mappingph}
\eea
with the phonon vertices $g^{\nu}$ and propagators $D_{\nu}(\omega)$:
\bea
g^{\nu(\sigma)}_{13} = \delta_{\sigma,+1}g^{\nu}_{13} + \delta_{\sigma,-1}g^{\nu\ast}_{31}, \ \ \ \ 
g^{\nu}_{13} = \sum\limits_{34}{\bar v}_{1234}\rho^{\nu}_{42}, \nonumber \\
D_{\nu}^{(\sigma)}(\omega) = \frac{\sigma}{\omega - \sigma(\omega_{\nu} - i\delta)}, \ \ \ \
\omega_{\nu} = E_{\nu} - E_0. \nonumber \\
\label{gDPVCph}
\eea 
Then it can be shown that
\be
\Sigma^{(r)ph}_{11'}(\omega) = -\sum\limits_{33'}\int\limits_{-\infty}^{\infty}\frac{d\varepsilon}{2\pi i} \Gamma^{ph}_{13',1'3}(\omega - \varepsilon)G_{33'}(\varepsilon).
\label{FISphe}
\ee
In Eqs. (\ref{FISrpp}-\ref{FISphe}) the index $\sigma = \pm1$ stands for the forward ("particle") and backward ("hole") components of the phonon propagators and vertices, and  the spectral representations (\ref{spgfspec},\ref{resppp}) along with the definitions (\ref{spgf},\ref{ppGF},\ref{phresp}) were applied.  Finally, the last term of the self-energy (\ref{SEirr2}) with only uncorrelated single-particle Green functions transforms as follows:
\bea
&\Sigma&^{(r)0}_{11'}(\omega) = -\sum\limits_{2342'3'4'} {\bar v}_{1234}\times \nonumber \\ &\times&\Bigl[\sum\limits_{mn'n''} \frac{\chi_{2'}^{m}\chi_2^{m\ast}\eta_3^{n'}\eta_{3'}^{n'\ast}\eta_4^{n''}\eta_{4'}^{n''\ast}}
{\omega - \varepsilon_{n'}^{(+)} - \varepsilon_{n''}^{(+)} - \varepsilon_{m}^{(-)} + i\delta} \nonumber \\
&+& \sum\limits_{nm'm''} \frac{\eta_{2'}^{n}\eta_{2}^{n\ast}\chi_3^{m'}\chi_{3'}^{m'\ast}\chi_4^{m''}\chi_{4'}^{m''\ast}}
{\omega + \varepsilon_{n}^{(+)} + \varepsilon_{m'}^{(-)} + \varepsilon_{m''}^{(-)} - i\delta} \Bigr] {\bar v}_{4'3'2'1'} = \nonumber \\
&=&-\sum\limits_{2342'3'4'}{\bar v}_{1234} {\tilde G}^{(3)0}_{432',23'4'}(\omega){\bar v}_{4'3'2'1'},
\\
&{\tilde G}&^{(3)0}_{432',23'4'} (\omega) = \nonumber \\
&=& -\int\limits_{-\infty}^{\infty}\frac{d\varepsilon d\varepsilon'}{(2\pi i)^2}  G_{44'}(\omega+\varepsilon'-\varepsilon)G_{33'}(\varepsilon)G_{2'2}(\varepsilon').
\eea

The complete dynamical part of the one-fermion self-energy (\ref{SEirr2}) is shown in Fig. \ref{SEdyn} in the diagrammatic form in terms of the particle-vibration coupling.
The first two terms on the right hand side are formed by the topologically similar one-loop diagrams which are analogous to the electron self-energy corrections  in quantum electrodynamics, 
where electron emits and reabsorbs a photon, or to the nucleonic self-energy of quantum hadrodynamics where a single nucleon emits and reabsorbs a meson. 
Here, they represent the effects of a strongly correlated medium, where a single fermion emits and reabsorbs a phonon of the particle-particle (first term) and the particle-hole (second term) nature, in addition to the emission and reabsorption of an uncorrelated two-particle-one-hole configuration (third term). The mappings established by Eqs. (\ref{mappingpp},\ref{mappingph}) and illustrated diagrammatically in Fig. \ref{PVCmap} explain the underlying mechanism of the induced in-medium interaction, where the emergent composite bosons are formed by quasibound fermionic pairs.

One may notice that the dynamical self-energy $\Sigma^{(r)}$ recast in the form of Eq. (\ref{SEirr1}) with the separation of fully correlated and non-correlated parts helps to relate the approach to the lowest-order perturbation theory and to assess the role of correlations. It is clear that in the case of weak coupling the uncorrelated term(s) play the leading role and the phonon-exchange interaction can be neglected, however, in the strong-coupling regime the phonon coupling dominates and the lowest-order uncorrelated term does not give the leading contribution. Indeed, in the major applications to nuclear systems only the first two terms are taken into account and, furthermore, coupling to the pairing phonons was found much less important than coupling to the normal particle-hole phonons. These approximations, however, were shown to be justified only within the methods based on the effective nucleon-nucleon interactions. Such interactions are typically obtained by fitting the bulk nuclear properties, such as their masses and radii, on the mean-field level, i.e., assuming that one-fermion self energy contains solely the static part (\ref{MF}) with only the one-body density, which is implicitly coupled to correlations in the dynamical part of the self-energy (\ref{SEirr2}). This coupling is essentially important as 
\be
\rho_{12} = -i\lim\limits_{t_2\to t_1+0} G(1,2),
\ee
which means that the one-fermion density matrix entering Eq. (\ref{MF}) is the equal-times limit of the full solution of Eq. (\ref{Dyson}). This fact is often expressed in terms of the density dependencies of the effective interactions while these dependencies are typically disconnected from a detailed analysis of Eq. (\ref{Dyson}) with the complete kernel $\Sigma(\omega)$.
Inevitably, existing versions of the PVC model which add the dynamical part on top of the effective interactions imply an additional subtraction procedure to remove the double counting of PVC which is contained in the static approximation in the parameters of the phenomenological mean field \cite{Tselyaev2013}. Such a subtraction turned out to be a very elegant way of avoiding the double counting, instead of the complicated refitting of the parameters of the mean field, and it is widely applied in calculations of two-body Green functions, in particular, the particle-hole response discussed below. However, an analogous procedure has not been formulated for the case of the one-body propagator.

Regardless what kind of the two-body interaction is used for calculations of the one-fermion Green function from Eq. (\ref{Dyson}), for an accurate solution beyond the static approximation to the interaction kernel the knowledge about the 
particle-hole response and particle-particle Green function is needed, as follows from Eqs. (\ref{Dyson},\ref{Somega},\ref{MF},\ref{SEirr2}). As we discuss in the next section following Refs. \cite{DukelskyRoepkeSchuck1998,Olevano2018},  it is possible to formulate the equation of motion for these two-fermion Green functions in a similar manner as for the one-fermion propagator and to obtain non-perturbative approximations to its closed form. Below we discuss the EOM approach to the particle-hole response function, while the EOM for the particle-particle response can be generated in complete analogy.

\section{Equation of motion for the particle-hole response}
\label{EOM2}
The equation of motion for the particle-hole response of Eq. (\ref{phresp}) can be generated by taking its time derivative with respect to $t$:
\bea
\partial_t R_{12,1'2'}(t-t') = 
-i\delta(t-t')\langle [{\psi^{\dagger}}_1\psi_2,{\psi^{\dagger}}_{2'}\psi_{1'}]\rangle + \nonumber \\ 
+ \langle T[H,{\psi^{\dagger}}_1\psi_2](t)({\psi^{\dagger}}_{2'}\psi_{1'})(t')\rangle. \nonumber\\ 
\label{dtG2}                           
\eea
After the evaluation of the first commutator and the commutator with the one-body part of the Hamiltonian, 
the EOM (\ref{dtG2}) takes the form
\bea
(i\partial_t &+& \varepsilon_{12})R_{12,1'2'}(t-t')
= \delta(t-t'){\cal N}_{121'2'} +
\nonumber \\
&+& i\langle T[V,{\psi^{\dagger}}_1\psi_2](t)({\psi^{\dagger}}_{2'}\psi_{1'})(t')\rangle
\label{dtG2b}                           
\eea
with the norm kernel ${\cal N}_{121'2'}$,
\bea
{\cal N}_{121'2'} = \langle[\psi^{\dagger}_{1}\psi_{2},\psi^{\dagger}_{2'}\psi_{1'}]\rangle =  \delta_{22'}\langle \psi^{\dagger}_{1}\psi_{1'} \rangle - 
\delta_{11'}\langle \psi^{\dagger}_{2'}\psi_{2} \rangle. \nonumber\\
\label{norm}
\eea
In the case when the one-body density matrix is diagonal, it takes the form ${\cal N}_{121'2'} = \delta_{11'}\delta_{22'}(n_1 - n_2) \equiv \delta_{11'}\delta_{22'}{\cal N}_{12}$, where
$n_1 = \langle{\psi^{\dagger}}_1\psi_1\rangle$ is the occupation number of the fermionic state $1$. In Eq. (\ref{dtG2b}) and in the following $\varepsilon_{12} = \varepsilon_{1} - \varepsilon_{2}$ is the difference between the matrix elements of the 
one-body Hamiltonian, or the single-particle energies, in the same representation. 

The second EOM is generated by differentiating the last term on the right hand side of Eq. (\ref{dtG2b}) with respect to $t'$:
\bea
i\langle T[V,{\psi^{\dagger}}_1\psi_2](t)({\psi^{\dagger}}_{2'}\psi_{1'})(t')\rangle(-i\overleftarrow{\partial_{t'}} - \varepsilon_{2'1'}) = \nonumber \\
= -\delta(t-t')
\langle [[V,{\psi^{\dagger}}_1\psi_2],{\psi^{\dagger}}_{2'}\psi_{1'}]\rangle + \nonumber \\ +i\langle T[V,{\psi^{\dagger}}_1\psi_2](t)[V,{\psi^{\dagger}}_{2'}\psi_{1'}](t')\rangle.
\label{dtG2c}
\eea
\begin{figure*}
\begin{center}
\includegraphics*[scale=0.7]{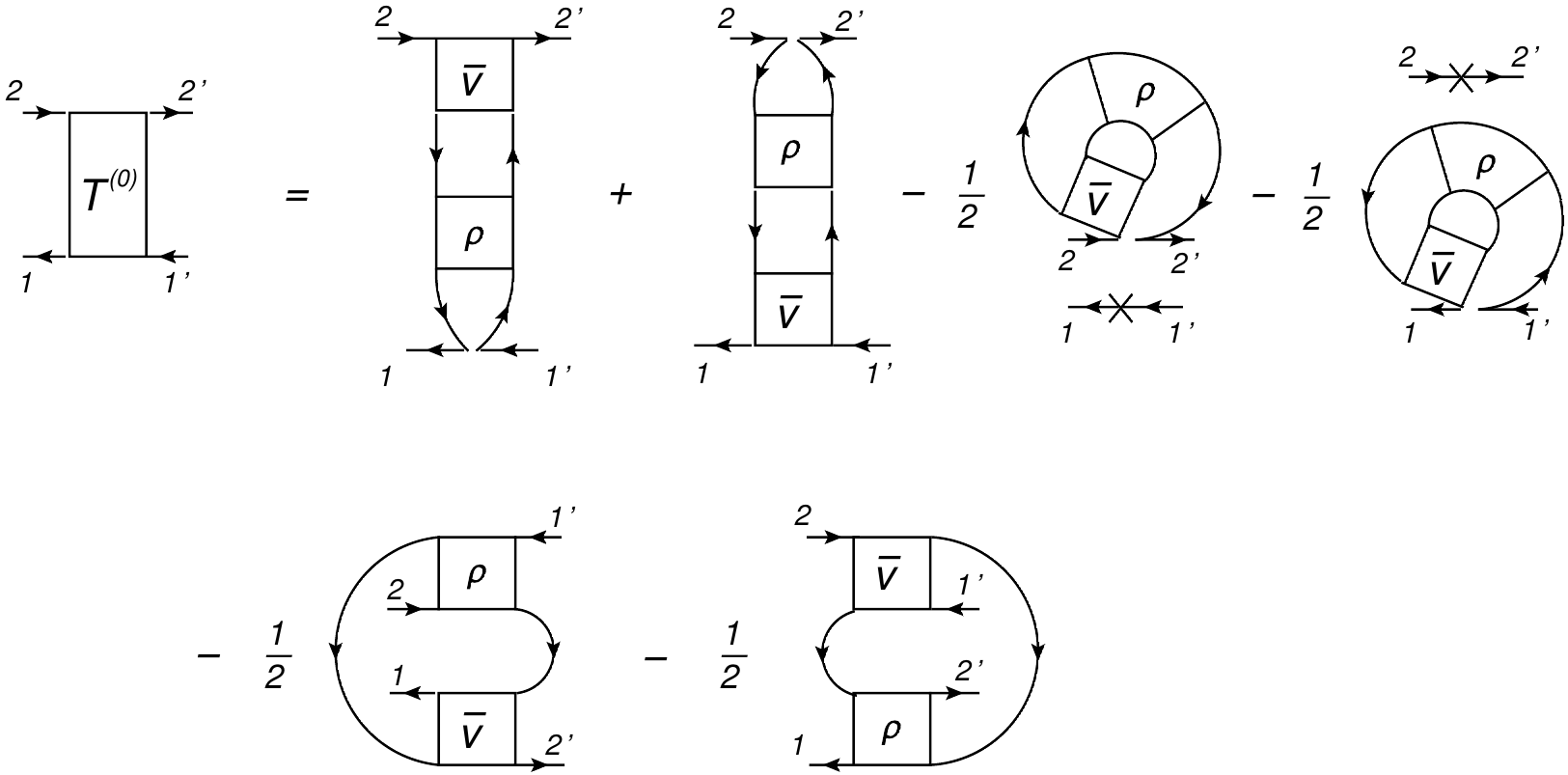}
\end{center}
\caption{Diagrammatic representation of the static part of the kernel $T^{(0)}_{12,1'2'}$ in terms of the two-body densities $\rho$ as in Eq. (\ref{Fstatic}).}
\label{SE2static}%
\end{figure*}
Combining it with the first EOM (\ref{dtG2b}) and performing the Fourier transformation to the energy domain, one obtains 
\bea
R_{12,1'2'}(\omega) &=& R^{(0)}_{12,1'2'}(\omega) + \nonumber \\
&+& \sum\limits_{343'4'} R^{(0)}_{12,34}(\omega)T_{34,3'4'}(\omega)R^{(0)}_{3'4',1'2'}(\omega),\nonumber \\
\label{2bgfb}
\eea
where the uncorrelated particle-hole response $R^{(0)}(\omega)$ is defined as
\be
R^{(0)}_{12,1'2'}(\omega) = \frac{{\cal N}_{121'2'}}{\omega - \varepsilon_{21}} = \delta_{11'}\delta_{22'}\frac{n_1 - n_2}{\omega - \varepsilon_{21}} .
\label{resp0}
\ee
The integral part is determined by the Fourier transform $T_{12,1'2'}(\omega)$ of the interaction kernel $T_{12,1'2'}(t-t')$  (not to be confused with the time ordering operator)
\bea
T_{12,1'2'}(t-t') &=& {\cal N}_{12}^{-1}\Bigl[-\delta(t-t')
\langle [[V,{\psi^{\dagger}}_1\psi_2],{\psi^{\dagger}}_{2'}\psi_{1'}]\rangle + \nonumber \\
&+& i\langle T[V,{\psi^{\dagger}}_1\psi_2](t)[V,{\psi^{\dagger}}_{2'}\psi_{1'}](t')\rangle \Bigr] {\cal N}_{1'2'}^{-1}, \nonumber\\
\label{F1}
\eea
which splits naturally into the instantaneous $T^{(0)}$ and the time-dependent $T^{(r)}$ parts
\bea
T_{12,1'2'}(t-t') &=& {\tilde{\cal N}}_{121'2'}^{-1}\Bigl(T^{(0)}_{12,1'2'}\delta(t-t') + T^{(r)}_{12,1'2'} (t-t')\Bigr), \nonumber \\
T^{(0)}_{12,1'2'} &=& -\langle [[V,{\psi^{\dagger}}_1\psi_2],{\psi^{\dagger}}_{2'}\psi_{1'}]\rangle, \nonumber \\
T^{(r)}_{12,1'2'}(t-t') &=&  i\langle T[V,{\psi^{\dagger}}_1\psi_2](t)[V,{\psi^{\dagger}}_{2'}\psi_{1'}](t')\rangle \nonumber\\
\label{Ft} 
\eea
with the shorthand notation for the product of the diagonal norm kernels ${\tilde{\cal N}}_{121'2'} = {\cal N}_{12}{\cal N}_{1'2'}$. 
In the operator form Eq. (\ref{2bgfb}) reads
\be
R(\omega) = R^{(0)}(\omega) + R^{(0)}(\omega)T(\omega)R^{(0)}(\omega).
\label{Tmatrix}
\ee
The latter equation can be further transformed to a formally closed equation for $R(\omega)$, similar to the Dyson equation for one-fermion 
Green function. The kernel of this new equation should be irreducible with respect to uncorrelated particle-hole response $R^{(0)}$, which means that
\be
R(\omega) = R^{(0)}(\omega) + R^{(0)}(\omega)K(\omega)R(\omega),
\label{Dyson2}
\ee
where
\be
T(\omega) = K(\omega) + K(\omega)R^{(0)}(\omega)T(\omega) 
\label{Kkernel}
\ee
or $K(\omega) = T^{irr}(\omega)$,
i.e. that $K(\omega)$ absorbs the irreducible contributions of $T(\omega)$.
 
Obviously, the kernel $K(\omega)$ can be also decomposed into the instantaneous (static) and  time-dependent (frequency-dependent) terms:
\bea
K(t-t') = {\tilde{\cal N}}^{-1}\Bigl(K^{(0)}\delta(t-t') + K^{(r)}(t-t')\Bigr), \nonumber \\ 
K^{(0)} = T^{(0)irr}, \ \ \ \ \ \  K^{(r)}(t-t') = T^{(r)irr}(t-t').
\label{Womega}
\eea 
In a complete analogy to the case of one-fermion EOM, the decomposition of the interaction kernel (\ref{Ft},\ref{Womega}) into the static and time(energy)-dependent, or dynamical, parts is a generic feature 
of the in-medium interaction in the particle-hole channel and the direct consequence of the time-independence of the bare interaction $V$ of Eq. (\ref{Hamiltonian2}). 

Evaluation of the static part of Eq. (\ref{F1}) with the help of the following commutator
\bea
[\psi^{\dagger}_i\psi^{\dagger}_j\psi_l\psi_k,\psi^{\dagger}_1\psi_2] = \nonumber \\
= -\delta_{2i}\psi^{\dagger}_1\psi^{\dagger}_j\psi_l\psi_k + \delta_{2j}\psi^{\dagger}_1\psi^{\dagger}_i\psi_l\psi_k - \nonumber \\
- \delta_{1l}\psi^{\dagger}_i\psi^{\dagger}_j\psi_k\psi_2 + \delta_{1k}\psi^{\dagger}_i\psi^{\dagger}_j\psi_l\psi_2,
\label{commutator24}
\eea
gives for the internal single commutator of $T^{(0)}$:
\bea
[V,{\psi^{\dagger}}_1\psi_2] &=& \frac{1}{2}\sum\limits_{jkl}{\bar v}_{j2kl}
 \psi^{\dagger}_1\psi^{\dagger}_j\psi_l\psi_k  + \nonumber \\
 &+&\frac{1}{2}\sum\limits_{ijk}{\bar v}_{ij1k} \psi^{\dagger}_i\psi^{\dagger}_j\psi_k\psi_2
\label{Vpdp}
\eea
and for the double commutator, using again the result of Eq. (\ref{commutator24}):
\bea
[[V,{\psi^{\dagger}}_1\psi_2],{\psi^{\dagger}}_{2'}\psi_{1'}] = \nonumber\\
= \frac{1}{2}\sum\limits_{jkl}{\bar v}_{2jkl} \bigl( \delta_{1'1}\psi^{\dagger}_{2'}\psi^{\dagger}_j\psi_l\psi_k -
\delta_{1'j}\psi^{\dagger}_{2'}\psi^{\dagger}_1\psi_l\psi_k + \nonumber \\
+ \delta_{2'l}\psi^{\dagger}_{1}\psi^{\dagger}_j\psi_k\psi_{1'} -
\delta_{2'k}\psi^{\dagger}_{1}\psi^{\dagger}_j\psi_l\psi_{1'}\bigr) + \nonumber\\
+\sum\limits_{ijk}{\bar v}_{ijk1} \bigl( \delta_{1'i}\psi^{\dagger}_{2'}\psi^{\dagger}_j\psi_k\psi_2 -
\delta_{1'j}\psi^{\dagger}_{2'}\psi^{\dagger}_i\psi_k\psi_2 + \nonumber \\
+ \delta_{2'k}\psi^{\dagger}_{i}\psi^{\dagger}_j\psi_2\psi_{1'} -
\delta_{2'2}\psi^{\dagger}_{i}\psi^{\dagger}_j\psi_k\psi_{1'}\bigr). \nonumber \\
\eea
Thus, the static part of Eq. (\ref{Ft}) reads:
\bea
T^{(0)}_{12,1'2'} =  \sum\limits_{jk}{\bar v}_{2j2'k} \rho_{1'k,1j} + \sum\limits_{jk}
{\bar v}_{1'k1j}\rho_{2j,2'k} - \nonumber \\
- \frac{1}{2}\delta_{11'}\sum\limits_{jkl}{\bar v}_{2jkl}\rho_{kl,2'j} 
- \frac{1}{2}\delta_{22'}\sum\limits_{ijk}{\bar v}_{ji1k}\rho_{1'k,ji} - \nonumber \\
- \frac{1}{2}\sum\limits_{ij}{\bar v}_{ij2'1}\rho_{1'2,ij} -
\frac{1}{2}\sum\limits_{kl}{\bar v}_{21'kl}\rho_{kl,12'},\nonumber \\
\label{Fstatic}
\eea
where we introduced the two-fermion density 
$\rho_{ij,kl}$,
\be
\rho_{ij,kl} = \langle\psi^{\dagger}_{k}\psi^{\dagger}_{l}\psi_{j}\psi_{i}\rangle = \rho_{ik}\rho_{jl} - \rho_{il}\rho_{jk} + \sigma^{(2)}_{ij,kl},
\ee
so that $\sigma^{(2)}_{ij,kl}$ represents its fully correlated part. After that, the static kernel takes the form:
\bea
K^{(0)}_{12,1'2'} =  {\cal N}_{12}{\bar v}_{21'12'}{\cal N}_{1'2'} + \nonumber \\
+ \sum\limits_{jk}{\bar v}_{2j2'k} \sigma^{(2)}_{1'k,1j} + \sum\limits_{jk}
{\bar v}_{1'k1j}\sigma^{(2)}_{2j,2'k} - \nonumber \\
- \frac{1}{2}\delta_{11'}\sum\limits_{jkl}{\bar v}_{2jkl}\sigma^{(2)}_{kl,2'j} 
- \frac{1}{2}\delta_{22'}\sum\limits_{ijk}{\bar v}_{ji1k}\sigma^{(2)}_{1'k,ji} - \nonumber \\
- \frac{1}{2}\sum\limits_{ij}{\bar v}_{ij2'1}\sigma^{(2)}_{1'2,ij} -
\frac{1}{2}\sum\limits_{kl}{\bar v}_{21'kl}\sigma^{(2)}_{kl,12'},\nonumber \\
\label{Fstatic1}
\eea
where the first term isolates the contribution from the bare interaction and the norm factor will be compensated by its inverse, according to Eq. (\ref{Ft}). In transforming Eq. (\ref{Fstatic}) to Eq. 
(\ref{Fstatic1}), the remaining terms with the single-particle mean field can be absorbed in the single-particle energies by replacing $\varepsilon_1 \to {\tilde{\varepsilon}}_1 = \varepsilon_1 + \Sigma^{(0)}_{11}$ in the uncorrelated response function of Eq. (\ref{resp0}) and redefining the working basis. From Eq. (\ref{Fstatic1}) it becomes clear that, in the absence  of correlations contained in the quantities $\sigma^{(2)}$ and $T^{(r)}$, the EOM (\ref{Dyson2}) takes the form of the well-known random phase approximation.
The complete static part of the effective two-fermion interaction is shown diagrammatically in Fig. \ref{SE2static} in terms of the full two-body densities $\rho$ represented by the rectangular blocks.

\begin{figure}
\begin{center}
\includegraphics[scale=0.57]{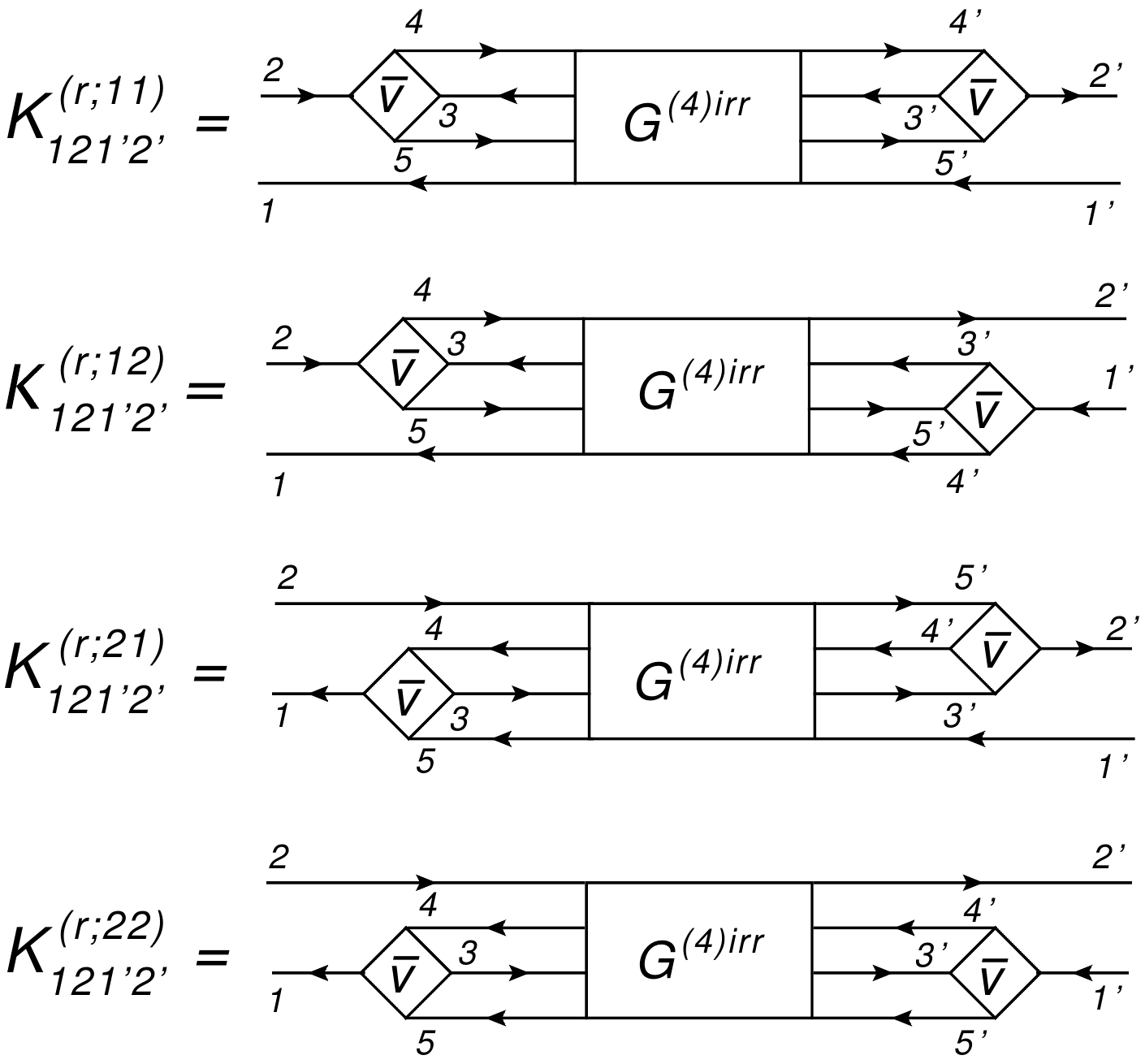}
\end{center}
\caption{Diagrammatic representation of the four components of the dynamical kernel $K^{(r)}_{12,1'2'}(t-t')=T^{(r)irr}_{12,1'2'}(t-t')$  of Eq. (\ref{Fcomponents22}).}
\label{SE2irrtot}%
\end{figure}

The time-dependent part $T^{(r)}$ of the kernel $T$ can be evaluated by making use of the commutator (\ref{Vpdp}):
\bea
&T&^{(r)}_{12,1'2'}(t-t') = i\langle T[V,{\psi^{\dagger}}_1\psi_2](t)[V,{\psi^{\dagger}}_{2'}\psi_{1'}](t')\rangle = \nonumber\\
 &=& \frac{i}{4}\langle T\Bigl(\sum\limits_{jkl}{\bar v}_{2jkl}
 \psi^{\dagger}_1\psi^{\dagger}_j\psi_l\psi_k  + 
\sum\limits_{ijk}{\bar v}_{ijk1} \psi^{\dagger}_i\psi^{\dagger}_j\psi_k\psi_2 \Bigr)(t) \times \nonumber \\
&\times& \Bigl(\sum\limits_{npq}{\bar v}_{1'npq}
 \psi^{\dagger}_{2'}\psi^{\dagger}_n\psi_q\psi_p  + 
\sum\limits_{mnp}{\bar v}_{mnp2'}\psi^{\dagger}_m\psi^{\dagger}_n\psi_p\psi_{1'} \Bigr)(t')\rangle =\nonumber\\
&=& T^{(r;11)}_{12,1'2'}(t-t') + T^{(r;12)}_{12,1'2'}(t-t') + \nonumber\\ &\ & \ \ \ \ \ \ \ \ \ \ \ \ \ \ \ \ \ \ \ +T^{(r;21)}_{12,1'2'}(t-t') + T^{(r;22)}_{12,1'2'}(t-t'),\nonumber\\
\label{Fr}
 \eea 
where we have decomposed the kernel $T^{(r)}_{12,1'2'}(t-t')$ into the four terms with different general structure: 
\bea
T^{(r;11)}_{12,1'2'}(t-t') =  
 -\frac{i}{4}\sum\limits_{jkl}{\bar v}_{j2kl}\langle T(\psi^{\dagger}_1\psi^{\dagger}_j\psi_l\psi_k)(t)
\times\nonumber \\
\times\sum\limits_{mnp}(\psi^{\dagger}_m\psi^{\dagger}_n\psi_p\psi_{1'})(t')\rangle {\bar v}_{nm2'p}\nonumber\\
T^{(r;12)}_{12,1'2'}(t-t') =  \frac{i}{4}\sum\limits_{jkl}
{\bar v}_{j2kl}\langle T (\psi^{\dagger}_1\psi^{\dagger}_j\psi_l\psi_k)(t) \times \nonumber\\
\times\sum\limits_{npq}(\psi^{\dagger}_{2'}\psi^{\dagger}_n\psi_q\psi_p)(t')\rangle{\bar v}_{n1'pq}\nonumber\\
T^{(r;21)}_{12,1'2'}(t-t') = 
\frac{i}{4}\sum\limits_{ijk}{\bar v}_{ji1k} \langle T(\psi^{\dagger}_i\psi^{\dagger}_j\psi_k\psi_2)(t)
\times\nonumber\\
\times\sum\limits_{mnp}(\psi^{\dagger}_m\psi^{\dagger}_n\psi_p\psi_{1'})(t')\rangle {\bar v}_{nm2'p}\nonumber\\
T^{(r;22)}_{12,1'2'}(t-t') =  
-\frac{i}{4}\sum\limits_{ijk}{\bar v}_{ji1k} \langle T(\psi^{\dagger}_i\psi^{\dagger}_j\psi_k\psi_2)(t)
\times\nonumber \\
\times\sum\limits_{npq}(\psi^{\dagger}_{2'}\psi^{\dagger}_n\psi_q\psi_p)(t')\rangle {\bar v}_{n1'pq}\nonumber\\
\eea
or, returning to the number indices, 
\bea
T^{(r;11)}_{12,1'2'}(t-t') = -\frac{i}{4}\sum\limits_{345}{\bar v}_{3245}\langle T({\psi^{\dagger}}_1\psi^{\dagger}_3\psi_5\psi_4)(t)\times\nonumber\\
\times\sum\limits_{3'4'5'}(\psi^{\dagger}_{4'}\psi^{\dagger}_{5'}\psi_{3'}\psi_{1'})(t')\rangle {\bar v}_{5'4'2'3'}= \nonumber \\
=  -\frac{i}{4}\sum\limits_{345}\sum\limits_{3'4'5'}{\bar v}_{3245}G(543'1',5'4'31){\bar v}_{5'4'2'3'} \nonumber
\label{Fcomponents11}
\\
T^{(r;12)}_{12,1'2'}(t-t') =  \frac{i}{4}\sum\limits_{345}
{\bar v}_{3245}\langle T (\psi^{\dagger}_1\psi^{\dagger}_3\psi_5\psi_4)(t) \times\nonumber\\
\times\sum\limits_{3'4'5'}(\psi^{\dagger}_{2'}\psi^{\dagger}_{5'}\psi_{4'}\psi_{3'})(t')\rangle{\bar v}_{5'1'3'4'} =\nonumber\\
= \frac{i}{4}\sum\limits_{345}\sum\limits_{3'4'5'}
{\bar v}_{3245}G(544'3',5'2'31){\bar v}_{5'1'3'4'} \nonumber\\
\label{Fcomponents12}
T^{(r;21)}_{12,1'2'}(t-t') = \frac{i}{4}\sum\limits_{345}{\bar v}_{4513} \langle T(\psi^{\dagger}_5\psi^{\dagger}_4\psi_3\psi_2)(t)\times\nonumber\\
\times\sum\limits_{3'4'5'}(\psi^{\dagger}_{3'}\psi^{\dagger}_{5'}\psi_{4'}\psi_{1'})(t')\rangle {\bar v}_{5'3'2'4'} = \nonumber\\
= \frac{i}{4}\sum\limits_{345}\sum\limits_{3'4'5'}{\bar v}_{4513}G(324'1',5'3'45){\bar v}_{5'3'2'4'} \nonumber \\
\label{Fcomponents21}
T^{(r;22)}_{12,1'2'}(t-t') = -\frac{i}{4}\sum\limits_{345}{\bar v}_{4513} \langle T(\psi^{\dagger}_5\psi^{\dagger}_4\psi_3\psi_2)(t)\times\nonumber\\
\times\sum\limits_{3'4'5'}(\psi^{\dagger}_{2'}\psi^{\dagger}_{3'}\psi_{5'}\psi_{4'})(t')\rangle {\bar v}_{3'1'4'5'} = \nonumber\\
-\frac{i}{4}\sum\limits_{345}\sum\limits_{3'4'5'}{\bar v}_{4513} G(325'4',3'2'45) {\bar v}_{3'1'4'5'}, \nonumber \\
\label{Fcomponents22}
\eea
where the two-time two-particle-two-hole (four-fermion) Green function $G(543'1',5'4'31)$ was introduced according to:
\bea
&G&(543'1',5'4'31) = \nonumber \\
&=& \langle T({\psi^{\dagger}}_1{\psi^{\dagger}}_3\psi_5\psi_4)(t)
({\psi^{\dagger}}_{4'}{\psi^{\dagger}}_{5'}\psi_{3'}\psi_{1'})(t')\rangle. \nonumber \\
\label{G4}
\eea
The components 
of the irreducible dynamical kernel $K^{(r)}_{12,1'2'}(t-t')=T^{(r)irr}_{12,1'2'}(t-t')$ are shown diagrammatically in Fig. \ref{SE2irrtot}, where we have omitted the factors $\pm i/4$ in front of each diagram.

 The general result of Eqs. (\ref{Fr}-\ref{Fcomponents22}) for the dynamical kernel is known in the literature, see, for instance, Refs. \cite{SchuckEthofer1973,DukelskyRoepkeSchuck1998}. One can see that the dynamical kernel of Eq. (\ref{Dyson2}) requires the knowledge about the two-time two-particle-two-hole (four-body) propagator $G$ of Eq. (\ref{G4}). An exact treatment of the time-dependent kernel would require generating EOM's for the three-fermion or four-fermion propagators, thus building a hierarchy of EOM's which is equivalent to that known as Bogoliubov-Born-Green-Kirkwood-Yvon (BBGKY) hierarchy. In this work we consider a truncation of this hierarchy on the level of two-fermion correlation functions, as proposed, in particular, in Refs. \cite{SchuckTohyama2016,Olevano2018}.
This approach avoids generating EOM's for the three-fermion and four-fermion propagators and, as we will show below, leads to a closed system of equations for the two-fermion correlation functions.

\begin{figure*}
\begin{center}
\includegraphics*[scale=0.75]{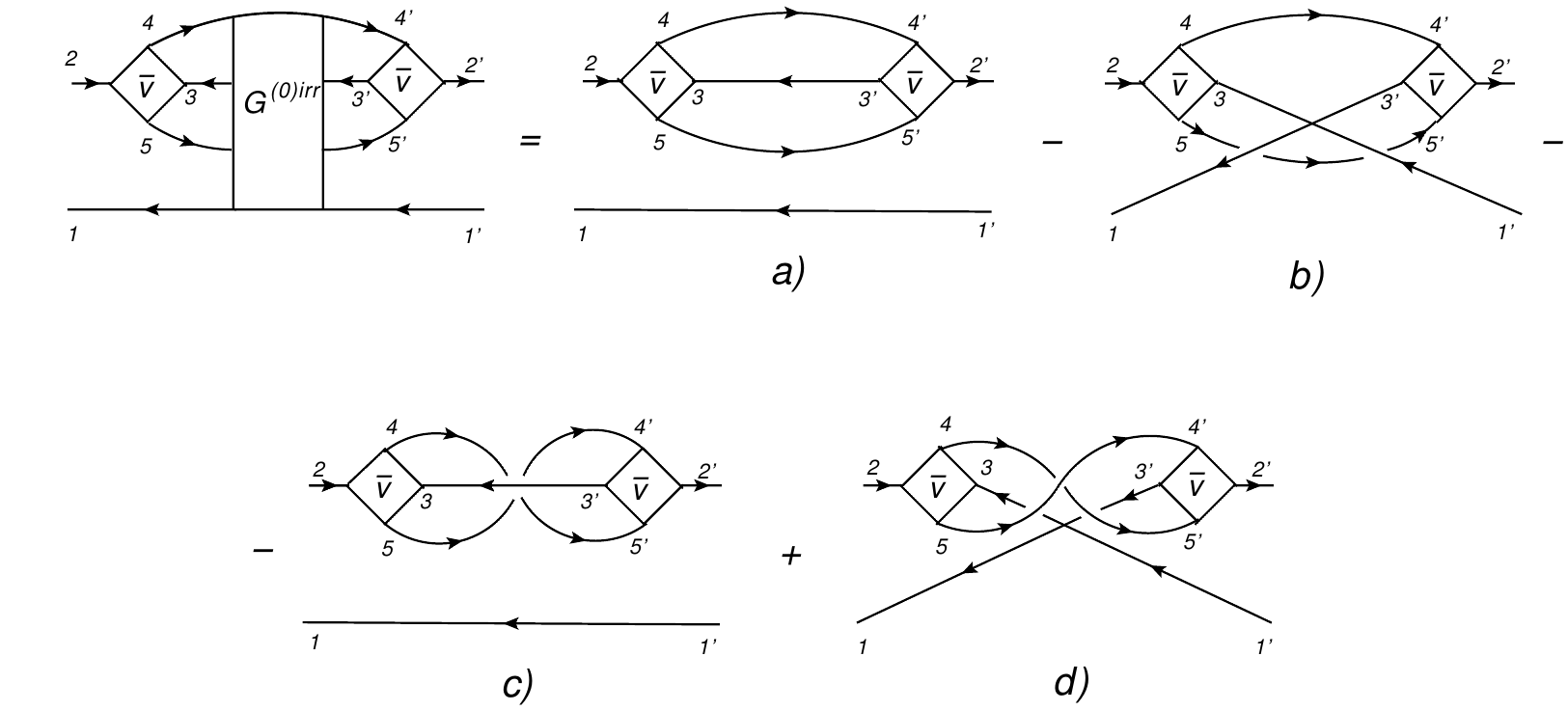}
\end{center}
\caption{Diagrammatic representation of the uncorrelated contributions of Eq. (\ref{G0irr}) to the dynamical kernel $K^{(r;11)}_{12,1'2'}(t-t')$ irreducible with respect to the particle-hole propagator.}
\label{SE2irr0}%
\end{figure*}

As all four components of $T^{(r)}(t-t')$  (\ref{Fcomponents22}) contain the same two-particle-two-hole propagator, it is sufficient to analyze in detail one of them, then the other three can be reconstructed straightforwardly. Let us consider $K^{(r;11)}(t-t')$ as the reference component keeping only the irreducible (with respect to the particle-hole uncorrelated propagator) contributions to $T^{(r)}(t-t')$.
The lowest-order approximation is determined by the uncorrelated irreducible part $G^{(0)irr}$ of the four-fermion propagator:
\bea
G^{(0)irr}(543'1',5'4'31) &=& \langle T({\psi^{\dagger}}_1{\psi^{\dagger}}_3)(t)({\psi}_{3'}{\psi}_{1'})(t')\rangle_0 \times \nonumber \\
&\times&\langle T({\psi}_5{\psi}_4)(t)({\psi^{\dagger}}_{4'}{\psi^{\dagger}}_{5'})(t')\rangle_0, \nonumber\\
\label{G0irr}
\eea
where
\bea
\langle T({\psi^{\dagger}}_1{\psi^{\dagger}}_3)(t)({\psi}_{3'}{\psi}_{1'})(t')\rangle_0 = \nonumber\\
= \langle T{\psi^{\dagger}}_1(t)\psi_{1'}(t')\rangle_0
\langle T{\psi^{\dagger}}_3(t)\psi_{3'}(t')\rangle_0 - \nonumber \\
- \langle T{\psi^{\dagger}}_1(t)\psi_{3'}(t')\rangle_0
\langle T{\psi^{\dagger}}_3(t)\psi_{1'}(t')\rangle_0.
\eea
Thus, the lowest-order approximation requires only the one-fermion (mean-field) Green functions:
\be
G^{(0)}_{11'}(t-t') = \delta_{11'} G^{(0)}_1(t-t') = -i\langle T {\psi}_1(t)\psi^{\dagger}_{1'}(t')\rangle_0
\ee
and their backward-going counterparts. The uncorrelated part $K^{(r;11)0}(t-t')$ of the irreducible kernel $K^{(r;11)}(t-t')$ is given diagrammatically in Fig. \ref{SE2irr0}.
Using the uncorrelated one-fermion Green function:
\be
G^{(0)}_{11'}(t-t') = -i\delta_{11'}\sigma_1\theta(\sigma_1t_{11'})e^{-i{\tilde\varepsilon}_1t_{11'}}
\ee
with $t_{11'} = t_1-t_{1'}$ and $\sigma_1 = \pm 1$ for particle/hole states, the Fourier transform of Eq. (\ref{G0irr}) can be calculated, so that
\be
K^{(r;11)0}_{12,1'2'}(\omega) = \frac{\delta_{11'}}{2}\sum\limits_{345}\frac{{\bar v}_{3245}{\bar v}_{542'3}}
{\omega - {\tilde\varepsilon}_4 - {\tilde\varepsilon}_5 + {\tilde\varepsilon}_1 + {\tilde\varepsilon}_3 + i\delta} - {\cal AS}, 
\ee
where $\delta \to +0$, the explicit term corresponds to the sum of a) and c) contributions, and the antisymmetrized term '$\cal AS$' to the sum of b) and d) contributions shown in Fig. \ref{SE2irr0}. 

The next step in building the dynamical part of the kernel is to include nonperturbative resummations between each pair of fermionic lines.
In this approximation, the correlated contributions to Eq. (\ref{G4}) are represented by products of the two-body correlated  
and uncorrelated propagators
\bea
&G&^{(c)irr}(543'1',5'4'31) =  \nonumber \\
&=& \langle T({\psi^{\dagger}}_1{\psi^{\dagger}}_3)(t)({\psi}_{3'}{\psi}_{1'})(t')\rangle \langle T({\psi}_5{\psi}_4)(t)({\psi^{\dagger}}_{4'}{\psi^{\dagger}}_{5'})(t')\rangle_0 \nonumber \\
&+& \langle T({\psi^{\dagger}}_1{\psi^{\dagger}}_3)(t)({\psi}_{3'}{\psi}_{1'})(t')\rangle_0 \langle T({\psi}_5{\psi}_4)(t)({\psi^{\dagger}}_{4'}{\psi^{\dagger}}_{5'})(t')\rangle  \nonumber \\
&+& \langle T({\psi^{\dagger}}_1{\psi}_5)(t)({\psi^{\dagger}}_{5'}{\psi}_{1'})(t')\rangle \langle T({\psi^{\dagger}}_3{\psi}_4)(t)({\psi^{\dagger}}_{4'}{\psi}_{3'})(t')\rangle_0  \nonumber\\
&+& \langle T({\psi^{\dagger}}_1{\psi}_5)(t)({\psi^{\dagger}}_{5'}{\psi}_{1'})(t')\rangle_0 \langle T({\psi^{\dagger}}_3{\psi}_4)(t)({\psi^{\dagger}}_{4'}{\psi}_{3'})(t')\rangle   \nonumber \\
&+& \langle T({\psi^{\dagger}}_3{\psi}_5)(t)({\psi^{\dagger}}_{5'}{\psi}_{3'})(t')\rangle \langle T({\psi^{\dagger}}_1{\psi}_4)(t)({\psi^{\dagger}}_{4'}{\psi}_{1'})(t')\rangle_0  \nonumber \\
&+& \langle T({\psi^{\dagger}}_3{\psi}_5)(t)({\psi^{\dagger}}_{5'}{\psi}_{3'})(t')\rangle_0  \langle T({\psi^{\dagger}}_1{\psi}_4)(t)({\psi^{\dagger}}_{4'}{\psi}_{1'})(t')\rangle   \nonumber\\
&-& {\cal AS},
\label{Gcirr}
\eea
where the term $\cal AS$ absorbs all possible antisymmetrizations and the index '0' marks an uncorrelated ground state. 
In turn, resummations within the remaining uncorrelated  fermionic pairs lead to the terms with two two-body correlators:
\bea
&G&^{(cc)irr}(543'1',5'4'31) =  \nonumber \\
&=& \langle T({\psi^{\dagger}}_1{\psi^{\dagger}}_3)(t)({\psi}_{3'}{\psi}_{1'})(t')\rangle \langle T({\psi}_5{\psi}_4)(t)({\psi^{\dagger}}_{4'}{\psi^{\dagger}}_{5'})(t')\rangle  \nonumber \\
&+& \langle T({\psi^{\dagger}}_1{\psi}_5)(t)({\psi^{\dagger}}_{5'}{\psi}_{1'})(t')\rangle \langle T({\psi^{\dagger}}_3{\psi}_4)(t)({\psi^{\dagger}}_{4'}{\psi}_{3'})(t')\rangle  \nonumber\\
&+& \langle T({\psi^{\dagger}}_3{\psi}_5)(t)({\psi^{\dagger}}_{5'}{\psi}_{3'})(t')\rangle \langle T({\psi^{\dagger}}_1{\psi}_4)(t)({\psi^{\dagger}}_{4'}{\psi}_{1'})(t')\rangle \nonumber \\
&-& {\cal AS}.
\label{Gccirr}
\eea
\begin{figure*}
\begin{center}
\includegraphics*[scale=0.75]{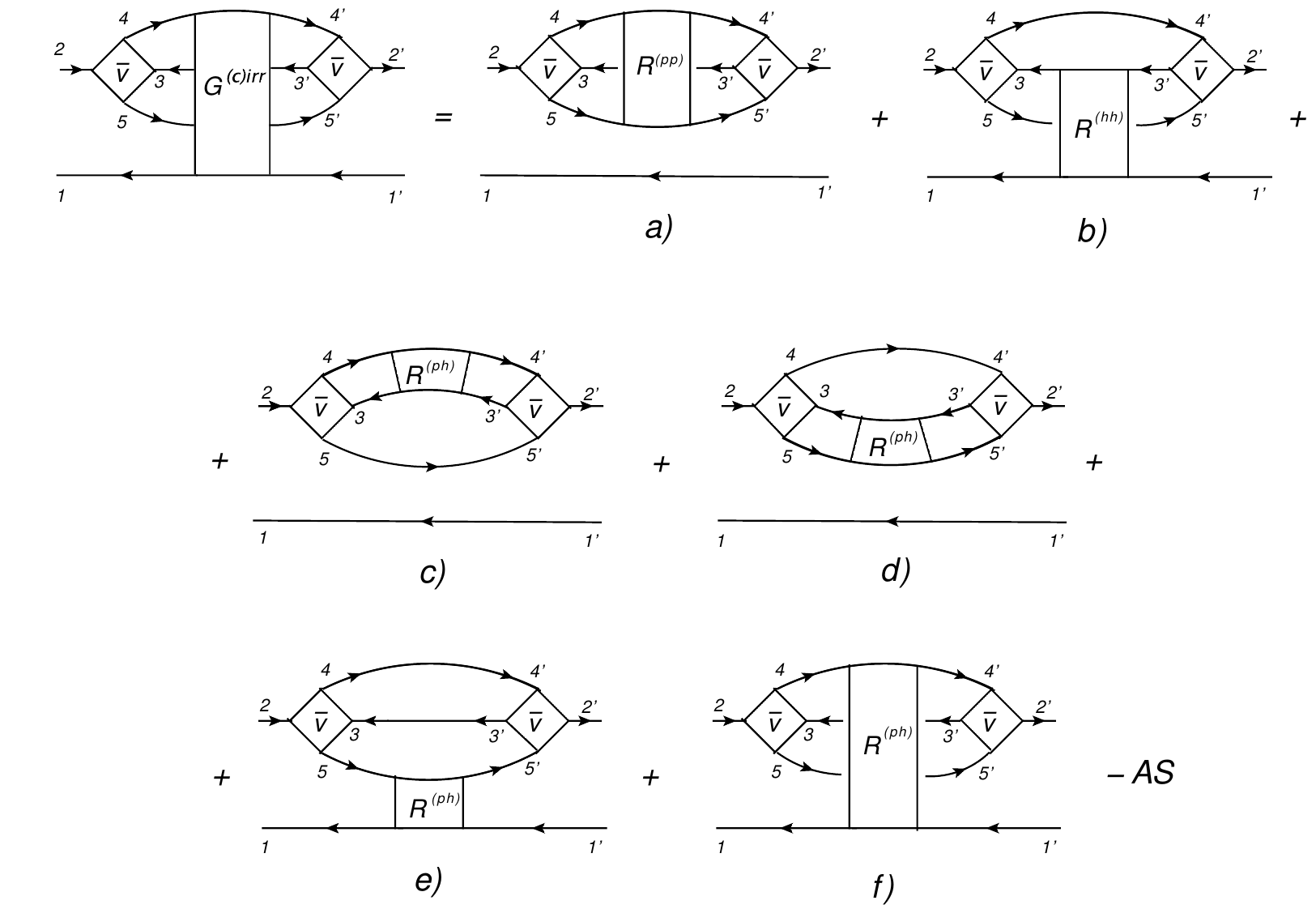}
\end{center}
\caption{Diagrammatic representation of the singly correlated approximation of Eq. (\ref{Gcirr}) to the dynamical kernel $K^{(r;11)}_{12,1'2'}(t-t')$ irreducible with respect to the particle-hole propagator. "AS" includes the antisymmetrized contributions, similar to that of the uncorrelated part, Fig. \ref{SE2irr0}.
}
\label{SE2irrc}%
\end{figure*}
\begin{figure*}
\begin{center}
\hspace{-2cm}\includegraphics*[scale=0.75]{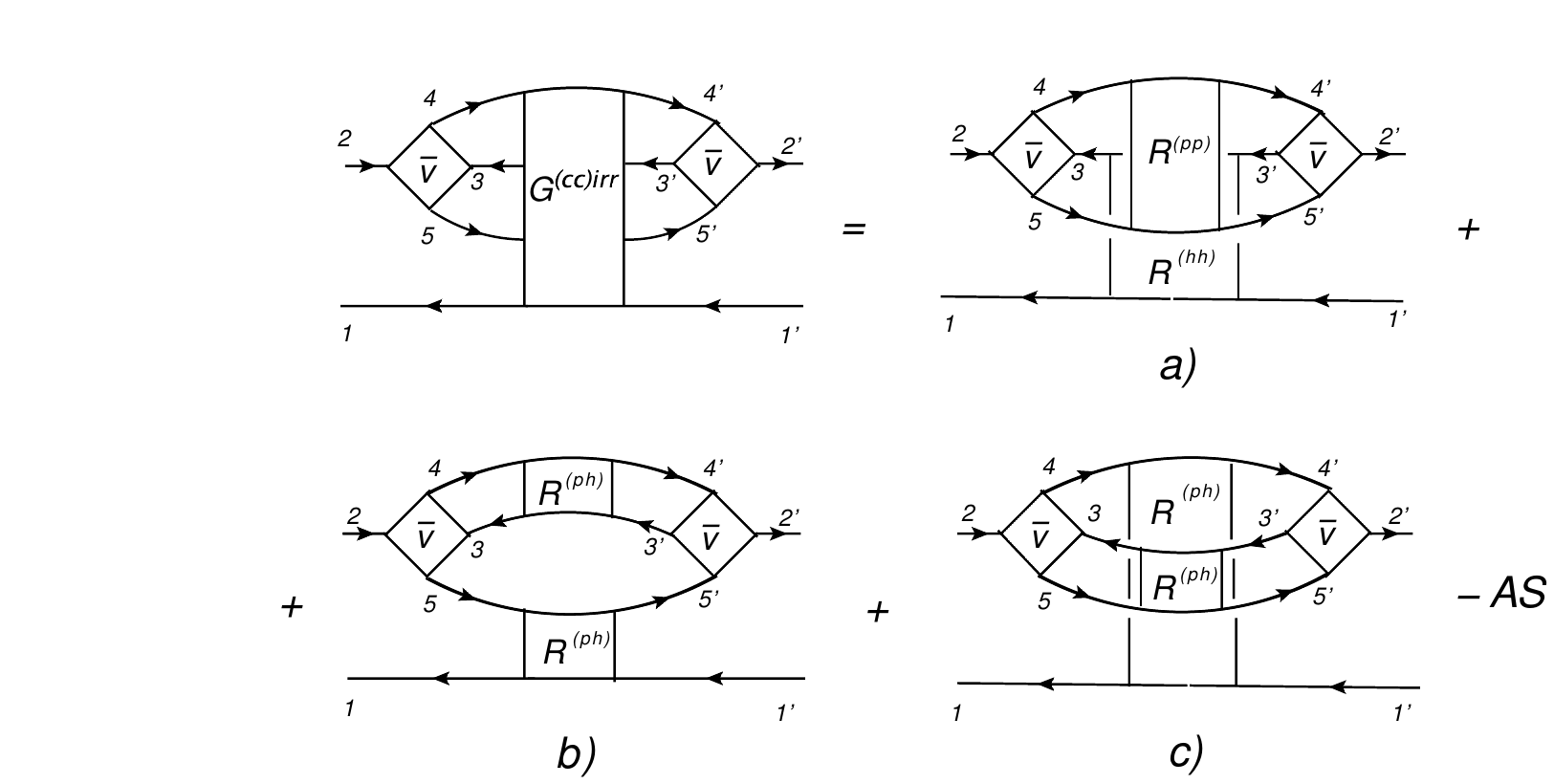}
\end{center}
\caption{Same as in Fig. \ref{SE2irrc}, but for the doubly correlated approximation of Eq. (\ref{Gccirr}).}
\label{SE2irrcc}%
\end{figure*}
%
The latter approach to the dynamical kernel of Eq. (\ref{Dyson2}) is, thus, the most complete one within the concept of the cluster-expansion truncation on the two-body level. The approximation of Eq. (\ref{Gccirr}) only neglects the additional uncorrelated contribution, which should be not very important in the strong-coupling regimes, as in the case of the one-fermionic kernel (\ref{SEirr2}), and the contributions with explicit correlation functions of three and more fermions.
One can see that  the nonperturbative approximations (\ref{Gcirr},\ref{Gccirr}) to the four-fermion propagator (\ref{G4}) contain the particle-hole response functions as well as the two-particle and two-hole Green functions.  
The singly correlated contributions from $G^{(c)irr}$ to the kernel $K^{(r;11)}$ are shown  in Fig. \ref{SE2irrc}, and the doubly correlated ones  in Fig. \ref{SE2irrcc}. 
At this stage it becomes clear that for the determination of the dynamical part of the interaction kernel one needs the knowledge about the two-body correlation functions
of the particle-hole and particle-particle (hole-hole) types, which enter Eqs. (\ref{Gcirr},\ref{Gccirr}).  To formalize this, we can complement the particle-hole propagator considered above, by the particle-particle and hole-hole ones. Thus, we can define
\bea
{\hat R} = \Bigl\{ R^{(ph)}, R^{(pp)}, R^{(hh)} \Bigr\}, \nonumber\\
{\hat R}^{(0)} = 
\Bigl\{ R^{(0; ph)}, R^{(0; pp)}, R^{(0; hh)} \Bigr\}.
\eea
Then, the EOM for ${\hat R}(\omega)$ can be written in a compact form:
\be
{\hat R}(\omega) = {\hat R}^{(0)}(\omega) + {\hat R}^{(0)}(\omega)K[{\hat R}(\omega)]{\hat R}(\omega).
\label{Dyson3}
\ee
Remarkably enough, all frequency (time) dependence of the kernel originates from the internal two-fermion propagators which are themselves the main variables. Such a closed approach has been discussed, in particular,  in the EOM language for the two-fermion 
Green functions \cite{AdachiSchuck1989,Danielewicz1994,DukelskyRoepkeSchuck1998} and time-dependent density matrices in Refs. \cite{SchuckTohyama2016,SchuckTohyama2016a}. A similar approach to the nuclear response has been developed over the years as the method of chronological decoupling of diagrams or the {\it time blocking approximation} \cite{Tselyaev1989,KamerdzhievTertychnyiTselyaev1997,LitvinovaTselyaev2007,Tselyaev2007,LitvinovaRingTselyaev2007,LitvinovaRingTselyaev2008}
which has become self-consistent in its later implementations based on meson-nucleon Lagrangians \cite{LitvinovaRingTselyaev2007,LitvinovaRingTselyaev2008,LitvinovaRingTselyaev2013,Litvinova2015,Litvinova2016,RobinLitvinova2016,RobinLitvinova2018}. This approach starts from the general Bethe-Salpeter equation for a four-time two-fermion Green function, but after applying a certain time projection technique reduces to the two-time or single-frequency equation of motion of the Dyson type with the kernel, which is topologically equivalent to selected components of the EOM singly-correlated kernel. This model is discussed in detail in the next Section. Analogous approaches of the nuclear field theory, although derived differently, also lead to single-frequency EOM's with frequency-dependent kernels containing couplings between single-fermion and phonon propagators \cite{Broglia1976,BortignonBrogliaBesEtAl1977,BertschBortignonBroglia1983,ColoBortignon2001,NiuColoBrennaEtAl2012}. 
All these methods are not related explicitly to the bare nucleon-nucleon interactions and based on phenomenological descriptions of the mean field, instead of the static part of the kernel (\ref{Fstatic},\ref{Fstatic1}), and effective in-medium nucleon-nucleon interactions. The EOM method for two-fermion response functions, taking into account both static and dynamical parts of the interaction kernel, is now being more applied in quantum chemistry \cite{Olevano2018}.


It is not likely that Eq. (\ref{Dyson3}) with the dynamical interaction kernels of Eqs. (\ref{Gcirr},\ref{Gccirr}) can be solved analytically except for some toy models \cite{SchuckTohyama2016,SchuckTohyama2016a}.
In practice, its self-consistent solution can be found iteratively.  To initialize an iterative algorithm, one would need a starting approximation to the response functions contained in the kernel $K[{\hat R}(\omega)]$. In the cases of phenomenological models discussed above and in the next Section, the starting-point response functions are approximated by the solution of Eq. (\ref{Dyson3}) keeping only the static term in the kernel, that corresponds to the random phase approximation with effective interactions. 

\section{Particle-vibration coupling (PVC) model  in the time blocking approximation (TBA)}
\label{PVC-TBA}

In this Section we revisit and investigate the physical content of the approach to the particle-hole response function of Eq. (\ref{Dyson2}),
known as particle-vibration coupling  model, in the context of the EOM method 
discussed above. 
In the Green function language, the PVC model was formulated in Refs.  \cite{Tselyaev1989,KamerdzhievTertychnyiTselyaev1997} within a method of chronological decoupling of diagrams. It emerged as an extension of the Migdal's theory of finite Fermi systems \cite{Migdal1967} beyond the quasiparticle random phase approximation (QRPA). 
The extended theory was generalized for the case of superfluid pairing in Refs. \cite{Tselyaev2007,LitvinovaTselyaev2007} and received a relativistic formulation and a fully self-consistent (parameter-free) implementation in Refs. \cite{LitvinovaRingTselyaev2007,LitvinovaRingTselyaev2008,MarketinLitvinovaVretenarEtAl2012,RobinLitvinova2016}. In the latter versions and numerous later developments and applications the method included the same idea of chronological decoupling of diagrams and was renamed to time blocking approximation (TBA). 

The final expressions of the PVC-TBA approach for the dynamical kernel of Eq. (\ref{Dyson2}) turned out to be equivalent to those of the NFT \cite{BertschBortignonBroglia1983,ColoBortignon2001} obtained by making use of the perturbation theory for the coupling between the quasiparticles and collective doorway states. Both approaches were based on the assumption about the dynamical part of the nucleonic self-energy in the EOM form of Eq. (\ref{SEirr2}), where the bare interaction is replaced by the effective interaction, and typically kept only the second term of it, which occurred to be dominant. The intermediate fermionic line in this term was approximated by the mean-field one-fermion Green function. The conventional PVC-TBA and NFT response theories were confined by the $ph\otimes phonon$ configurations following the idea of a small parameter hidden in the effective PVC vertices. Although PVC-TBA showed a considerable improvement of the description of nuclear excited states compared to QRPA, its unclear foundation and uncontrollable approximations remained the drawbacks preventing this approach from further development. Later, based on formal similarities with the EOM kernels \cite{Schuck1976}, the PVC-TBA method was generalized to the case of $2phonon$ configurations \cite{Tselyaev2007} implemented in the relativistic framework in Refs. \cite{LitvinovaRingTselyaev2010,LitvinovaRingTselyaev2013}. Further extensions were discussed in Refs. \cite{Litvinova2015,Tselyaev2018}, however, without a 
comparison to the EOM method and without a detailed analysis of the assumptions intuitively based on the EOM derivation.  Thus, the goals of this and subsequent Sections are (i) to explicitly compare the EOM with the dynamical kernel discussed in Section \ref{EOM2} with the PVC-TBA and, with the insights of this comparison, (ii) develop a systematic approach to the response functions with a non-perturbative treatment of higher configurations in a strongly correlated medium. 


Below we will follow the formalism and conventions of Refs. \cite{KamerdzhievTertychnyiTselyaev1997,LitvinovaRingTselyaev2007}, so that the index mapping will require the replacement $R(12,34)\to R(21,43)$ in order to compare the equations for the particle-hole response function with those of the previous section. Note that 
in the PVC-TBA model, one starts from the Bethe-Salpeter equation (BSE) for a more general four-time response \cite{Tselyaev1989,KamerdzhievTertychnyiTselyaev1997}: 
\bea
R(12,34)&=&G(1,3)G(4,2)-\nonumber \\
&-& i\sum\limits_{5678}G(1,5)G(6,2)V(56,78)R(78,34),\nonumber\\
\label{bse}%
\eea
where the summation over the number indices $1$, $2,\dots$ implies integration
over the respective time variables, and $V$ is the interaction amplitude irreducible in the $ph$-channel. This amplitude is, in general, a
variational derivative of the one-fermion self-energy $\Sigma$ with respect to the
exact single-particle Green function:
\begin{equation}
V(12,34)=i\frac{\delta\Sigma(1,2)}{\delta G(3,4)} \label{uampl}%
\end{equation}
and a four-time analog of the kernel of Eq. (\ref{Womega}).
Similarly, it is a sum of the
static effective interaction $\tilde{V}$ and the time-dependent (energy-dependent) one $V^{(e)}$:
\bea
V(12,34)&=&\tilde{V}(12,34)+V^{(e)}(12,34) \nonumber \\
\tilde{V}(12,34)&=&\tilde{V}_{1234}\delta(t_{31})\delta
(t_{21})\delta(t_{34}) \label{V-static} \nonumber\\
V^{(e)}(12,34)&=&i\frac{\delta{\Sigma^{(e)}(1,2)}}{\delta G(3,4)}, \label{dcons}%
\label{effint}
\eea
where we implied that $t_{12}=t_{1}-t_{2}$.
Despite the fact that in this approach one starts from the formally exact BSE (\ref{bse}) 
and employs the decomposition (\ref{dcons}) of the interaction kernel derived by the EOM method, 
it does not imply a connection with the underlying bare interaction.
Instead, the static part of the self-energy is adjusted to experimental data, for instance, the data
on nuclear binding energies and radii assuming that they depend only on one-body density
or fitted to reproduce the lowest single-particle excitations obtained in knock-out or transfer reactions.
Thus, in this theory,
\bea
\Sigma_{11'}(\varepsilon)  &=&  {\tilde\Sigma_{11'}} + \Sigma^{(e)}_{11'}(\varepsilon) \nonumber\\
{\tilde \Sigma}_{11'} &=& \sum\limits_{22'}{\tilde V}_{11'22'}\rho^{(0)}_{22'}\nonumber\\
\Sigma_{11'}^{(e)}(\varepsilon)&=&\sum\limits_{\nu,2}\frac{g_{12}^{\nu
(\sigma_{2})}\,g_{1'2}^{\nu(\sigma_{2})\ast}}{\varepsilon-{\tilde\varepsilon
_{2}}-\sigma_{2}(\Omega_{\nu}-i\delta)}, \ \ \ \delta \to +0,
\label{PVCsigma}
\eea
where the dynamical part of the self energy is borrowed from that of the EOM (\ref{SEirr2}), neglecting the uncorrelated part, performing the mapping 
(\ref{mappingph},\ref{gDPVCph}) with the replacement $\bar v \to {\tilde V}$ and Fourier transformation to the energy domain. In Eq. (\ref{PVCsigma}) ${\tilde\varepsilon}_{{1}}$ are the eigenvalues 
 of the phenomenological one-body part of the Hamiltonian.
Usually the PVC models do not include coupling to pairing phonons, but in principle it can be added as a term similar to the last line of Eq. (\ref{PVCsigma}).
Notice that, although both the self-energy (\ref{PVCsigma}) and the interaction kernel (\ref{effint}) are again decomposed into the static and dynamical parts, in the effective theory these parts 
are not equivalent to those in the EOM starting from the bare interaction. However, we can call them {\it topologically equivalent} as they have the same internal propagator structure.

\begin{figure}[ptb]
\begin{center}
\includegraphics*[scale=0.43]{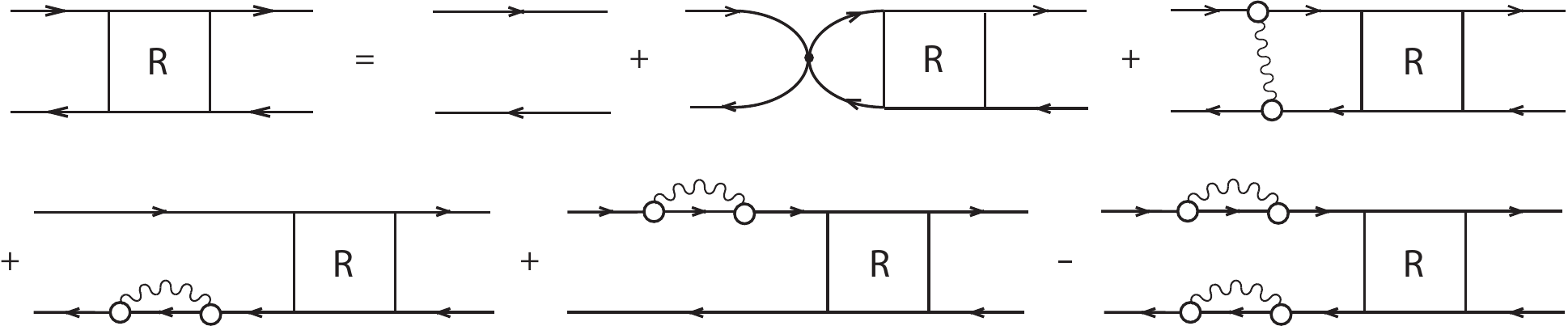}
\end{center}
\caption{Bethe-Salpeter equation for the $ph$-response function $R$ corresponding to Eq.~(\ref{bseop1}) with the interaction kernel defined by  Eqs. (\ref{wampl},\ref{wampl1})
in graphical representation (not time-ordered). Phonon vertices and propagators are defined in Fig.~\ref{PVCmap} and
the small black circle stands for the static part of the residual effective
$ph$-interaction $\tilde V$. }%
\label{f1}%
\end{figure}

In terms of the free response, 
which is, in the time domain, a product of two one-fermion propagators 
$R^{0}(12,34)=G(1,3)G(4,2)$, the BSE
(\ref{bse}) can be written in the operator form as
\begin{equation}
R=R^{0}-iR^{0}VR.
\label{bseop}
\end{equation}
Since the static mean-field part of the interaction kernel is fixed by fitting the global characteristics of the many-body system to data,
it is convenient for further analysis
to eliminate the exact Green function $G$ and rewrite it
in terms of the mean field Green function $\tilde{G}$, such as
\be
{\tilde G}_{11'}(\varepsilon) 
= G^{(0)}_{11'}(\varepsilon) + 
\sum\limits_{22'}G^{(0)}_{12}(\varepsilon){\tilde\Sigma}_{22'}(\varepsilon){\tilde G}_{2'1'}(\varepsilon),
\label{spEOM4}
\ee
with
\bea
{\tilde{G}}(1,2)&=&-i\sigma_{{1}}\delta_{{1}{2}}\theta(\sigma_{{1}}%
\tau)e^{-i{\tilde\varepsilon}_{{1}}\tau},\ \ \ \ \ \tau=t_{1}-t_{2}, \\
{\tilde{G}}_{{1}{2}}(\varepsilon)&=&\frac{\delta_{{1}{2}}}%
{\varepsilon-{\tilde\varepsilon}_{{1}}+i\sigma_{{1}}\delta}, \ \ \ \ \ \sigma_1 = \text{sign}({\tilde\varepsilon}_1),  
\label{G-tilde}%
\eea
where $\delta \to +0$. 
Using the connection between $\tilde{G}$ and $G$ in the Nambu form
\begin{equation}
{\tilde{G}}^{-1}(1,2)=G^{-1}(1,2)+\Sigma^{(e)}(1,2),
\end{equation}
one can rewrite Eq.~(\ref{bseop}) as follows:
\begin{equation}
R={\tilde{R}}^{0}-i\tilde{R}^{0}WR, 
\label{bseop1}%
\end{equation}
with the uncorrelated mean-field particle-hole 
 response function
${\tilde{R}}^{0}(12,34)={\tilde{G}}(1,3){\tilde{G}}(4,2)$ and $W$ as a
new interaction kernel of the form
\begin{equation}
W=\tilde{V}+W^{(e)},
\label{wampl}%
\end{equation}
where
\bea
W^{(e)}(12,34)=V^{(e)}(12,34)+i\Sigma^{(e)}(1,3){\tilde{G}}^{-1}(4,2)+\nonumber \\
+ i{\tilde{G}%
}^{-1}(1,3)\Sigma^{(e)}(4,2)-i\Sigma^{(e)}(1,3)\Sigma^{e}(4,2). \nonumber \\
\label{wampl1}%
\eea
The BSE in this form is more convenient to consider since the mean-field Green function ${\tilde{G}}$ 
 is well defined. However, the interaction kernel $W$ in Eq.~(\ref{wampl}) becomes more
complicated. The graphical representation of the Eq.~(\ref{bseop1}) with the interaction kernel defined by  Eqs. (\ref{wampl},\ref{wampl1}) is
shown in Fig.~\ref{f1}.
Thus, one can see that the dynamical part of the kernel takes the form, which is similar to the one obtained in the EOM,
but still depends on four times. The last term $i\Sigma ^{e}(1,3)\Sigma^{e}(4,2)$ of Eq. (\ref{wampl1}) can be still related to one of the doubly correlated contributions of 
the EOM, but in the present context it looks very different from the others. As it is discussed in Ref. \cite{KamerdzhievTertychnyiTselyaev1997}, this term
is of a higher order and compensates multiple counting of the particle-phonon coupling
arising from the two previous terms, if the entire dynamical kernel is still unrestricted by only particle-hole pairs of indices.  
Indeed, terms with backward-going correlation functions in the dynamical kernel are also possible, but they require a
special consideration. They have been analyzed, in particular, within a phenomenological approach and included in calculations of nuclear neutral excitations in
Refs. \cite{Tselyaev1989,Kamerdzhiev1991,KamerdzhievTertychnyiTselyaev1997} as well as more recently within a self-consistent relativistic framework applied to charge-exchange excitations in Ref. \cite{Robin2019}. 

In the leading resonant time blocking approximation, which is discussed below, the possibility of having particle-particle and hole-hole states as well as the connection between particle-hole and hole-particle states in the dynamical kernel is neglected. This corresponds to the absence of ground state correlations more complex than the particle-hole ones.
In this case multiple counting does not take place and  the term $i\Sigma^{e}(1,3)\Sigma^{e}(4,2)$ can be omitted. 

Let us consider the Fourier transformation of the Eq.
(\ref{bseop1}) to the energy domain. The response function formally depends on four time variables, but, in fact, on three time differences, because of the time translational invariance.
Thus, a triple Fourier transform is needed to translate the BSE (\ref{bseop1}) into the equation with respect to energy variables. In order to obtain the spectral representation of the response, two of them have to be integrated out. These operations lead to the following equation:
\bea
R_{12,34}(\omega,\varepsilon)  &=& {\tilde{G}}_{13}%
(\varepsilon+\omega){\tilde{G}}_{42}(\varepsilon) \nonumber\\
+\sum\limits_{5678}{\tilde{G}}_{15}(\varepsilon
&+&\omega){\tilde{G}}_{62}(\varepsilon)
\int\limits_{-\infty}^{\infty
}\frac{d\varepsilon^{\prime}}{2\pi i}
W_{56,78}(\omega
,\varepsilon,\varepsilon^{\prime})R_{78,34}(\omega
,\varepsilon^{\prime}) \nonumber \\
\label{bsee}%
\eea
with the subsequent integration over $\varepsilon$:
\be
R_{12,34}(\omega)  = \int\limits_{-\infty}^{\infty
}\frac{d\varepsilon}{2\pi i}R_{12,34}(\omega,\varepsilon).
\ee
One can notice, however, that both the solution of Eq. (\ref{bsee})  $R$ and its kernel
$W$ are singular with respect to the energy variables. This is related to the 
fact that Eq.~(\ref{bseop1}) contains integrations
over all time points of the intermediate states. This means that
many configurations which are actually more complex than 
$1p1h\otimes$phonon are contained in the exact response function. In
Ref.~\cite{Tselyaev1989} a special time-projection technique was
introduced to block the $ph$-propagation through these complex
intermediate states. It has been shown that for this type of
response it is possible to reduce the integral equation (\ref{bsee})
to a relatively simple algebraic equation. Below we will see that this approximation corresponds to retaining certain part of the terms
specified in Eq. (\ref{Gcirr}) in the internal Green functions
propagators 
of the dynamical two-time kernel in the EOM method.
\begin{figure*}
\begin{center}
\includegraphics*[scale=0.92]{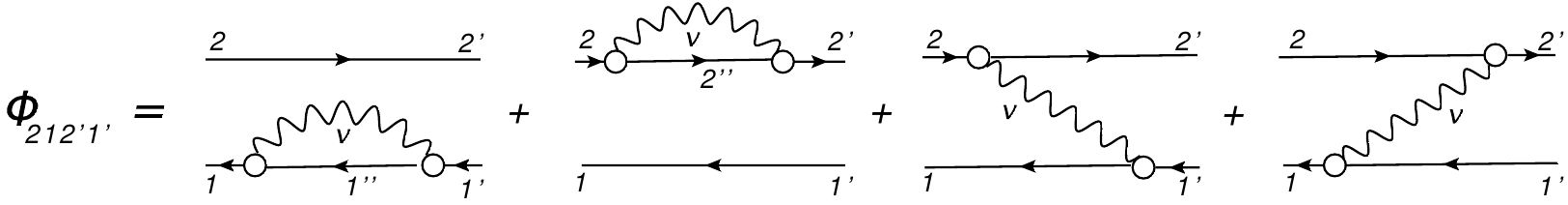}
\end{center}
\caption{Dynamical kernel (time-ordered) of the PVC-TBA equation for the particle-hole response function in the resonant approximation  of Eq. (\ref{Phikernel}).}
\label{Phi_PVC}%
\end{figure*}

The Bethe-Salpeter equation (\ref{bseop1}) can be solved in two steps. First one can calculate the
correlated particle-hole response $R^{(e)}$, which describes the particle-hole propagation under
the influence of the interaction $W^{(e)}$
\begin{equation}
R^{(e)}={\tilde{R}}^{0}-i{\tilde{R}}^{0}W^{(e)}R^{(e)} \label{RE}%
\end{equation}
and contains all the effects of particle-phonon coupling and all the
singularities of the integral part of the main BSE (\ref{bseop1}). Second, the remaining equation for the full response function $R$
\begin{equation}
R=R^{(e)}-iR^{(e)}\tilde{V}R, \label{respre}%
\end{equation}
which contains only the static effective interaction
$\tilde{V}$, can be easily solved when $R^{(e)}$ is known. Thus, the
main problem to address singularities is to calculate the correlated particle-hole response $R^{(e)}$.
The latter can be represented  as an infinite series of graphs with uncorrelated
$ph$-propagators alternated with single interaction events:
\bea
R^{(e)}  &  ={\tilde{R}}^{0}-i{\tilde{R}}^{0}{\Gamma}^{(e)}{\tilde{R}}%
^{0},\label{re}\\
{\Gamma}^{(e)}  &  =W^{(e)}-iW^{(e)}{\tilde{R}}^{0}{\Gamma}^{(e)}, \label{gamma}%
\eea
where ${\Gamma}^{(e)}$ is, thus, a reducible analog of  $W^{(e)}$ containing 
correlated two-particle-two-hole blocks connected by the uncorrelated
$ph$-propagators. Then, to avoid higher-complexity configurations, 
a time-projection operator
\begin{equation}
\Theta(12,34)=\delta_{\sigma_{{1}},-\sigma_{{2}}}
\delta_{{1}{3}}\delta_{{2}{4}}
\theta(\sigma_{{1}}t_{14})\theta(\sigma_{{1}}t_{23})
\label{theta}%
\end{equation}
with the Heaviside type functions $\theta(t)$ is introduced into the integral part of Eq.~(\ref{gamma}) 
\cite{Tselyaev1989} according to:
\begin{equation}
\tilde{R}^{0}(12,34)\rightarrow\tilde{R}^{0}(12,34)\Theta(12,34),
\end{equation}
so that
\bea
{\Gamma}^{(e)}(12,34)=W^{(e)}(12,34)-\nonumber\\ 
-i\sum\limits_{5678}W^{(e)}%
(12,56){\tilde{R}}^{0}(56,78)\Theta(56,78){\Gamma}^{(e)}(78,34). \label{gamma1}\nonumber \\
\eea
After the Fourier transformation in time we restrict ourselves to the two-time response
function $R_{12,34}(\omega)$, because it has to be subsequently contracted with equal-times external
field operators: 
\bea
R_{12,34}(\omega)&=&-i\int\limits_{-\infty}^{\infty}dt_{1}%
dt_{2}dt_{3}dt_{4}\delta(t_{1}-t_{2})\delta(t_{3}-t_{4})\times\nonumber\\
&\times& \delta(t_{4}%
)e^{i\omega t_{13}}R(12,34), \label{tbaresp}%
\eea
which depends only on one energy variable $\omega$.
As a result, we obtain an
algebraic equation for the spectral representation of the particle-hole response:
\bea
R_{12,34}(\omega)=\tilde{R}_{12,34}^{0}%
(\omega)+\nonumber\\
+\sum\limits_{5678}\tilde{R}_{12,56}^{0}(\omega)(\tilde{V}_{56,78}%
+\Phi_{56,78}(\omega))R_{78,34}%
(\omega),\nonumber\\ 
\label{respdir}%
\eea
where
\bea
{\tilde{R}}_{12,34}^{0}(\omega)={\tilde{R}}_{12}(\omega)\delta_{13}\delta_{24} \label{R0}\\
{\tilde{R}}_{12}(\omega) = \frac{n_2 - n_1}{\omega -{\tilde\varepsilon}_{12}},%
\eea
${\tilde\varepsilon}_{12}={\tilde\varepsilon}_{1}-{\tilde\varepsilon}_{2}$ and $\Phi(\omega)$ is the
particle-phonon coupling amplitude with the following components:
$ph-ph$ matrix element has the form
\bea
\Phi_{12,1'2'}^{(ph,ph)}(\omega) = \sum\limits_{\nu}\Bigl[  
\delta_{22'}\sum\limits_{1''}\frac{g^{\nu}_{11''}g^{\nu*}_{1'1''}}{\omega - {\tilde\varepsilon}_{1''} + {\tilde\varepsilon}_{2} - \Omega_{\nu}}\nonumber \\
+ \delta_{11'}\sum\limits_{2''}\frac{g^{\nu}_{2''2}g^{{\nu}*}_{2''2'}}{\omega - {\tilde\varepsilon}_1 + {\tilde\varepsilon}_{2''} - \Omega_{\nu}} 
- \frac{g^{\nu}_{11'}g^{{\nu}*}_{22'}}{\omega - {\tilde\varepsilon}_{1'} + {\tilde\varepsilon}_{2} - \Omega_{\nu}} - \nonumber \\ -
\frac{g^{{\nu}*}_{1'1}g^{{\nu}}_{2'2}}{\omega - {\tilde\varepsilon}_{1} + {\tilde\varepsilon}_{2'} - \Omega_{\nu}} \Bigr],\nonumber \\
\label{Phikernel}
\eea
where $\{1,1',1''\}$ are the particle states with ${\tilde\varepsilon}_{1},{\tilde\varepsilon}_{1'},{\tilde\varepsilon}_{1''}  > \varepsilon_F$, $\{2,2',2''\}$ are the hole states with ${\tilde\varepsilon}_{2},{\tilde\varepsilon}_{2'},{\tilde\varepsilon}_{2''} \leq \varepsilon_F$ and $\varepsilon_F$ is the Fermi energy. The hp-hp matrix elements $\Phi_{12,1'2'}^{(hp,hp)}(\omega)$ are obtained by Hermitian conjugation and time-reversal transformation $\omega \to -\omega$.  The diagrammatic representation of $\Phi (\omega)$ is shown in Fig. \ref{Phi_PVC}.
\begin{figure}
\begin{center}
\includegraphics[scale=0.5]{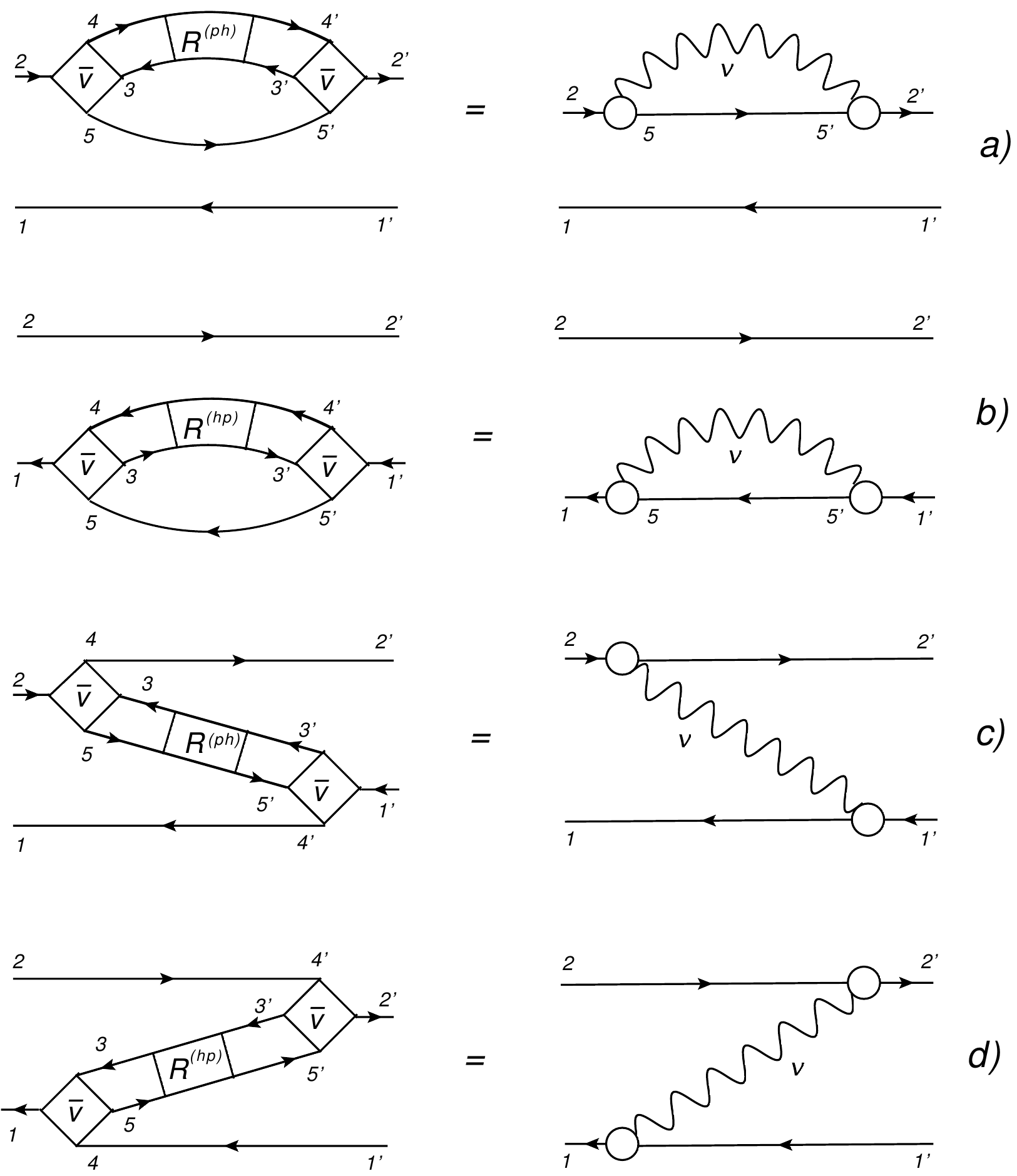}
\end{center}
\caption{Diagrammatic mapping of the singly correlated terms of $K^{(r)}$ containing $R^{(ph)}$ to the PVC kernel, according to Eq. (\ref{mappingph}). }
\label{SE2_PVC_ph}%
\end{figure}

\begin{figure}
\begin{center}
\includegraphics[scale=0.5]{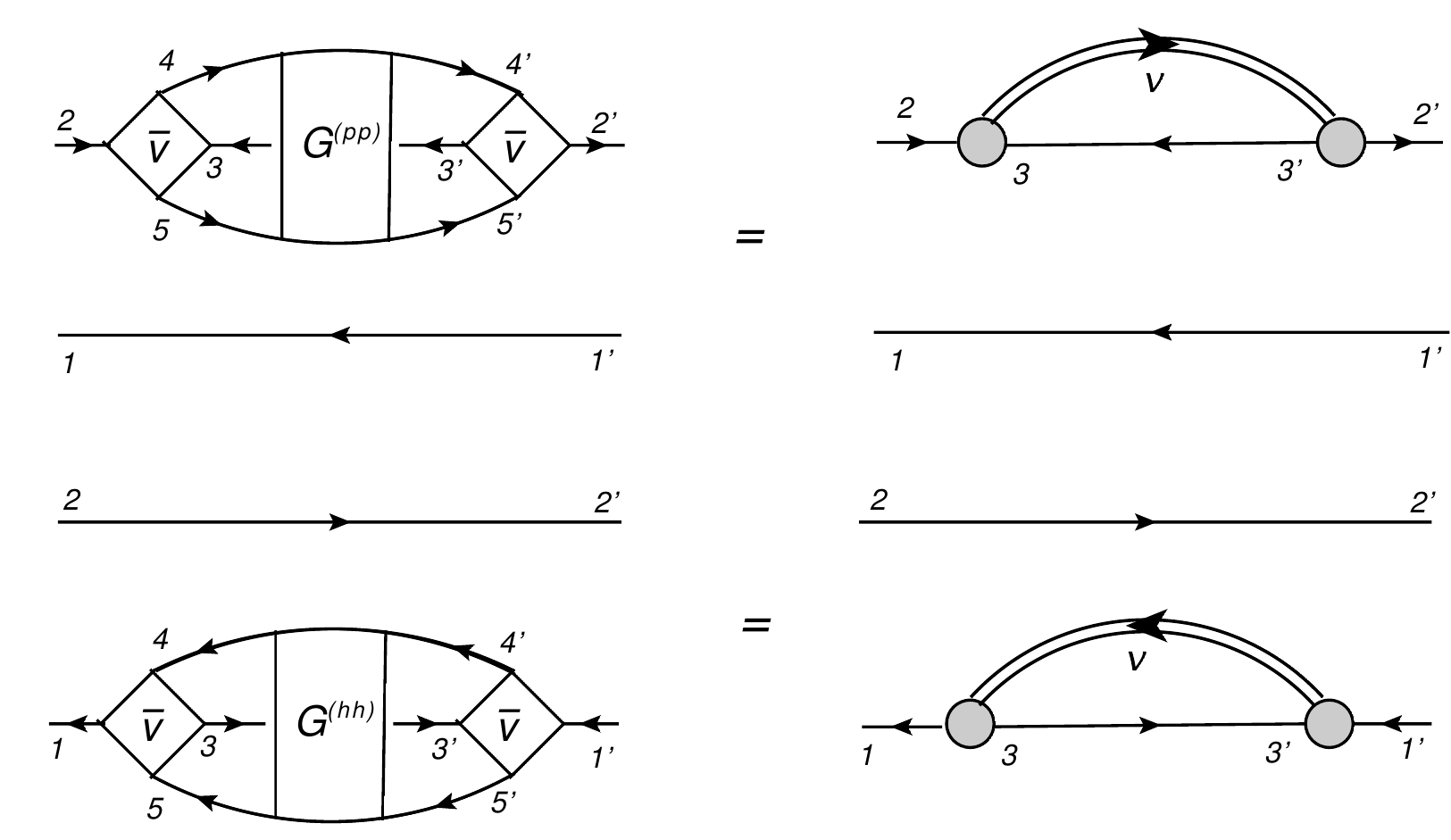}
\end{center}
\caption{Diagrammatic mapping of the singly-correlated terms of $K^{(r)}$ containing $G^{(pp)}$ and $G^{(hh)}$ from $K^{(r;11)}$ (top) and $K^{(r;22)}$ (bottom), respectively, to the PVC kernel, according to Eq. (\ref{mappingpp}). 
}
\label{SE2_PVC_pp}%
\end{figure}

Thus, we have obtained the expression for the dynamical interaction kernel which can be compared to the kernel of the EOM method. Indeed, a complete matching can be revealed by looking at the parts of the latter kernel associated with the singly-correlated terms and performing the mapping defined by Eq. (\ref{gDPVCph}). The matching is illustrated in Figs. \ref{SE2_PVC_ph},\ref{SE2_PVC_pp} and can be additionally verified by taking the Fourier transform of the dynamical kernel with the  internal Green functions of the type (\ref{Gcirr}) \cite{Olevano2018}. One should be, however, aware of the 
differences between ${\bar v}$ of the ab-initio theory and ${\tilde V}$ used in effective theories as well as of the double-counting removal  correction needed in the latter case.
%
In the framework of an effective theory, indeed, an important correction has to be done to the Eq.~(\ref{respdir}).
Being adjusted to experimental data, the static interaction $\tilde V$ contains, in principle, all beyond-mean-field
correlations, in particular, those which are explicitly included into the dynamical kernel ${\tilde V}^{(e)}$, in the static approximation. 
Therefore, a direct addition of the dynamical interaction leads to a double counting of
the static correlations contained in $\Phi(\omega)$.  In order to avoid this, the dynamical kernel has to be corrected as follows \cite{Tselyaev2013}:
\begin{equation}
\Phi(\omega)\rightarrow\delta\Phi(\omega)=\Phi(\omega)-\Phi(0)\mathbf{.}
\label{subtraction-procedure}%
\end{equation}
The physical meaning of this subtraction is clear: the average value
of the particle-vibration coupling amplitude $\Phi$ in the ground
state is supposed to be contained already in the residual effective
interaction $\tilde{V}$, therefore, we should take into account only
the additional energy dependence, i.e.,
$\delta\Phi(\omega)=\Phi(\omega)-\Phi(0)$, on top of this effective
interaction. Instead of Eq.~(\ref{respdir}), we finally
solve the following response equation:
\begin{equation}
R=\tilde{R}^{0}+\tilde{R}^{0}[\tilde{V}+\delta\Phi]R. 
\label{resp-subtr}%
\end{equation}

In early applications of PVC-TBA \cite{Tselyaev1989,KamerdzhievTertychnyiTselyaev1997} the phonon coupling vertices $g^{\nu}$ were calculated based on experimental information about deformation parameters for the lowest collective excitations. Their experimental energies were taken as the $\Omega_{\nu}$ values. The common practice was to include only very few collective phonons, to justify neglecting the uncorrelated term in the dynamical kernel.
In later applications, such as in Refs. \cite{LitvinovaTselyaev2007,LitvinovaRingTselyaev2007,LitvinovaRingTselyaev2008,Lyutorovich2008,LitvinovaRingTselyaev2010,NiuColoBrennaEtAl2012,LitvinovaRingTselyaev2013,NiuColoVigezziEtAl2014,LyutorovichTselyaevSpethEtAl2015,Tselyaev2016,RobinLitvinova2016}, the phonon spectra were extracted from the RPA solutions, i.e., the solutions of Eq. (\ref{respdir}) without the dynamical kernel (more specifically, keeping only the first term $\tilde V$ in the interaction kernel of Eq. (\ref{respdir})). These works adopted larger phonon spaces and the subtraction of Eq. (\ref{subtraction-procedure}) elaborated in detail in Ref. \cite{Tselyaev2013}, thus, representing an important step toward  a closed calculation scheme for the particle-hole response function containing dynamical medium effects in nuclei. 

The first fully self-consistent version of the PVC-TBA was implemented in the relativistic framework based on the effective meson-nucleon Lagrangian of quantum hadrodynamics \cite{LitvinovaRingTselyaev2007}. In this approach, the most important particle-hole
phonons again correspond to the solutions of the same equation (\ref{respdir}) without the dynamical kernel. In a more complete version also particle-particle and hole-hole (pairing) phonons should be included as well as the phonons appeared in the form of proton-neutron (charge-exchange) correlated pairs.
Contribution of the charge-exchange phonons was investigated within the relativistic formalism in Refs. \cite{Litvinova2016,RobinLitvinova2018}, where their role was found somewhat smaller than that of the regular like-particle ones, but sizable enough to modify the single-particle structure \cite{Litvinova2016} and spin-isospin strength distributions \cite{RobinLitvinova2018}. The contribution of pairing vibrations to the dynamical kernel of Eq. (\ref{Dyson2}) is commonly neglected, however, a systematic study of those effects is highly desirable to properly assess their role. 

But even without the inclusion of the pairing channels, accurate solutions of Eqs. (\ref{Dyson2},\ref{respdir}) require an iterative scheme, as discussed in Section \ref{EOM2}. This was not realized in the conventional PVC-TBA, however, the idea of iterative solutions is straightforward and can be, in principle, implemented numerically. In the calculations based on the effective interaction derived from a reasonably good density functional theory the most important phonons are described relatively well already in RPA.
Their characteristics do not change significantly (except for acquiring fragmentation and larger widths by the high-frequency and soft modes, which is expected to give second-order effects to the dynamical kernel) in the calculations beyond RPA. This has been verified in Ref. \cite{Tselyaev2018} by direct calculations, where the phonons obtained as full solutions of Eq. (\ref{respdir}) were recycled and reused in the dynamical PVC-TBA kernel (\ref{respdir}) to compute the dipole response of some medium-mass and heavy nuclei, which remained almost unaffected by these non-linear effects. At the same time, it follows from numerous studies performed with this type of the dynamical kernel that there is a clear need of extensions beyond its configuration complexity. Indeed, as we have shown in Section \ref{Propagators}, the response function should describe, in principle, the complete excitation spectrum of a quantum many-body system, while in numerical applications of the conventional PVC-TBA method with $1p1h\otimes phonon$ configurations the deficiency of the level density is quite obvious after the comparison to data \cite{EndresLitvinovaSavranEtAl2010,Oezel-TashenovEndersLenskeEtAl2014}. 


In this context it becomes interesting to look at the terms which are missing in the conventional PVC-TBA, but present in the EOM of Section \ref{EOM2}.
One can see after performing the matching shown in Fig. \ref{SE2_PVC_ph} that the singly-correlated terms  of $K^{(r;11)}$, such as b), e) and f) of Fig. \ref{SE2irrc} and their counterparts from $K^{(r;12)}$, $K^{(r;21)}$, and $K^{(r;22)}$ as well as all doubly correlated terms of Fig. \ref{SE2irrcc} and their respective 
counterparts are not included in the original PVC-TBA. A way to include the terms of such structure to some extent was formulated in \cite{Tselyaev2007} as a two-phonon version of the approach based on the EOM \cite{Schuck1976}, where both phonon correlation functions in the dynamical kernel of the type (\ref{Gccirr}) were taken in the random phase approximation. 
Such a possibility has been also discussed in the context of a formal comparison between the models where the internal Green functions in the dynamical kernel are confined by only the uncorrelated terms of the type (\ref{G0irr}), the singly-correlated terms (\ref{Gcirr}) and those of Eq.  (\ref{Gccirr}) \cite{Olevano2018}.
In the PVC-TBA framework, the two-phonon approach was implemented numerically for the dipole nuclear response and analyzed in Refs. \cite{LitvinovaRingTselyaev2010,LitvinovaRingTselyaev2013}. In the latter work, a quantitative comparison between the results obtained within the approximations of Eq. (\ref{Phikernel}) (the kernel of Fig. \ref{SE2_PVC_ph})  and Eq. (\ref{Gccirr}) with two coupled RPA phonons has been made, which showed some improvements of the description of nuclear strength functions, when the doubly-correlated dynamical kernel is used. However, the major deficiencies, which occur because of the too small overall number of excited states in the resulting response function (\ref{respspec}) remained unsolved, because this approach does not go beyond the correlated $2p-2h$ configurations.

\section{Doubly correlated dynamical kernel beyond $2p-2h$ configurations}
\label{PVCext}

The direct way to overcome the latter problem is to calculate the dynamical kernel beyond $2p-2h$ configurations. Indeed, the  Green function of Eq. (\ref{Gccirr}) entering the dynamical kernel consists of coupled two-fermion correlation functions, however, it is not limited by any approximation to these correlation functions. Ideally, they should provide a convergent solution of Eq. (\ref{Dyson3}) and contain, in principle, the entire excitation spectrum.  We have already mentioned two attempts to go beyond the conventional PVC-TBA kernel of Eq. (\ref{Phikernel}) and Fig. \ref{Phi_PVC}, which were investigated in Refs. \cite{LitvinovaRingTselyaev2010} and \cite{Tselyaev2018}. The latter uses the phonons computed beyond RPA as solutions of the full Eq. (\ref{respdir}) and the former employs the kernel of Eq. (\ref{Gccirr}) with two RPA response functions. As mentioned above, both models still left some room for improvement. 

\begin{figure*}[ptb]
\vspace{-0.5cm}
\begin{center}
\includegraphics*[scale=0.95]{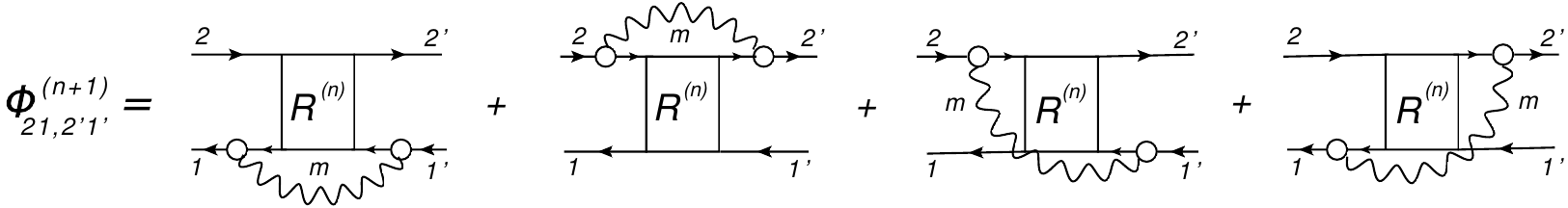}
\end{center}
\vspace{-0.5cm}
\caption{Dynamical kernel of the generalized PVC-TBA,  according to Eqs. (\ref{phires2ex},\ref{respn}).}
\label{Phiiter}
\end{figure*}

\begin{figure}[ptb]
\vspace{-0.2cm}
\begin{center}
\includegraphics*[scale=0.49]{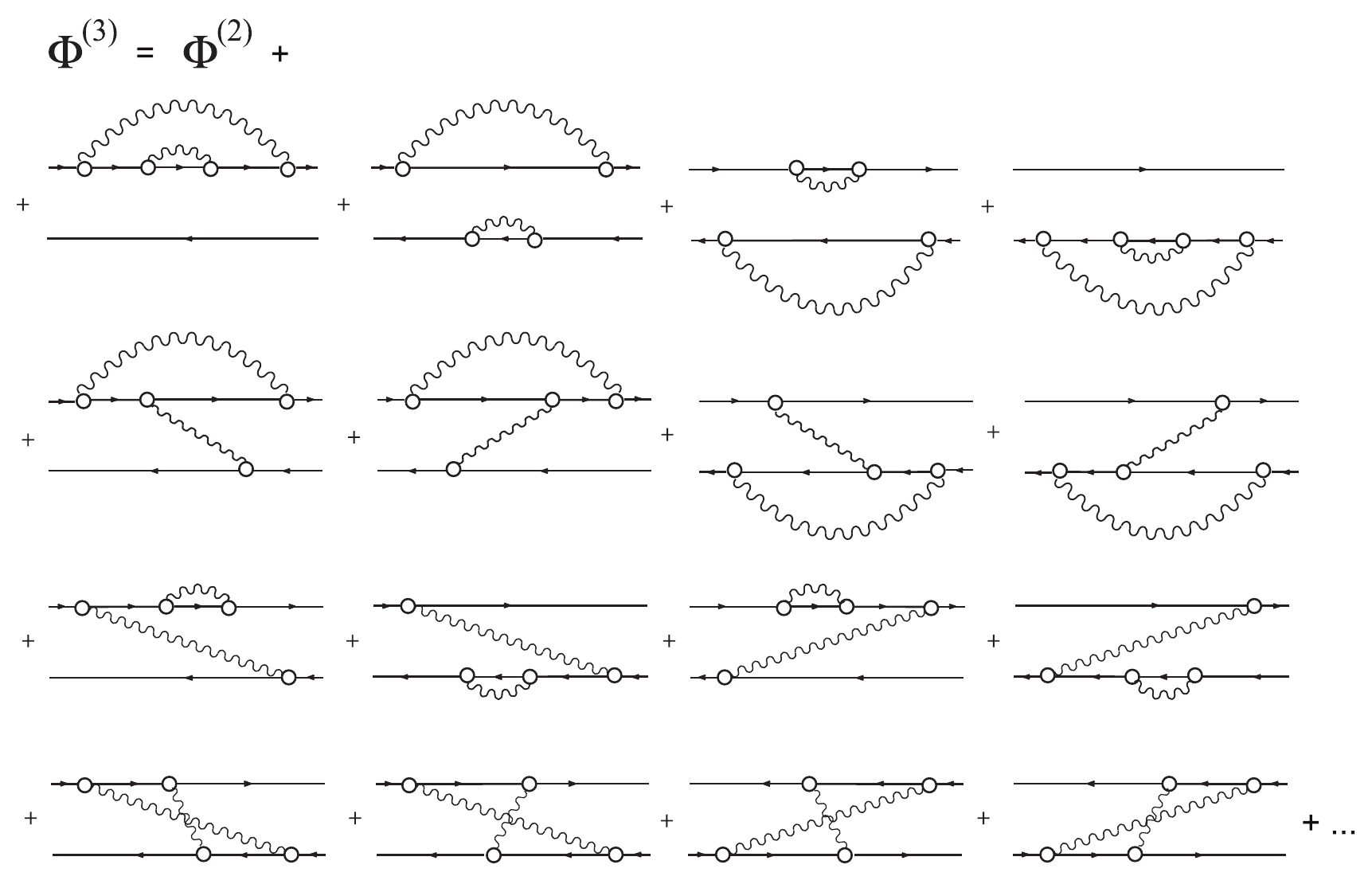}
\end{center}
\vspace{-0.5cm}
\caption{The simplest time-ordered $2q\otimes 2phonon$ diagrams taken into account in 
EOM/RQTBA$^3$. The ellipsis stands for multiple PVC exchange and self-energy contributions as well as for the correlated particle-hole configurations in the internal particle-hole propagators.}
\label{3ph}%
\end{figure}

A different strategy has been outlined in Ref. \cite{Litvinova2015} as a generalized PVC-TBA, which proposes an iterative algorithm for the solution of Eq. (\ref{respdir}). After determining the RPA phonons, like in the conventional PVC-TBA method, we can solve the equation of motion (\ref{respdir}) for the particle-hole response function $R(\omega)$,
which contains configurations of the $ph\otimes phonon$  or two-quasiparticles coupled to phonon ($2q\otimes phonon$) type, for various multipolarities, and after that reiterate the dynamical kernel 
as follows:
\be 
{\Phi}^{(3)\eta}_{12,34} (\omega) =
\sum\limits_{56,{5^{\prime}}{6^{\prime}}\nu}
\zeta^{\,\nu\eta}_{12;56}
R^{\eta}_{56,5^{\prime}6^{\prime}} (\omega -
\eta\,\Omega_{\nu})\zeta^{\,\nu\eta
*}_{34;5^{\prime}6^{\prime}},
\label{phires3} \ee
where the quantities $\zeta$ are the phonon vertex matrices:
\bea
\zeta^{\,\nu\eta}_{12,56} =
\delta^{\vphantom{(+)}}_{15}\,g^{\nu(\eta)}_{62}
-g^{\nu(\eta)}_{15}\delta^{\vphantom{(+)}}_{62},
%
%
\eea
%
so that the resulting four terms 
\bea {\Phi}^{(3)\eta}_{12,34} (\omega) &=&
\sum\limits_{{1'}{3'}\nu}g^{\nu(\eta)}_{1{1'}}R^{\eta}_{{1'}2,{3'}4}(\omega -
\eta\,\Omega_{\nu})
g^{\nu(\eta)\ast}_{3{3'}} + \nonumber\\
&+&
\sum\limits_{2'{4'}\nu}g^{\nu(\eta)}_{{2'}2}R^{\eta}_{1{2'},3{4'}}(\omega -
\eta\,\Omega_{\nu})
g^{\nu(\eta)\ast}_{{4'}4} - \nonumber \\
&-&
\sum\limits_{{1'}{4'}\nu}g^{\nu(\eta)}_{1{1'}}R^{\eta}_{{1'}2,3{4'}}(\omega -
\eta\,\Omega_{\nu})
g^{\nu(\eta)\ast}_{{4'}4} - \nonumber\\
&-&
\sum\limits_{{2'}{3'}\nu}g^{\nu(\eta)}_{{2'}2}R^{\eta}_{1{2'},{3'}4}(\omega -
\eta\,\Omega_{\nu})
g^{\nu(\eta)\ast}_{3{3'}} 
\label{phires2ex} 
\eea
correspond to the four diagrams in 
Fig. \ref{Phiiter} with $n=2$. The index $\eta=\pm 1$ denotes upper and lower components in the quasiparticle space, see Refs. \cite{Litvinova2015,Tselyaev2007} for more details. 
Thus, the amplitude $\Phi^{(3)}$ contains the contributions of
the graphs shown in Fig. \ref{3ph} in all orders with respect to the internal propagators. However, the proposed procedure  allows calculations of their contribution without
explicit evaluation of the diagrams of Fig. \ref{3ph}. It is straightforward
to see that these terms contain $2q\otimes 2phonon$ configurations
and thereby represent the next, three-particle-three-hole ($3p3h$),
level of the configuration complexity, as compared to the original PVC-TBA, which in this implementation 
for superfluid systems in the relativistic framework was named relativistic quasiparticle time blocking approximation (RQTBA).
Thus, we adopt EOM/R(Q)TBA$^3$ as a working name for the approach of Eq. (\ref{phires2ex}).
The amplitude $\Phi^{(3)}$ forms the dynamical interaction kernel for the
correlated particle-hole propagator $R^{(e)(3)}$ taking into account 3p3h
correlations (to be compared to $R^{(e)(2)} \equiv R^{(e)}$ which includes
2p2h ones):
\bea
R^{(e)(3)\eta}_{12,34}(\omega) &=& {\tilde
R}^{(0)\eta}_{12}(\omega)\delta_{13}\delta_{24} + \nonumber \\
&+& {\tilde R}^{(0)\eta}_{12}(\omega)
\sum\limits_{56}{\Phi}^{(3)\eta}_{12,56}(\omega)
R^{(e)(3)\eta}_{56,34}(\omega). \nonumber \\
\label{respe3}
\eea
%
Analogously to the conventional PVC-TBA and RQTBA, the remaining equation for the full
response function is formulated in terms of the correlated
 particle-hole propagator $R^{(e)(3)}(\omega)$ as a free term and the static effective
interaction as a kernel: \bea
R_{12,34}^{(3)\eta\eta^{\prime}}(\omega) =
R^{(e)(3)\eta}_{12,34}(\omega)\delta^{\eta\eta^{\prime}} +
\sum\limits_{56}
R^{(e)(3)\eta}_{12,56}(\omega)\times \nonumber\\
\times\sum\limits_{78\eta^{\prime\prime}}
\bigl[{\tilde V}_{56,78}^{\eta\eta^{\prime\prime}}- {\Phi}^{(3)\eta}_{56,78}(0)\delta^{\eta\eta^{\prime\prime}}\bigr]R_{78,34}%
^{(3)\eta^{\prime\prime}\eta^{\prime}}(\omega),\nonumber\\
\label{respdir3}%
\eea
where the superscript '$(3)$' indicates that this response function
takes into account $3p-3h$ configurations. Analogously to the
$2q\otimes phonon$ RQTBA, the subtraction of the amplitude
$\Phi^{(3)}$ at zero frequency from the effective interaction in Eq.
(\ref{respdir3}) eliminates the double counting of the static
contribution of the phonon coupling effects. It also improves convergence of the sums in Eq. (\ref{phires2ex}) with respect to the phonon and quasiparticle energies of the intermediate complex configurations
\cite{Tselyaev2013}. 
Finally, the observed strength function is a (model-independent) double convolution of the response function with the
external field operator:
\be
S(E) = -\frac{1}{2\pi}\lim_{\Delta \to 0} \text{Im} \sum\limits_{1234} V^{(0)\eta}_{12}R^{\eta\eta^{\prime}}_{12,34}(E+i\Delta) V^{(0)\eta^{\prime}\ast}_{34}.
\label{strength}
\ee
 The imaginary part $\Delta$ of the energy variable in Eq. (\ref{strength}) is often used in the numerical calculations as a smearing parameter. 
It provides a smooth envelope of the strength distribution and averages over complex configurations which are not taken into account explicitly.
This parameter can also mimic the experimental resolution of the data invoked for the comparison. 

Calculations presented below  were performed within the approach of Eqs. (\ref{respdir},\ref{phires3}-\ref{strength}), however, in 
principle, the iteration procedure can be continued until convergence is achieved. The initial steps are characterized as follows:
\bea
\Phi^{(1)}_{12,34}(\omega) &=& 0 \nonumber\\
\Phi^{(2)}_{12,34}(\omega) &=& \Phi_{12,34}(\omega)\nonumber\\
R^{(e)(1)}_{12,34}(\omega) &=& {\tilde
R}^{(0)}_{12}(\omega)\delta_{31}\delta_{42}\nonumber\\
R^{(e)(2)}_{12,34}(\omega) &=& R^{(e)}_{12,34}(\omega)\nonumber \\
R^{(1)}_{12,34}(\omega) &=& {\tilde
R}^{(0)}_{12}(\omega)\delta_{31}\delta_{42}\nonumber \\
R^{(2)}_{12,34}(\omega) &=& R_{12,34}(\omega) 
\eea 
Then, the chain of operator equations for the correlated particle-hole propagator
$R^{(e)(n)}$, the phonon coupling amplitude $\Phi^{(n)}$ and the response
function $R^{(n)}$ reads
\bea 
R^{(e)(1)}(\omega) &=& {\tilde R}^{(0)}(\omega) \nonumber \\
R^{(e)(n)}(\omega) &=& {\tilde R}^{(0)}(\omega) + {\tilde R}^{(0)}(\omega)
\Phi^{(n)}(\omega) R^{(e)(n)}(\omega) \nonumber \\
 {\Phi}^{(n)} (\omega) &=&
\sum\limits_{\nu}
\zeta^{\nu} R^{(n-1)} (\omega -\Omega_{\nu})
\zeta^{\nu\ast} 
\label{Phi_gen}
\nonumber \\
R^{(n)}(\omega) &=& R^{(e)(n)}(\omega) + R^{(e)(n)}(\omega)
\Bigl[{\tilde V} - \Phi^{(n)}(0) \Bigr]R^{(n)}(\omega),\nonumber\\
\label{respn}
%
%
%
\eea
where the index '$(n)$' with $n\geq 2$ indicates the iteration step. 
 The phonon vertex matrices $\zeta^{\nu}$ can be, in principle, recalculated on each step using the spectral representation of the response function (\ref{respspec}) while, as discussed above, the study of Ref. \cite{Tselyaev2018} showed that the effect of such corrections may be small. 
In the last equation (\ref{respn}) $\Phi^{(n)}(0)$ is subtracted from the static interaction kernel to avoid double counting effects for the case of calculations based on effective interactions. In ab-initio calculations this subtraction is absent. The proposed iterative method allows one to obtain the contributions of the three, four , and higher PVC-loop diagrams 
without calculating them explicitly. The generalized dynamical kernel is illustrated diagrammatically in Fig. \ref{Phiiter}. 

There are various factors which favor the convergence of the iterative procedure with respect to the internal response functions (\ref{respn}). 
In an analogy to Eliashberg theory for electron-phonon interaction in solids \cite{Eliashberg1960},
in finite nuclei a certain smallness is contained in the phonon vertices which enter the dynamical kernel, that serves as a foundation for the phenomenological PVC approaches \cite{BohrMottelson1975}. 
 Indeed, in spherical nuclei the PVC vertices enter the dynamical kernel in combinations ${\bar g}^{\nu}_{12} \sim \langle 1 ||g^{\nu}|| 2\rangle /(\Omega_{\nu}\sqrt{2j_1+1})$ \cite{BohrMottelson1975,TselyaevPhD,KamerdzhievTertychnyiTselyaev1997}. Thus, on each iteration, a factor of $\sim|{\bar g}^{\nu}|^2 \ll 1$
 suppresses higher-order contributions.
 The equation for the response function is commonly solved for certain angular momentum and parity, which correspond to those quantum numbers of the external field. For this purpose, Eqs. (\ref{phires3}-\ref{respn}) have been formulated in the coupled form in Ref. \cite{Litvinova2015}, where it is shown that the kernel $\Phi^{(n)}(\omega)$ contains a geometrical factor which enters $\Phi^{(n)}(\omega)$ as a multiplier on each iteration. This factor shows up as a product of two 6j-symbols that contains a smallness originated from the coupling of the angular momentum of the external channel with the angular momenta of the two internal correlation functions.
As it is shown below, in the numerical calculations the fragmentation effects generated by the $2q\otimes phonon$ configurations on the two-quasiparticle R(Q)RPA states are considerably stronger than those caused by the subsequent addition of the $2q\otimes 2phonon$ configurations, that can serve as a clear indication of convergence.

\section{Numerical details and results}
\label{Results}

The EOM/RQTBA$^3$ model, which was originally proposed in Ref. \cite{Litvinova2015} and briefly revisited in the previous section, has been implemented numerically and tested in calculations of nuclear dipole response. This type of response is known to dominate nuclear spectra and  associated with the largest corpus of available experimental data. The original as well as the evaluated data on the giant dipole resonance (GDR), the high-energy part of the dipole response, can be found in Ref. \cite{nndc}. Typically the GDR above the particle emission threshold and its low-energy counterpart are measured with different methods, although the newer techniques, such as inelastic proton scattering \cite{TamiiPoltoratskaNeumannEtAl2011,PoltoratskaFearickKrumbholzEtAl2014}, allow for unified high-quality measurements. The knowledge about dipole strength distributions in nuclei is crucial for many applications in nuclear sciences and astrophysics, see more details in Refs. \cite{PaarVretenarKhanEtAl2007,SavranAumannZilges2013}. Overall, testing both the high- and low-energy dipole strength distributions is the best benchmark for newly developed many-body models. 

As in the base RQTBA model \cite{LitvinovaRingTselyaev2008}, we implement a multi-step parameter-free calculation scheme, but with a few more steps now: (i)  the closed set of the relativistic mean field (RMF) Hartree-Bogoliubov equations \cite{SerotWalecka1986a,Ring1996,VretenarAfanasjevLalazissisEtAl2005} are solved using the NL3 parametrization of Refs. \cite{BogutaBodmer1977,Lalazissis1997} for the non-linear sigma-meson model and monopole pairing forces adjusted to reproduce empirical pairing gaps. The obtained single-particle Dirac spinors and the corresponding single-nucleon energies,  being the eigenstates and eigenvalues of the relativistic mean-field Hamiltonian, formed the working basis for subsequent calculations of the response, where the same effective NL3 meson-exchange interaction $\tilde V$ is also adopted;
(ii) the relativistic quasiparticle random phase approximation (RQRPA) equation \cite{PaarRingNiksicEtAl2003}, which is equivalent to Eq. (\ref{respdir}) without the dynamical kernel, is solved to obtain the phonon vertices $g^{\nu}$ and their frequencies $\Omega_{\nu}$. 
The set of phonons with the $J_{\nu}^{\pi_{\nu}}$ = 2$^+$, 3$^-$, 4$^+$, 5$^-$, 6$^+$ and frequencies $\Omega_{\nu} \leq$ 15 MeV,
together with the RMF single-particle basis, forms the $2q\otimes phonon$ configurations for the particle-phonon coupling amplitude ${\Phi}(\omega)$,  while the phonon space was additionally truncated according to the values of the reduced transition probabilities of the corresponding electromagnetic transitions: the modes with the values of the reduced transition probabilities $B(EL)$ less than 5\% of the maximal one (for each $J_{\nu}^{\pi_{\nu}}$) were neglected. These are the common truncation criteria for the PVC models based on the RMF, see, for instance \cite{Litvinova2016}, where a convergence study was presented;
(iii) Eq. (\ref{respdir}) for the response function is solved in the truncated configuration space, which includes excitations within the energy window of interest 0-25 MeV, for spins and parities $J^{\pi}$ = 0$^{\pm}$ - 6$^{\pm}$ (in this first application we, however, neglected the static part of the kernel of Eq. (\ref{respdir}), just to study the power of the correlated internal particle-hole propagator to induce additional fragmentation - in further applications the full kernel will be taken into account);
(iv) the obtained response functions, together with the previously obtained phonon characteristics, are used to  compute the dynamical kernel of Eq. (\ref{phires2ex}); (v) the correlated particle-hole propagator with $J^{\pi} = 1^-$ is obtained according to Eq. (\ref{respe3}) in the same truncated configuration space; (vi) the full dipole response function  is computed by solving Eq. (\ref{respdir3}) in the momentum space, as described in \cite{LitvinovaRingTselyaev2007,LitvinovaRingTselyaev2008}; (vii) finally, the strength function 
is found according to Eq. (\ref{strength}) with the external field operator of the electromagnetic dipole (EME1) character:
\bea
V^{(0)EM}_{1M} &=& \frac{eN}{A}\sum\limits_{i=1}^Z r_iY_{1M}({\hat{\bf r}}_i) - \frac{eZ}{A}\sum\limits_{i=1}^N r_iY_{1M}({\hat{\bf r}}_i),  \nonumber \\
\label{opE1}
\eea 
where $Z$ and $N$ are the numbers of protons and neutrons, respectively, $A = N+Z$, and $e$ is the proton charge. The sums in Eq. (\ref{opE1}) are performed over the corresponding nucleonic degrees of freedom.  While the response to this operator is often attributed to electromagnetic probes,  hadronic probes are associated predominantly \cite{LanzaVitturiLitvinovaEtAl2014} with the response to the isoscalar dipole (ISE1) operator 
\bea
V^{(0)IS}_{1M} &=& \sum\limits_{i=1}^A (r^3_i - \eta r_i)Y_{1M}({\hat{\bf r}}_i),  \nonumber \\
\label{opISE1}
\eea 
where $\eta = 5\langle r^2\rangle/3$ and the second term in the brackets eliminates the spurious translational  mode \cite{Garg2018}.

\begin{figure}[ptb]
\begin{center}
\includegraphics[scale=0.36]{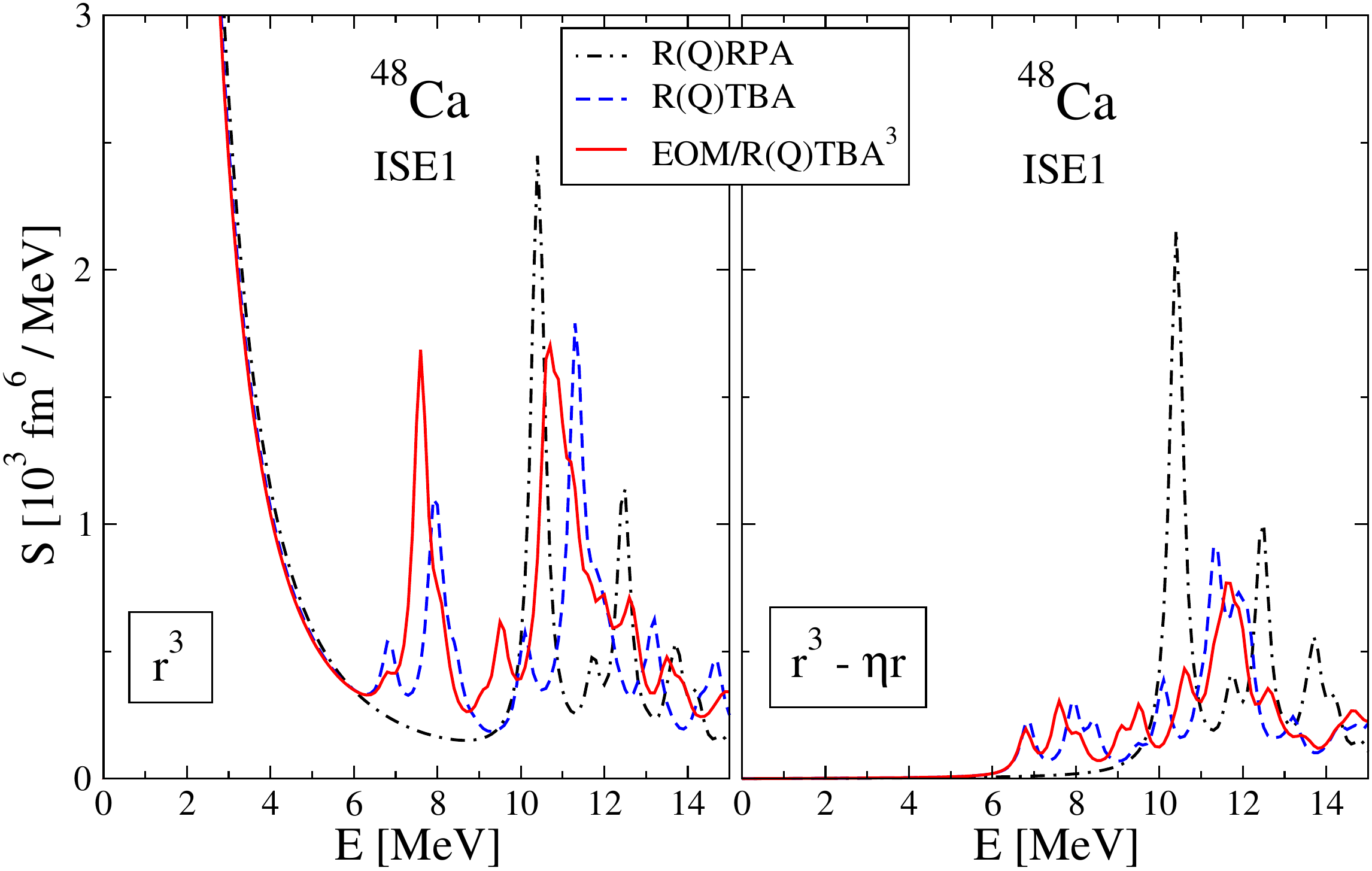}
\end{center}
\vspace{-0.5cm}
\caption{The low-energy isoscalar dipole strength distributions in $^{48}$Ca calculated in R(Q)RPA, R(Q)TBA and  EOM/R(Q)TBA$^3$ with $\Delta$ = 200 keV for the uncorrected (left) and corrected (right) for the spurious translational mode operators of Eq. (\ref{opISE1}).} 
\label{Translational}
\end{figure}

A sufficiently large quasiparticle basis in both Fermi (particle) and Dirac (antiparticle) sectors should be used in solving Eqs. (\ref{respdir}, \ref{respdir3}).
Although the dynamical kernels $\Phi(\omega)$, which induce fragmentation of two-quasiparticle configurations, may be cut off outside a window confined by the energy of interest, the static kernel $\tilde V$ has to be included in the complete or nearly complete two-quasiparticle space \cite{LitvinovaRingTselyaev2008}. The latter kernel is responsible for the correct location of the simple (R(Q)RPA) modes of the strength distribution and associated mainly with the medium-range correlations, while the former kernels introduce the long-range effects causing the redistribution of the strength.
Here the completeness means that  the two-quasiparticle basis, in which Eqs. (\ref{respdir}, \ref{respdir3}) are solved, should include all the single-quasiparticle states which participate in the RMF self-consistent procedure. In our case the basis spans the single-quasiparticle states with the angular momenta up to  $41/2$ in both Fermi and Dirac sectors, the same range where the parameters of the Lagrangian have been fitted. The respective energy range of the two-quasiparticle excitations is confined by $\sim$250 MeV in the Fermi sector and by $\sim$-1950 MeV in its Dirac counterpart. Keeping the complete two-quasiparticle basis in 
Eqs. (\ref{respdir}, \ref{respdir3}) guarantees full self-consistency, in particular, the proper decoupling of the dipole translational mode from the intrinsic dipole excitations in R(Q)RPA \cite{RingSchuck1980,PaarRingNiksicEtAl2003}. This fact can be verified numerically, for instance, by calculating the isoscalar dipole strength distribution produced by the response to the operator of Eq. (\ref{opISE1}), where the radial form factor is corrected for the center of mass motion. Without this correction one typically sees a dominant peak located at zero energy, 
as we show in the left panel of Fig. \ref{Translational} in comparison to the right panel, where the response to the corrected isoscalar operator is displayed, for $^{48}$Ca. One can see that in the latter case the zero-energy translational mode is suppressed. 
Moreover, this property is kept in the extended R(Q)TBA and EOM/R(Q)TBA$^3$ models - indeed, the subtraction procedure of Eqs. (\ref{subtraction-procedure}, \ref{respdir3}) leads to the purely R(Q)RPA kernel in the $\omega \to 0$ limit.  This feature is known since early implementations of the time blocking method with the subtraction \cite{LitvinovaTselyaev2007}. 
However, in R(Q)TBA and EOM/R(Q)TBA$^3$ the translational mode may be fragmented because, like the physical states, it can be coupled to the phonons. Although, due to the subtraction procedure, the main peak of the translational mode remains at zero energy, its fragments may spread around it. As the excited states calculated with the uncorrected and corrected isoscalar dipole operators look different in both the R(Q)TBA and EOM/R(Q)TBA$^3$, in the present implementations these models do not guarantee complete decoupling of the spurious mode from the physical states. A solution to this problem was proposed in Ref. \cite{Tselyaev2014}  in the form of a projection operator applied to the dynamical kernel, that prevents the coupling of complex configurations to the translational mode and, thus, removes its admixture to the physical states. Performing this transformation is beyond the scope of the present article, but will be considered in future work.
The sensitivity of our present implementation to the two-quasiparticle basis incompleteness was inspected and revealed that an energy cut-off of this basis by $\sim$100 MeV in the Fermi sector and $\sim$-1800 MeV in its Dirac counterpart does not introduce noticeable  changes in the excitation spectra, that can be used for more economical calculations.

\begin{figure}[ptb]
\begin{center}
\includegraphics[scale=0.37]{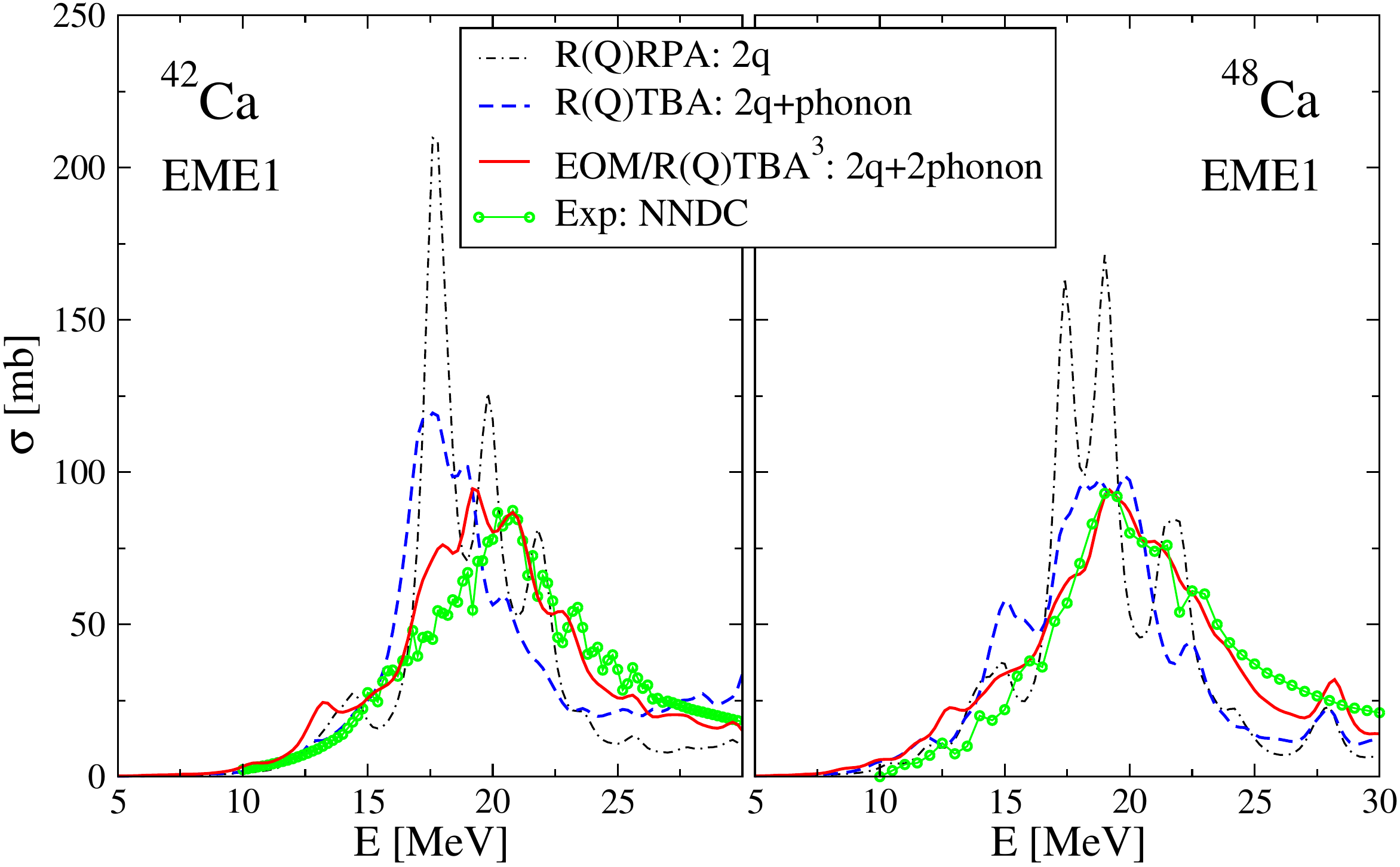}
\end{center}
\caption{Giant dipole resonance in $^{42,48}$Ca nuclei calculated within R(Q)RPA, R(Q)TBA and EOM/R(Q)TBA$^3$ with $\Delta$ = 500 keV, in comparison to experimental data of Ref. \cite{Erokhova2003,nndc}. } 
\label{42-48-Ca}
\end{figure}

The results of calculations for the electromagnetic dipole response in $^{42,48}$Ca are displayed in Fig. \ref{42-48-Ca}. The strength distribution obtained within EOM/RQTBA$^3$  (red solid curves) is plotted against the results of RQRPA (black dot-dashed curves) and RQTBA (blue dashed curves) and compared to experimental data (green curves and circles) of Ref. \cite{nndc} in terms of the dipole photoabsorption cross section
\begin{equation}
\sigma_{E1}(E)={\frac{{16\pi^{3}e^{2}}}{{9\hbar c}}}E~S_{E1}(E),
\end{equation}
i.e. with an additional energy factor in front of the strength distribution. These cross sections were investigated in the RQTBA framework in Ref. \cite{EgorovaLitvinova2016}  in order to establish the role of the $2q\otimes phonon$ configurations in the formation of the spreading width of the GDR in the chain of calcium isotopes. We found that these configurations indeed result in the significant spreading width improving considerably the agreement to data as compared to RQRPA, while the latter approach provides a very good description of the GDR's position and the total strength. However, although we used a fairly large model space of the $2q\otimes phonon$ configurations, the total width of the RQTBA dipole strength distribution still underestimates its experimental value.  Another major shortcoming was found on the high-energy shoulder of the GDR above its centroid, where the cross sections were systematically underestimated. A similar situation was reported in Ref. \cite{Tselyaev2016}, where a systematic downshift of the non-relativistic QTBA strength distributions, as compared to the RPA ones, was revealed in calculations with various Skyrme forces for monopole, dipole and quadrupole resonances in both light and heavy nuclei.  Now, when EOM/RQTBA$^3$ with more complex $2q\otimes 2phonon$ configurations becomes available, we can see that it shows the potential of resolving those problems. Indeed, Fig. \ref{42-48-Ca} shows that the new configurations present in EOM/RQTBA$^3$ induce a stronger fragmentation of the GDR and its additional spreading toward both higher and lower energies. Another new effect is a relatively small, but a visible shift of the main peak toward higher energies. These features of changing the high-energy behaviour of the strength distributions can be directly related to the appearance of the new higher-energy complex configurations and, consequently, the higher-energy poles in the resulting response functions, that has the power to rearrange the energy balance of the overall strength distribution. This change is, however, modest enough to conserve the energy-weighted sum rule (EWSR) which varies quantitatively only within a few percent as compared to the RQRPA and RQTBA ones in the investigated energy region.

 As it has been shown in Ref. \cite{Tselyaev2007}, in the resonant time blocking approximation, where the high-energy behaviour of the dynamical kernel is $\Phi(\omega) \sim 1/\omega $ at $\omega \to \infty$, the EWSR is exactly the same as in the (Q)RPA. In other words, the presence of the dynamical kernel of Eq. (\ref{Phikernel}) with such a behaviour does not violate the EWSR of RQRPA. It can be seen from Eqs. (\ref{phresp},\ref{phires2ex}) that the iterations of the dynamical kernel do not change its high-energy asymptotics as it is determined by the internal particle-hole propagators satisfying the model-independent Eq. (\ref{phresp}).   
The presence of subtraction in Eq. (\ref{respe3}), however, changes the static part of the kernel and, thus, introduces a small violation of the EWSR as in the conventional time blocking approximation of Eqs. (\ref{subtraction-procedure},\ref{resp-subtr}) \cite{LitvinovaTselyaev2007,LitvinovaRingTselyaev2007,LitvinovaRingTselyaev2008}. As it is known in the literature, in extended models with the ground state correlations caused by PVC another sum rule violation occurs,  as well as its restoration \cite{AdachiLipparini1988,Tselyaev2007}. The formalism of Ref. \cite{DelionSchuckTohyama2016} proves that the self-consistent RPA (SCRPA) and possibly also the second SCRPA keeps all desirable qualities of the standard RPA intact, that includes also the sum rules.
The complex ground state correlations are not included in the present work, but were recently addressed in other developments \cite{Robin2019}.

The obtained change of the high-energy behaviour of the strength distributions in EOM/RQTBA$^3$ may also provide some new arguments to the discussion of fitting the nuclear energy density functionals. In particular, using the position of the GDR as one of the reference observables may become difficult because of its model dependence.

The low-energy behaviour of the dipole strength distribution in nuclei has been a topic of an intense research during the last couple of decades. A strong astrophysics connection of this type of strength, in particular, to the r-process nucleosynthesis, attracted much of interest from both experimental and theoretical sides. In this context, both RQRPA and RQTBA models were examined for their performance and  
for their potential of describing the low-energy dipole strength associated with the pygmy dipole resonance (PDR), or the soft dipole mode \cite{Litvinova2007,LitvinovaRingTselyaevEtAl2009,LitvinovaLoensLangankeEtAl2009,EndresLitvinovaSavranEtAl2010,EndresSavranButlerEtAl2012,MassarczykSchwengnerDoenauEtAl2012,LitvinovaRingTselyaev2013,Oezel-TashenovEndersLenskeEtAl2014}. Similarly to the GDR case, RQTBA provided an improved description of the PDR showing a considerably richer spectral structure because of the fragmentation effects induced by the $2q\otimes phonon$ and $2 phonon$ configurations. However, in many cases it became clear that even calculations with quite large model spaces reveal a deficiency of configuration complexity that lead to too low level density in the discrete and quasicontinuum energy sectors below the particle emission threshold. Therefore, it is interesting to examine the newly-developed EOM/RQTBA$^3$ for its performance in the low-energy regime. 

\begin{figure}[ptb]
\begin{center}
\includegraphics[scale=0.38]{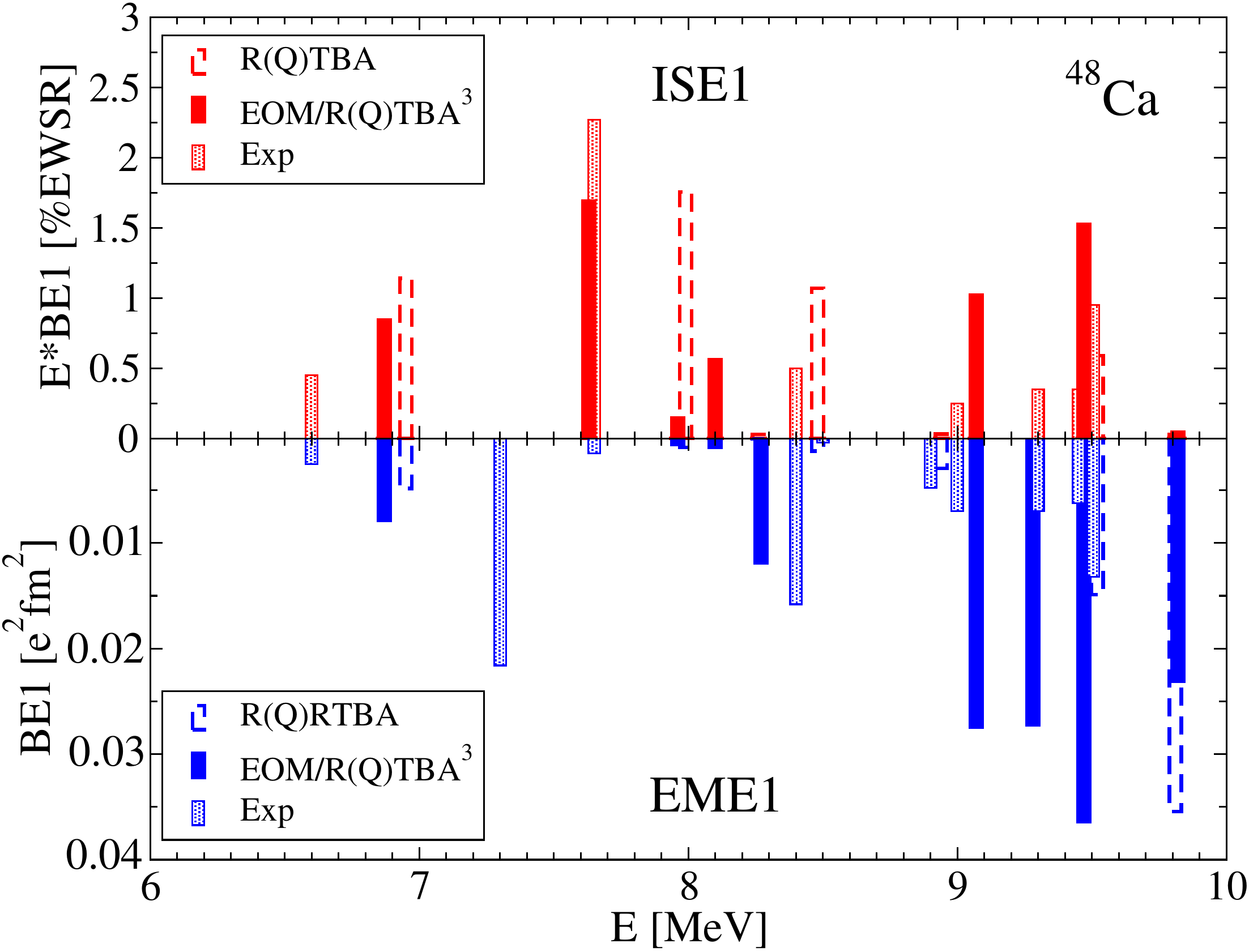}
\end{center}
\vspace{-0.5cm}
\caption{The low-energy dipole strength of $^{48}$Ca nucleus calculated within R(Q)RPA, R(Q)TBA and EOM/R(Q)TBA$^3$. The reduced transition probabilities are shown for  the isoscalar (top) and electromagnetic (bottom) dipole operators in comparison to experimental data of Ref. \cite{Derya2014}.  } 
\label{48Ca_emis}
\end{figure}

 Fig. \ref{48Ca_emis} illustrates the performance of $R(Q)TBA$ and EOM/R(Q)TBA$^3$ in the description of the isoscalar (top) and isovector (bottom) low-energy dipole response of $^{48}$Ca. The experimental particle emission threshold of $^{48}$Ca is $\sim$9.95 MeV \cite{nndc}, so that the observed spectrum below this energy has a discrete character. Two kinds of experiments, namely the inelastic photon scattering ($\gamma,\gamma^{\prime}$) and ($\alpha,\alpha^{\prime}\gamma$) coincidence dominated by the strong interaction have been performed and discussed in Refs. \cite{HartmannBabilonKamerdzhievEtAl2004,Derya2014}. It was found, in particular, that the sub-threshold dipole response of $^{48}$Ca does not show an isospin splitting as observed in heavier neutron-rich nuclei \cite{EndresLitvinovaSavranEtAl2010}, where the lowest-lying states are mostly isoscalar in
nature and the states at higher energy are of the mixed isoscalar-isovector character.  Here we do not intend to go into the details of the underlying structure of the  states, which can be essentially model dependent, but rather look at the potential of our many-body models to reproduce the experimental strength distribution. For this purpose, we have extracted the reduced transition probabilities from the computed strength distributions by making use the simple relationship: $B_{\nu} = \pi\Delta S(\omega_{\nu})$, which follows from Eq. (\ref{strength}) and which is valid for sufficiently small values of $\Delta$ (here we used $\Delta$ = 20 keV) in the vicinity of the peak $\omega_{\nu}$.
The first observation here is that R(Q)RPA does not produce any excited states in this energy region and gives its first solution just above 10 MeV. The inclusion of the 2q$\otimes$phonon configurations in R(Q)TBA improves the picture considerably: due to the fragmentation effects caused by the coupling of the R(Q)RPA modes to these configurations some strength comes down below the neutron threshold.  The cumulative EME1 strength below 10 MeV agrees with the experimental  data quite well as it was discussed in detail in Ref. \cite{EgorovaLitvinova2016}, however, the major fraction of the R(Q)TBA strength remains in the threshold area. On the other hand, the ISE1 strength distribution in R(Q)TBA shows the pattern which is rather close to the experimental one (here we plot the energy-weighted values in order to compare with Ref. \cite{Derya2014}). It can be also seen that adding more complex configurations within the EOM/R(Q)TBA$^3$ further improves the description:   it  (i) slightly rearranges the ISE1 strength bringing it even closer to the observed one and (ii) gives two distinct EME1 states below 9 MeV, which can be approximately matched the experimental ones, and brings more strength from the higher-energy region to the threshold area,  where also several experimental levels are found at about the same energies. The experimentally extracted BE1 values of the states in the threshold area may strongly depend on the methods employed in the data analysis \cite{Tonchev2010}, so that those BE1 values may not be a good benchmark for the theory.

\begin{figure}[ptb]
\begin{center}
\includegraphics[scale=0.35]{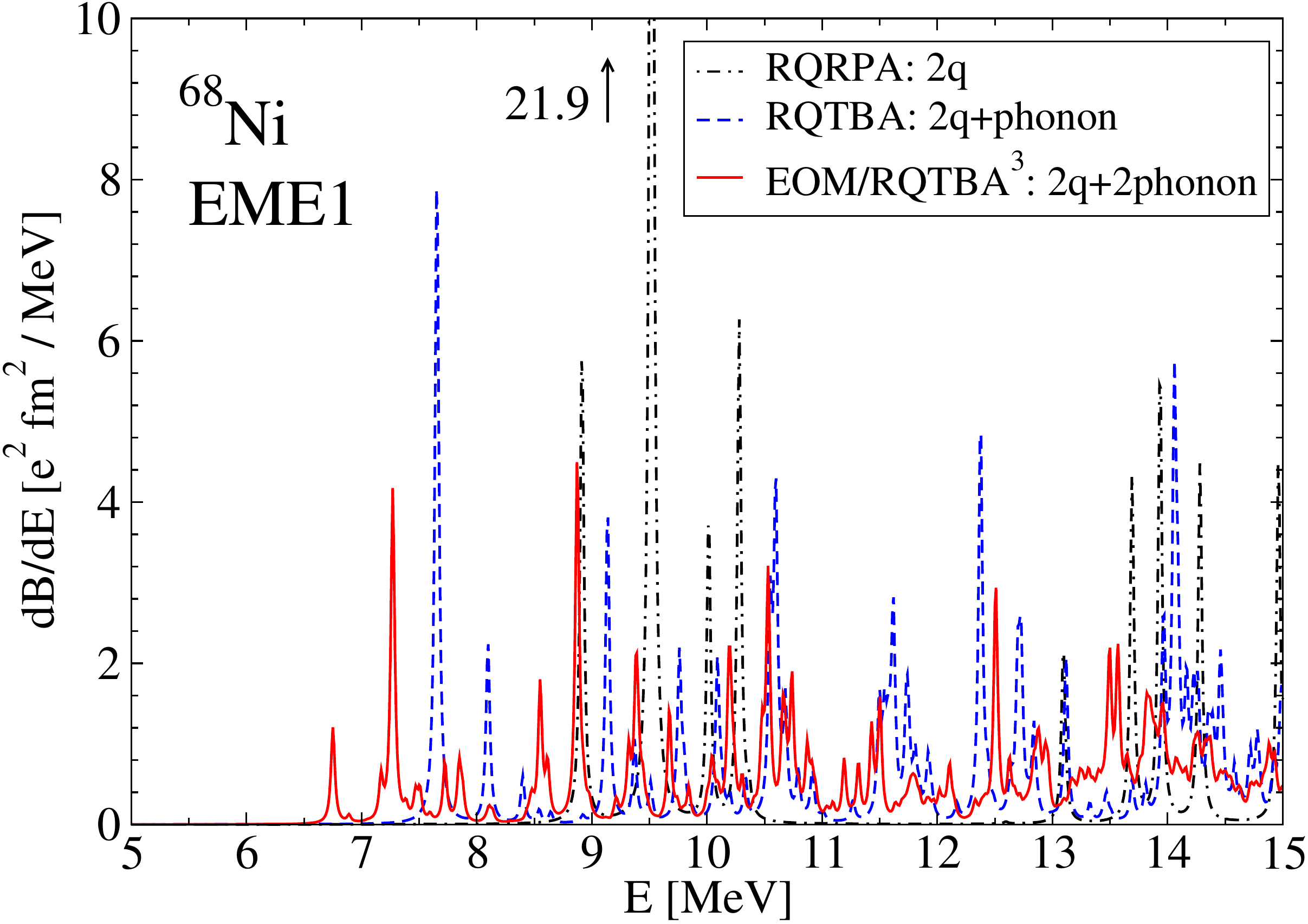}
\end{center}
\vspace{-0.5cm}
\caption{The low-energy dipole spectrum $S(E) = dB(E)/dE$ of $^{68}$Ni nucleus calculated within RQRPA, RQTBA and EOM/RQTBA$^3$ with $\Delta$ = 20 keV. } 
\label{68Ni}
\end{figure}

\begin{figure}[ptb]
\begin{center}
\includegraphics[scale=0.36]{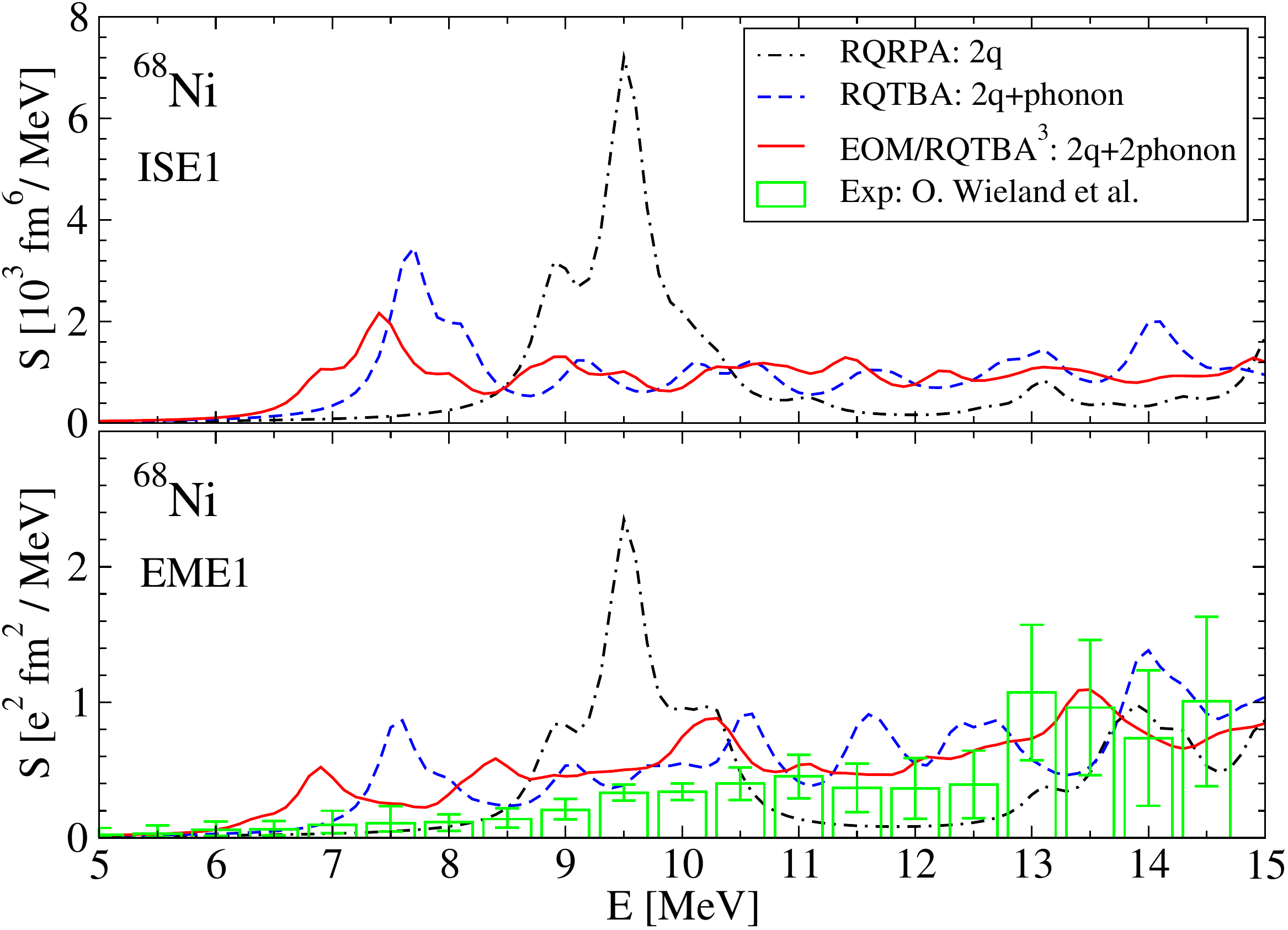}
\end{center}
\vspace{-0.5cm}
\caption{The low-energy dipole spectrum of $^{68}$Ni nucleus calculated within RQRPA, RQTBA and EOM/RQTBA$^3$ with $\Delta$ = 200 keV. Top: the isoscalar dipole strength distribution, bottom: the electromagnetic dipole strength distribution in comparison to experimental data of Ref. \cite{Wieland2018}. } 
\label{68Ni_1}
\end{figure}

Another interesting case is the low-energy dipole response of the unstable neutron-rich $^{68}$Ni, which has been investigated both experimentally \cite{Wieland2009,RossiAdrichAksouhEtAl2013,Wieland2018} and theoretically \cite{LitvinovaRingTselyaev2010,LitvinovaRingTselyaev2013,Papakonstantinou2015}. 
The results of our calculations for the low-energy dipole response of $^{68}$Ni are displayed in Figs. \ref{68Ni}, \ref{68Ni_1}. As in the case of calcium isotopes, we compare the results of the three models, RQRPA, RQTBA and EOM/RQTBA$^3$, with the same curve- and color-coding as in Fig. \ref{42-48-Ca}.  Fig. \ref{68Ni} shows the strength functions calculated with a small value of the smearing parameter $\Delta = $20 keV, thus allowing for a clear illustration of the fragmentation mechanism in a parameter-free many-body theory with an effective static interaction. In particular, one can see how the spectrum of RQTBA emerges from a relatively poor one of RQRPA, which is essentially the single strong and relatively collective state at $~9.5$ MeV. The addition of $2q\otimes phonon$ configurations of RQTBA results in the blue curve which is obviously the fragmented major RQRPA state spread over a broader energy region. Remarkably, the RQTBA strength no longer shows any clear  dominance of a single state, but is rather spread uniformly over the $7-15$ MeV energy interval. Finally, when we add the EOM/RQTBA$^3$ strength distribution with an additional higher configuration complexity $2q\otimes 2phonon$, the fragmentation effect is reinforced again. One can notice, in particular, the appearance of excited states at lower energies and the overall even more uniform strength redistribution, compared to RQTBA.  Thus, the three models with the increasing complexity of the dynamical kernel form a hierarchy which translates to the hierarchy of spectral functions with increasing richness of their fine structure, as it was predicted in Ref. \cite{Litvinova2015}. Now this purely theoretical conjecture is confirmed by direct calculations. 


 The bottom panel of Fig. \ref{68Ni_1} displays the same strength functions as Fig. \ref{68Ni}, but computed with a larger smearing parameter $\Delta =$ 200 keV, in order to confront them with the experimental data of Ref. \cite{Wieland2018}.  In contrast to the case of $^{48}$Ca, the major part of the investigated strength below 15 MeV, which is often associated with the pygmy dipole resonance and the neutron skin oscillation, lies above the neutron emission threshold that is located experimentally at $\sim$7.8 MeV \cite{nndc}. Thus, this strength forms mostly a continuous spectrum.
It can be seen that RQRPA provides a too poor description of the observed strength:  it gives a distinct peak at about 9.5 MeV while the experimental strength of Ref. \cite{Wieland2018} shows a nearly flat distribution with a slow growth toward higher energies up to 13 MeV, where it relatively sharply increases by a factor of two. RQTBA approach produces a significant improvement of the description of the strength: the main RQRPA peak is fragmented and the overall strength distribution comes out much flatter following better the experimental trend. The EOM/RQTBA$^3$, in turn, smoothes the strength distribution further improving the agreement with data. The only remaining drawback is that the total strength between 6 and 10 MeV is somewhat overestimated. Further refinement of the model should clarify whether more spreading toward lower energies can be induced by more complex configurations and the exact treatment of the continuum \cite{LitvinovaTselyaev2007,Kamerdzhiev1998} or the static effective interaction of the NL3 type,  the minimal RMF parametrization with only 6 parameters, employed for these calculations is responsible for the remaining discrepancy.

In the top panel of Fig. \ref{68Ni_1} we show the ISE1 counterpart of the low-energy dipole strength in $^{68}$Ni. Remarkably, the coarse-grain pattern of the isoscalar dipole strength is very similar to that of the electromagnetic one. A similar sharp peak appears in RQRPA at about 9.5 MeV and similar fragmentation effects are induced by the 2q$\otimes$phonon and 2q$\otimes$2phonon configurations. In the final EOM/RQTBA$^3$ calculation a relatively distinct peak at approximately 7.5 MeV remains on the background of the flat isoscalar strength distribution, that is not the case for the EME1 strength. While there is no experimental data for the ISE1 strength in  $^{68}$Ni, some theoretical studies are available. In particular, Ref. \cite{Papakonstantinou2015} provides RPA and QRPA calculations of the isoscalar dipole strength for a chain of nickel isotopes including $^{68}$Ni. In the low-energy region both QRPA based on the Gogny D1S forces and continuum RPA with the SLy4 Skyrme interaction give a dominant peak around 10.5 MeV, that agrees reasonably well with our RQRPA calculation. For the EME1 strength the authors of Ref.  \cite{Papakonstantinou2015} obtain a two-peak structure at the energies corresponding to the major and a minor peaks of their ISE1 strength distribution. However, fragmentation effects, if they were added beyond R(Q)RPA, would, probably, change those patterns, as it typically occurs in various implementations of the PVC mechanism. The insights about the exact continuum effects provided in this work are very important and point out to the necessity of an accurate continuum treatment.

Other types of interactions may be also considered in a future work. Density-dependent parametrizations of the meson-exchange interaction \cite{Typel1999,Lalazissis2005} or point-coupling \cite{Burvenich2002,Niksic2008} should provide a better performance in the description of the modes related to the symmetry energy as they imply more careful fits of the isovector sector \cite{Meng2016}. Ideally, the realization of the presented approach should be based on a 
microscopic interaction, in order to increase the predictive power. 
Numerical implementations based on 
microscopic interactions should provide a reasonable approximation to the two-body density matrix at the starting point. There can be various strategies, such as the Similarity Renormalization Group \cite{Hergert2016}, Br\"uckner G-matrix \cite{McIlroy2018,Shen2017}  or the Unitary Correlation Operator Method \cite{PapakonstantinouRoth2009,Papakonstantinou2010} 
with subsequent solution of the RPA equations and extracting the two-body densities.  
The capabilities of various potentials describing the nucleon-nucleon scattering data to successfully perform within the presented approach will be also addressed by future effort.
 
\section{Summary and outlook}
\label{summary}

In this article we revisit, compare and advance non-perturbative approaches to the quantum many-body problem. The equation of motion method is reviewed for the one-fermion and two-time two-fermion  Green functions in a strongly-correlated medium. The dynamical kernels of the final EOM's containing three- and four-body propagators are approximated by the non-perturbative cluster expansions truncated on the two-body level. The resulting EOM's form a closed set of equations for one- and two-fermion propagators, where the latter include the particle-hole, particle-particle, and hole-hole components.  

This approach  is confronted with another class of closely related methods developed originally as extensions of the Landau-Migdal Fermi-liquid theory by the particle-vibration coupling and time blocking techniques, PVC-TBA. We showed that, in fact, the latter methods employed the EOM's dynamical mass operator, whose structure can be mapped to the coupling between the single-fermion and emergent phonon degrees of freedom. These phonons are built from the correlated fermionic pairs present in the EOM's dynamical kernels and, in the simplest non-perturbative random phase approximation, acquire a character of harmonic vibrations.  To address the description of the particle-hole response the PVC-TBA method starts, in contrast to the EOM,  from the general Bethe-Salpeter equation for the four-time particle-hole propagator, however, the 
final PVC-TBA equation is reduced to the two-time dependence by a time projection method and, thereby, to the one-energy variable equation for the spectral image. Eventually, the dynamical kernel of the resonant PVC-TBA is found to be topologically equivalent to a part of the EOM's kernel  containing terms with single two-fermion correlation functions. However, while PVC-TBA is based on the effective description of the static part of the interaction kernel (which enters, in turn, its dynamical part), the EOM method provides an accurate derivation of both the static and dynamical kernels from the single underlying bare interaction. 

The insights revealed throughout this work allowed for further developing the nuclear response theory beyond the previous content and capabilities of the PVC-TBA. For this purpose,  we followed the opportunities offered by the EOM, first of all, in advancing its dynamical interaction kernel beyond the $2p-2h$ level and discussed a possible iterative algorithm on the way to a highly-accurate approach to the two-fermion correlation functions. We performed a numerical implementation of the approach with the $3p-3h$ dynamical kernel and investigated the dipole response of medium-mass nuclei $^{42,48}$Ca and $^{68}$Ni. The obtained results showed some important refinements in both high- and low-energy  sectors of the dipole response and indicated that the approach is indeed systematically improvable and converging. The possibility of a continued iterative algorithm opens the way to a highly-accurate nuclear many-body approach of the shell-model quality, but without prohibitive limitations on the excitation energy and mass.

 An important aspect of the EOM method, which is not a feature of the PVC-TBA, is its direct connection to the bare interaction between fermions. In fact, the EOM derivation of the response theory together with the theory for the one-fermion propagator is based solely on the knowledge about this interaction. In contrast, the existing PVC-TBA and its nuclear field theory analogs imply an assumption about the existence of the underlying energy density functional, which provides information about the static part of the interaction kernel. The parameters of this functional are fitted to data for bulk nuclear properties and nuclear matter and, thus, disconnected from the bare nucleon-nucleon interaction. This feature occurs to be rather a drawback, because it lowers the predictive power of the theory. Therefore, an approach based on the bare interaction, which would also include higher configurations in the dynamical kernel in a manner discussed in this work, appears as a desirable solution. Such a theory would further clarify the mechanisms of emergent collective phenomena, superfluidity and other dynamical aspects of strongly-correlated many-body systems.

\section*{Acknowledgements}
Very fruitful discussions with Hans Feldmeier, Peter Ring, Caroline Robin, and Herlik Wibowo are gratefully acknowledged. We thank Oliver Wieland for providing insights and numerical data for the experimental low-energy dipole strength in $^{68}$Ni. 
This work is supported by the US-NSF Career Grant PHY-1654379.
%

\bibliography{Bibliography_Jun2019}
\end{document}